\documentclass[%
 reprint,
 amsmath,amssymb,
 aps,
]{revtex4-2}
\usepackage{graphicx}
\usepackage{dcolumn}
\usepackage{bm}
\usepackage{xcolor}

\newcommand{\ilm}{Univ Lyon, Univ Claude Bernard Lyon 1, CNRS, Institut Lumi\`ere Mati\`ere, F-69622, VILLEURBANNE, France}
\renewcommand{\selectlanguage}[1]{}
\DeclareUnicodeCharacter{2212}{-}
\usepackage{booktabs}
\begin{document}

\preprint{APS/123-QED}

\title{Enhanced thermal conductance at interfaces between gold and amorphous silicon and amorphous silica }

\author{Julien El Hajj}
\email{julien.el-hajj@univ-lyon1.fr}
\author{Christophe Adessi}
\author{Michaël de San Feliciano}
\author{Gilles Ledoux}
\author{Samy Merabia}
\email{samy.merabia@univ-lyon1.fr}
\affiliation{\ilm}

\date{\today}

\begin{abstract}
Heat transfer at the interface between two materials is becoming increasingly important as the size of electronic devices shrinks. Most studies concentrate on the interfacial thermal conductance between either crystalline-crystalline or amorphous-amorphous materials. Here, we investigate the interfacial thermal conductance at crystalline-amorphous interfaces using non-equilibrium molecular dynamics simulations. Specifically, gold and two different materials, silicon and silica, in both their crystalline and amorphous structures, have been considered. The findings reveal that the interfacial thermal conductance between amorphous structures and gold is significantly higher as compared to crystalline structures for both planar and rough interfaces ($\approx$ 152 MW/(m$^2$K) for gold-amorphous silicon and $\approx$ 56 MW/(m$^2$K) for gold-crystalline silicon). \textcolor{black}{We explain this increase by two factors~:~the relative commensurability between amorphous silicon/silica and gold leads to enhanced bonding and cross-correlations of atomic displacements at the interface, contributing to enhance phonon elastic transmission. Inelastic phonon transmission is also enhanced due to the relative larger degree of anharmonicity
characterizing gold-amorphous silicon/silica. We also show that all the vibrational modes that participate to interfacial heat transfer are delocalized and use the Ioffe-Regel (IR) criterion to separate the contributions of propagating~(propagons) and non-propagating modes~(diffusons). In particular, we demonstrate that, while at gold-amorphous silicon interfaces elastic phonon scattering involves propagons and inelastic phonon scattering involves a mixture of propagons and diffusons, in gold-amorphous silica, all modes transmitting energy at the interface are diffusons.} This study calls for the systematic experimental determination of the interfacial thermal conductance between amorphous materials and metals.

\end{abstract}

\maketitle

\section{Introduction}
The sharp decrease in the size of electronic devices has led to a situation where the mean-free paths of energy carriers like electrons and phonons are increasingly similar to or even exceed the characteristic length scales seen in heterogeneous materials. As the size of these materials shrinks, the limited space for heat dissipation can cause localized heating effects, affecting the performance and reliability of nanoscale devices and structures. Thermal transport across interfaces is a critical factor governing effective heat transfer, which is especially true in devices with a high density of interfaces or reduced dimensions. Therefore, thermal transport across interfaces severely impedes heat dissipation, and has a significant impact on the functionality and failure point of a device \cite{shrinkdevices1,shrinkdevices2,shrinkdevices3}. \\ 
The interfacial thermal conductance (ITC) is an important thermal quantity that determines the ability to dissipate or confine energy more efficiently at interfaces between two materials. In 1941, Kapitza measured the ITC for the first time quantitatively between solid copper and liquid Helium \cite{kapitza}. As a result, in literature, ITC and "Kapitza conductance" are frequently used interchangeably. ITC can be calculated using a variety of experimental, theoretical, and computational methods. In the past few years, ITC between various metal/dielectric and dielectric/dielectric solids has been characterized thanks to the development of ultrafast measurement techniques \cite{ITCbyexp1,ITCbyexp2}. Although it is less common, ultrafast laser spectroscopy can also be used to determine the ITC between a metal and an amorphous material \cite{expITCamorphous}. However, analytical predictions of the interfacial thermal transport are extremely challenging due partly to the anharmonicities of the interatomic forces at the interface. One of the main reasons for the discrepancy between the predictions of the diffuse mismatch model (DMM) \cite{IntroDMM} and acoustic mismatch model (AMM) \cite{introAMM} and experimental measurements at room temperature, are the exclusion of these anharmonic interactions \cite{Saaskilahti_prb2014}. However, molecular dynamics (MD) simulations are now recognized as one of the most effective methods for predicting ITC \cite{introMD1,intromd2,merabia-2012} especially for amorphous interfaces, as it makes no assumptions other than the classical nature of the energy carriers, which is a reasonable assumption close to and above the Debye temperature of the softer material. \\ 
Many studies have been conducted on the investigation of ITC between a variety of crystalline-crystalline interfaces \cite{chen-2022}. In contrast, ITC across amorphous-amorphous interfaces has not received as much attention. The low thermal conductivity of amorphous structures results in a negligible temperature drop across amorphous-amorphous interfaces, making it difficult to estimate ITC both experimentally and computationally \cite{highITC1}. In crystalline materials, heat is transported by propagating lattice waves known as phonons. In amorphous solids, heat carriers are categorized into propagons, diffusons, and locons based on the localization of atomic vibrations and their mean free paths \cite{Allen_1999}. 
Locons are localized vibrational modes, confined to small regions and contributing minimally to heat conduction. Propagons and diffusons, however, are delocalized. Propagons efficiently propagate heat through the material, while diffusons transfer heat more locally, resulting in less efficient conduction \cite{Fledman-PhysRevB.48.12589}. The Ioffe-Regel criterion, based on the lifetime of vibrational modes, differentiates propagons from diffusons \cite{ioffe1960non}. Giri et al. and Gordiz et al. have reported high ITC of the order of GW/m$^2$K at amorphous Si/Ge interfaces using MD simulations \cite{highITC1,highITC2}. In the case of their crystalline counterparts, MD simulations show a significant reduction in ITC \cite{introSi/GeMD1,introGe/Simd2}. The main mechanisms for heat transport in these crystalline and amorphous states exhibit substantial differences. Diffusons that are delocalized and non-propagating phonon modes in the Si and Ge layers are the main heat carriers in amorphous superlattices \cite{gordizHopkins}. While in contrast, for crystalline-crystalline interfaces, thermal conductance is primarily controlled by spatially extended phonon modes. As a result, the vibrational characteristics at amorphous-amorphous interfaces differ noticeably from those at crystalline-crystalline interfaces, providing insight into the significant contrast seen between the two types of interfaces. Notwithstanding the remarkable distinctions in ITC highlighted (between these two types of interfaces), it is noteworthy to mention that there have only been a very small number of studies that have computed and analyzed ITC for crystalline-amorphous interfaces \cite{France-Lanord_2014}. \\
In this article, the ITC between gold and silicon and gold and silica, respectively, is examined for both their crystalline and amorphous structural forms. We report a twofold to threefold increase in ITC when comparing the crystalline-crystalline to the crystalline-amorphous interfaces. 
The ITC between gold-amorphous silicon and gold-amorphous silica systems has not been thoroughly examined although these systems have high significance and widespread use, for instance for heat transfer around core-shell nanoparticles \cite{alkurdi-2020}. Let us mention, however, one recent study \cite{Raj-goldsilica} where the authors investigated ITC at the interfaces of gold and silica, and reported a double increase in ITC when comparing gold/amorphous silica to gold crystalline silica interfaces. The authors attributed this increase to a higher overlap of the vibrational density of states between amorphous silica and gold as compared to crystalline silica and gold. Here, we show that a moderate change of bonding strength yields strong correlations of the atomic displacements at the interface resulting in a large increase of elastic phonon scattering. Bonding is known to influence interfacial heat transfer~\cite[]{zhang-2022,zong-2023}. \textcolor{black}{Inelastic phonon scattering is also shown to be enhanced at gold-amorphous silicon/silica interfaces due to relatively higher amplitude of the anharmonic interaction. We also characterize the nature of energy carriers transmitted at the interface, and conclude that the huge majority of vibrational modes are not localized close to the interface. Finally, our simulations reveal that elastic phonon scattering at gold-amorphous silicon interfaces is governed by propagons while inelastic phonon scattering by a mixture of propagons and diffusons. By contrast, only diffusons participate in heat transfer at gold-amorphous silica interfaces.} 
\begin{figure*}
    \begin{center}
    \includegraphics[width=0.45\textwidth]{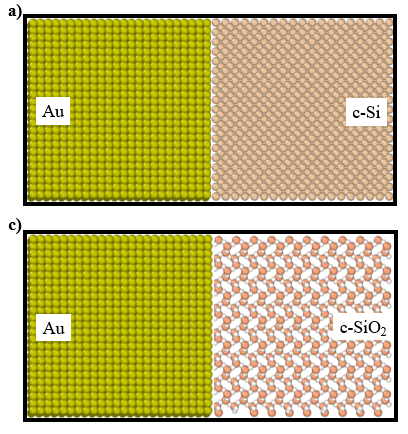}
    \includegraphics[width=0.45\textwidth]{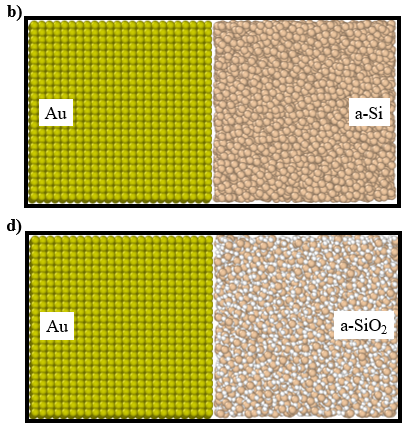}
    \caption{Illustration of the different systems considered: a) gold-crystalline silicon interface, b) gold-amorphous silicon interface, c) gold-crystalline silica interface, d) gold-amorphous silica interface.}
    \label{fig:1}
    \end{center}
\end{figure*}
\section{Theory} \label{sec:theory}
\textcolor{black}{
In this section, we present a new theoretical framework to quantify the effect of bonding at the interface to the thermal flux spectrum. This framework will be useful to interpret the role of bonding on interfacial phonon scattering when we will analyze later MD simulation results.
To relate the atomic bonding and vibrational displacements to the heat flux, we start from the expression of the heat flux spectrum in an out of equilibrium situation\cite{Saaskilahti_prb2014}:}
\begin{equation}
    \left\langle\tilde{Q}_{B \rightarrow A}(\omega)\right\rangle=\sum_{\substack{i \in A \\ j \in B}}\left\langle\tilde{\Vec{F}}_{i j}(\omega) \cdot \tilde{\vec{V}}_{i}^{*}(\omega)\right\rangle
    \label{eq:thermal_spectrum}
\end{equation}
where $\sim$ denotes Fourier transform, $\langle\ldots\rangle$ an ensemble average, $\tilde{F}_{i j}$ is the Fourier transform of the interatomic force between atoms \textit{i} and \textit{j} that belong to material $A$ and $B$ respectively, and * denotes complex conjugate. \\
\textcolor{black}{Note that the thermal spectrum, eq.~\ref{eq:thermal_spectrum} involves both elastic and inelastic phonon scattering at the interface. Elastic phonon scattering corresponds to the transmission of phonon modes preserving their frequencies,{\em i.e,} $\omega \rightarrow \omega$}, \textcolor{black}{while inelastic phonon scattering typically involves interactions between multiple phonon modes at the interface. A dominant process is the three-phonon interaction, which can be sketched as $\omega+\omega' \rightarrow \omega"=\omega+\omega'$.} 
In this study, we will be primarily interested in thermal transfer between gold and silicon,
and because silicon has a higher Debye temperature than gold, inelastic scattering processes are more likely to involve the interaction of two phonon modes coming from the gold side.

\textcolor{black}{Elastic phonon scattering is mediated by harmonic forces at the interfaces, while inelastic phonon scattering is mediated by anharmonic forces. In the following, we focus on elastic phonon scattering and the corresponding thermal flux spectrum  $\widetilde{Q}_{B \rightarrow A}^{\rm el}$ obtained by computing the forces in eq.\ref{eq:thermal_spectrum} in the harmonic approximation.}

\noindent In the harmonic approximation, one can write:
\begin{equation}
    \tilde{\vec{F}}_{i j}=-\overleftrightarrow{K}_{i j} \cdot\left(\tilde{\vec{u}}_{i}(\omega)-\tilde{\vec{u}}_{j}(\omega)\right)
\end{equation} 

\noindent where $\left.\overleftrightarrow{K}_{i j}=\frac{\partial^{2} V}{\partial \vec{r}_{i} \partial \Vec{r}_{j}}\right)_{\vec{r}_{i}, \vec{r}_{j}}$ is the Hessian of the interaction which is estimated at the equilibrium positions of the atoms, and $\vec{r}_{i}(\omega)=\vec{r}_{i}^{0}+\vec{{u}}_{i}(\omega)$, with $\vec{r}_{i}^{0}$ denotes the equilibrium positions and $\vec{{u}}_{i}(\omega)$ the relative displacement.

\noindent In the harmonic approximation, one has then:
\begin{equation}
\begin{gathered}
    \left\langle\widetilde{Q}_{B \rightarrow A}^{\rm el}(\omega)\right\rangle = \sum_{\substack{i \in A \\ j \in B}} \overleftrightarrow{K}_{i j} \cdot \left\langle\left(\tilde{\vec{u}}_{i}(\omega)-\tilde{\vec{u}}_{j}(\omega)\right) \right. \\
    \hspace*{5em} \left. \cdot\left(-i \omega \tilde{u}_{i}^{*}(\omega)\right) \right\rangle
\end{gathered}
\end{equation}

\noindent Since $\left\langle\widetilde{Q}_{B \rightarrow A}^{\rm el}(\omega)\right\rangle=-\left\langle\widetilde{Q}_{A \rightarrow B}^{\rm el}(\omega)\right\rangle$, one can obtain a symmetrized version of the thermal flux:
\begin{equation}
    \langle\tilde{Q}_{\rm el}(\omega)\rangle=\frac{1}{2}\left(\left\langle\tilde{Q}_{B \rightarrow A}^{\rm el}(\omega)\right\rangle-\left\langle\tilde{Q}_{A \rightarrow B}^{\rm el}(\omega)\right\rangle\right)
\end{equation}

\noindent With: \begin{equation}
\tilde{Q}_{A \rightarrow B}^{\rm el}(\omega)= -\sum_{\substack{i \in A \\
j \in B}} \overleftrightarrow{K}_{i j} \cdot\left(\tilde{\vec{u}}_{i}(\omega)-\tilde{\vec{u}}_{j}(\omega)\right) \cdot (-i \omega \tilde{\vec{u}}_{j}^{*}(\omega))
\end{equation}

\noindent Leading to: \begin{equation} \label{eq:dev}
\begin{split}
\langle\tilde{Q}_{\rm el}(\omega)\rangle = & \frac{1}{2} \sum_{\substack{i \in A \\
j \in B}}\overleftrightarrow{K}_{ij} \cdot \Bigg\langle\left(\tilde{\vec{u}}_{i}(\omega)-\tilde{\vec{u}}_{j}(\omega)\right) \\
& . \left(-i \omega\left(\tilde{\vec{u}}_{i}^*(\omega)+\tilde{\vec{u}}_{j}^*(\omega)\right)\right) \Bigg\rangle
\end{split}
\end{equation}
\noindent After developing Equation~\ref{eq:dev} the thermal flux can be expressed as:
\begin{equation} \label{Qomega}
     \langle\tilde{Q}_{\rm el}(\omega)\rangle  = \frac{-i \omega}{2} [\sum_{i \in A} \sum_{j \in B} \overleftrightarrow K_{ij} \cdot\langle \Tilde{\vec \Delta}_{ij}(\omega) \cdot \Tilde{\vec \Delta}_{ij}^*(\omega)  \rangle] 
\end{equation}
\noindent with:  \begin{equation*}
    \Tilde{\vec \Delta}_{ij}(\omega) = \Tilde{\vec u}_{i}(\omega) - \Tilde{\vec u}_{j}(\omega)
\end{equation*} 
\textcolor{black}{As a result, we have established a relation between the elastic thermal flux, the interfacial bonding  and the cross-correlation between atomic vibrational displacements close to the interface. This approach helps clarify the intricate mechanisms governing elastic phonon scattering at the interface.}
\textcolor{black}{In the following, it would be helpful to define the cumulative integral of the elastic thermal spectrum~:} 
\begin{equation} \label{eq:th-int-el}
    ITC_{\rm el} = \frac{1}{\Delta T} \int_{0}^{\omega_{\rm max}} \langle\tilde{Q}_{\rm el}(\omega)\rangle  d\omega'
\end{equation}
\textcolor{black}{with $\Delta T$, the temperature jump at the interface and $\omega_{\rm max}$, the maximum phonon frequency. $ITC_{\rm el}$ represents the contribution of elastic phonon scattering processes to the interfacial thermal conductance between the two materials considered.}
\section{Methodology} \label{sec:meto}
\begin{figure*}
    \centering \vspace*{1cm}
    \includegraphics[width=0.42\textwidth]{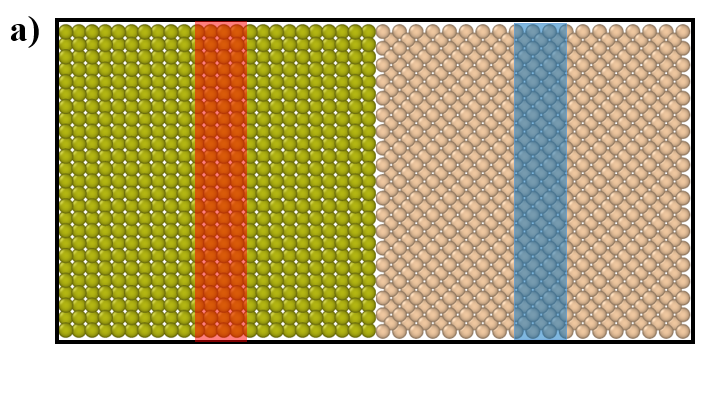}
    \includegraphics[width=0.5\textwidth]{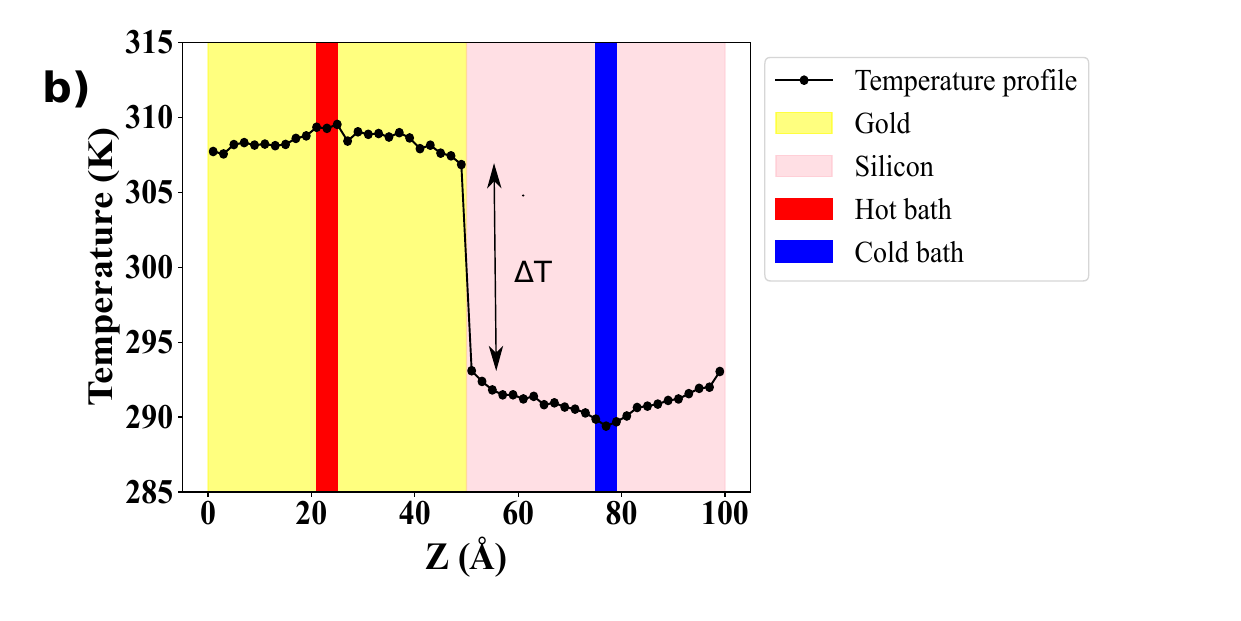}
    \includegraphics[width=0.44\textwidth]{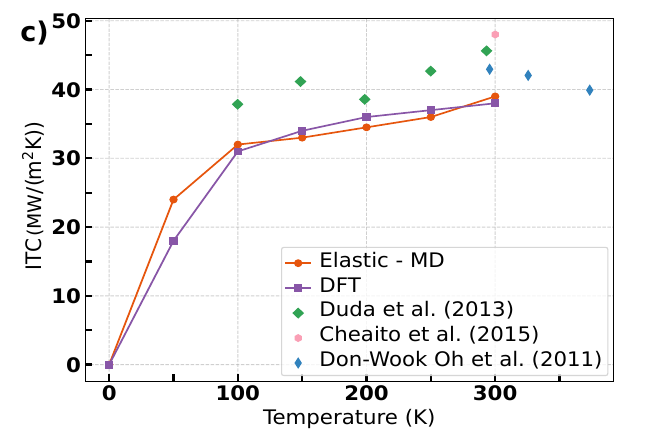}

    \caption{a) The gold-crystalline silicon interface with its respective thermal baths used in the NEMD simulations to calculate ITC (hot bath in red and cold bath in blue). 
    b) Temperature profile averaged from the NEMD simulation used to estimate the temperature jump at the gold-crystalline silicon interface. \textcolor{black} { c) Comparison between MD elastic thermal conductance calculated from eq.\ref{Qomega} and NEGF-DFT calculations for the gold - crystalline silicon interface. The unlinked symbols correspond to experimental measurements \cite[]{duda,Au-cSi_expITC,oh-dong-exp3}}.}
    \label{fig:2}
\end{figure*} 
\subsection{Modelling set up}

Molecular dynamics (MD) simulations are performed using the Large-scale Atomic/Molecular Massively Parallel Simulator (LAMMPS) \cite[]{LAMMPS}. The atomic positions are integrated using the velocity Verlet algorithm with a $0.5$ fs time step for gold-silcion systems and $0.25$ fs for gold-silica systems. The images of the simulated systems are generated using ovito software \cite[]{ovito}. To model the interactions between silicon atoms in both crystalline and amorphous silicon structures, the Stillinger-Weber potential is employed \cite[]{SW}. This potential accurately reproduces both structural and vibrational properties in both structural forms. For both crystalline and amorphous silica structures, the Tersoff potential function with Munetoh et al.'s parameters \cite[]{Tersoff-param} is used to describe the interaction between silicon and oxygen atoms. A 12 - 6 Lennard-Jones potential is employed to model the gold-gold, gold-silicon and gold-oxygen interactions with the specified parameters from Heinz et al. \cite[]{gold-heinz} for gold, and Rape et al. \cite[]{LJ-interfaces} for silicon and oxygen. The Lennard-Jones coefficients used at the interface are: $\epsilon_{Au-Si}= 62.5$ meV and  $\sigma_{Au-Si}= 3.39$ {\AA} for gold-silicon interactions and $\epsilon_{Au-O}= 42.9$ meV and  $\sigma_{Au-O}= 3.056$ {\AA} for gold-oxygen interactions. 
\textcolor{black}{The embedded atom method (EAM) potential, with parameters from \cite[]{EAM-gold}, has been also considered for gold. We calculate that it has a negligible effect on the presented results, with changes lower than 4\% in the ITC. }
\textcolor{black} {The accuracy of the Lennard-Jones cross-interaction will be assessed in Section IV through a comparison with ab-initio density functional calculations (DFT). }

\subsection{Sample generation and amorphization}
The interfaces under consideration are composed of gold and crystalline/amorphous silicon or silica. For crystalline silicon samples, we use periodic boundary conditions to generate systems with a diamond lattice of $a_0=5.43$ {\AA}, and for crystalline silica, we use the $\alpha$-Cristobalite structure with space group symmetry P$4_12_12$ and cell dimensions of a=$4.99$ {\AA} and c=$6.93$ {\AA}. Amorphous structures are generated using the melt-quench technique \cite[]{France-Lanord_2014}. Starting with the respective crystalline structure, at $300$ K, the structure is first equilibrated in the canonical ensemble. Then, within $1$ ns, it is heated to $5000$ K in the liquid, and held at that temperature for another $1$ ns until any memory of its initial configuration is lost. The temperature is then decreased back to $300$ K with a $5$ K/ps rate. The resulting amorphous structures are then stabilized for $1$ ns at $300$ K in the canonical ensemble and the Radial Distribution Function (RDF) is computed, showing good agreement with previous simulation results in the literature \cite[]{asi-sio2conductivity} and with experimental measurements as discussed \textcolor{black}{in the Supplemental Material \cite{supplement} (see also references \cite{silicon-exp-rdf, ELorch_1969} therein)}. To further validate the modelling of the amorphous structures, the bulk thermal conductivity of each system is calculated and compared with prior results. The bulk thermal conductivity of amorphous silicon is found equal to $1.44$ $\pm$ $0.10$ W/(m.K) which is in good agreement with values reported in the literature \cite[]{asi-sio2conductivity}; while the thermal conductivity of amorphous silica is found equal to $0.98$ $\pm$ $0.09$ W/(m.K) which is also in good agreement with values from other studies \cite[]{cond2silica,sio2cond_1}.

\subsection{Interfacial thermal conductance}
The calculation of ITC at the gold-crystalline/amorphous interfaces is performed using non-equilibrium molecular dynamics (NEMD) by applying periodic boundary conditions along the x, y and z directions. The Nose–Hoover equations of motion are used to integrate the atomic trajectories during all NEMD simulations \cite[]{nosee,HOOver}. Figure \ref{fig:1} shows an example of the studied gold-crystalline silicon interfaces (a), gold-amorphous silicon interfaces (b), gold-crystalline silica interfaces (c) and gold-amorphous silica interfaces (d). Each system has a total length of 100 {\AA} along the z direction perpendicular to the interface, \textcolor{black}{where the interface is defined as the region within a cutoff radius distance of 10 {\AA} from each side. We have checked that extending this region by 2 {\AA} will result in a change of less than $5 \%$ of the temperature jump at the interface, which corresponds to a variation of the ITC smaller than the error bars, which are calculated based on $5$ independent simulations.} 
\begin{figure*}[htbp!]
    \centering
    \includegraphics[width=0.96\textwidth]{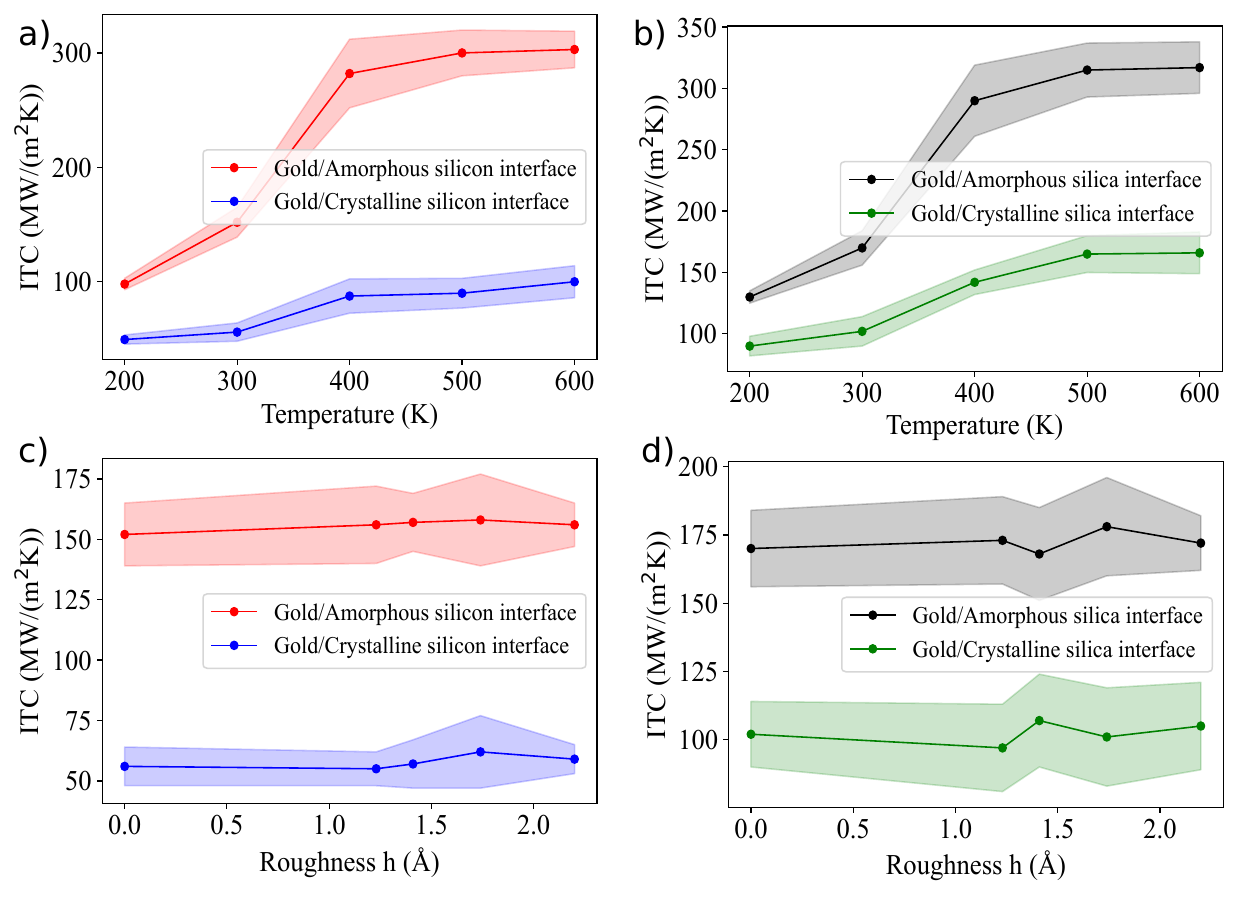}
    \caption{The interfacial thermal conductance as a function of the system temperature for: a) gold-crystalline silicon and gold-amorphous silicon systems, \textcolor{black} { b) gold-crystalline silica and gold-amorphous silica systems}. The interfacial thermal conductance as a function of interfacial roughness for: c) gold-crystalline silicon and gold-amorphous silicon systems, \textcolor{black} {d) gold-crystalline silica and gold-amorphous silica systems}. Here the roughness is quantified by the RMS of the height distribution $h$ at the (x,y) surface along the $z$ direction at $300$ K.}
    \label{fig:3}
\end{figure*} 
In order to calculate ITC, hot and cold baths are placed in the middle of the left and right sections as shown in Figure \ref{fig:2}a). Once the system has reached a steady state as discussed in the Supplemental Material \cite{supplement}, the heat flux (q) and the temperature profiles are averaged over $1$ ns for ITC calculations, and the heat flux is calculated as :
$$
q=\frac{\frac{dE}{dt}}{2 A}
$$
with $\frac{dE}{dt}$, the slope of the energy supplied/extracted from the system as a function of time, and $A$ is the surface area. ITC, expressed in $\mathrm{~W} \mathrm{~m}^{-2} \mathrm{~k}^{-1}$, computed using the ratio between the heat flux and the temperature jump at the interface ($\Delta T$): \begin{equation} \label{eq:direct}
    ITC=\frac{q}{\Delta T}
\end{equation} 
An example of a temperature profile from which the temperature jump at the gold-crystalline silicon interface is extracted is shown in Figure \ref{fig:2}b).

\section{Results and Discussion}
Five different simulations are run for each interface and the results for ITC are shown in table \ref{tab:crys-am-table}. These findings indicate that: (1) ITC is significantly higher at gold-amorphous interfaces: we report a threefold increase in the case of silicon and twofold increase for silica. (2) The calculated ITC at the gold/crystalline silicon interface is in good agreement with previous reported experimental value of $\approx$ 50 MW/(m$^2$K) \cite[]{duda,Au-cSi_expITC}.

\begin{table}[h]
    \centering
    \begin{tabular}{cc}
    \toprule
         & \textbf{ITC (MW/(m$^2$K))}\\
        \midrule
         \textbf{Gold-Amorphous Silicon} & 152  $\pm$ 13\\ 
         \textbf{Gold-Crystalline Silicon}&  56 $\pm$ 8\\
         - - - - - - - - \\
          \textbf{Gold-Amorphous Silica} & 170  $\pm$ 14\\ 
         \textbf{Gold-Crystalline Silica}&  102 $\pm$ 12\\
    
    \bottomrule
    \end{tabular}
    \caption{Values of the interfacial thermal conductance for gold/crystalline-amorphous systems at $300$ K. The thermal conductance is much higher at the gold-amorphous systems interfaces as compared to gold-crystalline interfaces.}
    \label{tab:crys-am-table}
\end{table} 

\noindent We have checked that size effects are small. According to the calculated results, doubling the system length increases ITC by only $5-6$\% for all simulated interfaces as discussed in the Supplemental Material \cite{supplement}. Therefore, the results are almost size independent.

\textcolor{black} {It should be emphasized at this point that the Diffusive Mismatch Model (DMM) predicts close values for the ITC of the gold-amorphous and gold-crystalline structures, as discussed in the Supplemental Material \cite{supplement} (see also references \cite{DMM, landau,dynasor} therein). The failure of the DMM model can be traced back to the fact that the vibrational density of states of silicon close to the interfaces are not significantly different in the amorphous and crystalline structures.}

\textcolor{black} {Finally, to assess the accuracy of the Lennard-Jones potential to describe heat transfer at interfaces, we calculated the interfacial thermal conductance at the gold-crystalline silicon interface using density functional non-equilibrium Green's function calculation (NEGF-DFT) as discussed} \textcolor{black}{in the Supplemental Material \cite{supplement} (see also references \cite{siesta, perdew-prl-96, troullier-prb-91, phonopy, adessi-pccp-2020} therein)}. Figure \ref{fig:2}c) compares the calculated values at a selected temperature range of $50$ to $300$ K for both NEGF-DFT and elastic MD derived from eq.~\ref{eq:th-int-el} as well as  experimental measurements. Note that we compute only the elastic scattering contribution in MD to be consistent with the NEGF framework which considers only harmonic interactions.
The results of the elastic MD and NEGF-DFT are fairly consistent. Note also that the values of the interfacial thermal conductance predicted by either MD or NEGF-DFT are close to experimental data. This has the important consequence that electron-phonon interfacial couplings which are absent from both MD and NEGF calculations should not play a significant role, at least for gold-silicon interfaces.

\subsection{Effect of temperature}
The ITC values previously reported have been calculated at $300$ K. \textcolor{black}{As a result, variations in the system temperature may have an impact on the results obtained and, in turn, in the threefold (twofold) increase observed at the gold-silicon (silica) interfaces. In order to characterize how the system average temperature affects ITC, all the interfaces are simulated at different temperatures ranging between $200$ and $600$ K. The results are shown in Figure \ref{fig:3}a) for gold-silicon interfaces and in Figure \ref{fig:3}b) for gold-silica interfaces. The relative increase in ITC observed at $300$ K when comparing gold-amorphous interfaces to the gold-crystalline ones is found to be constant throughout the temperature range under consideration.} ITC consistently escalates as temperatures increase. The effect of anharmonic phonon scattering at higher temperatures increases phonon transmission at the interface, resulting in higher ITC \cite[]{Saaskilahti_prb2014}.

\subsection{Effect of interfacial roughness}
\textcolor{black}{Real interfaces may display roughness, consequently, it is critical to address the effect of surface roughness on interfacial heat transfer \cite[]{merabia-2014} as it can significantly impact the threefold (twofold) increase observed when comparing ITC for gold/amorphous silicon (silica) to gold/crystalline silicon (silica).} We considered a roughness along the (x,y) surface perpendicular to the z direction using a self-affine scaling transformation and the root mean square (RMS) of a height distribution $h$ \cite[]{pyrough}. \newline
\noindent \textcolor{black}{The ITC is represented in Figure \ref{fig:3}c) for gold-silicon interfaces and in Figure \ref{fig:3}d) for gold-silica interfaces as a function of the RMS of the height distribution $h$ at $300$ K.} The results show unequivocally that surface roughness has little effect on the enhancement of ITC. It is clear that the differences observed were not significant when contrasting simulations with planar interfaces with those that consider a rough interface. This result shows that the behavior of the investigated interfaces was not significantly affected by the presence of surface roughness.

\subsection{Frequency dependent thermal spectrum}
\begin{figure*}[htbp!]
    \centering
    \includegraphics[width=0.96\textwidth]{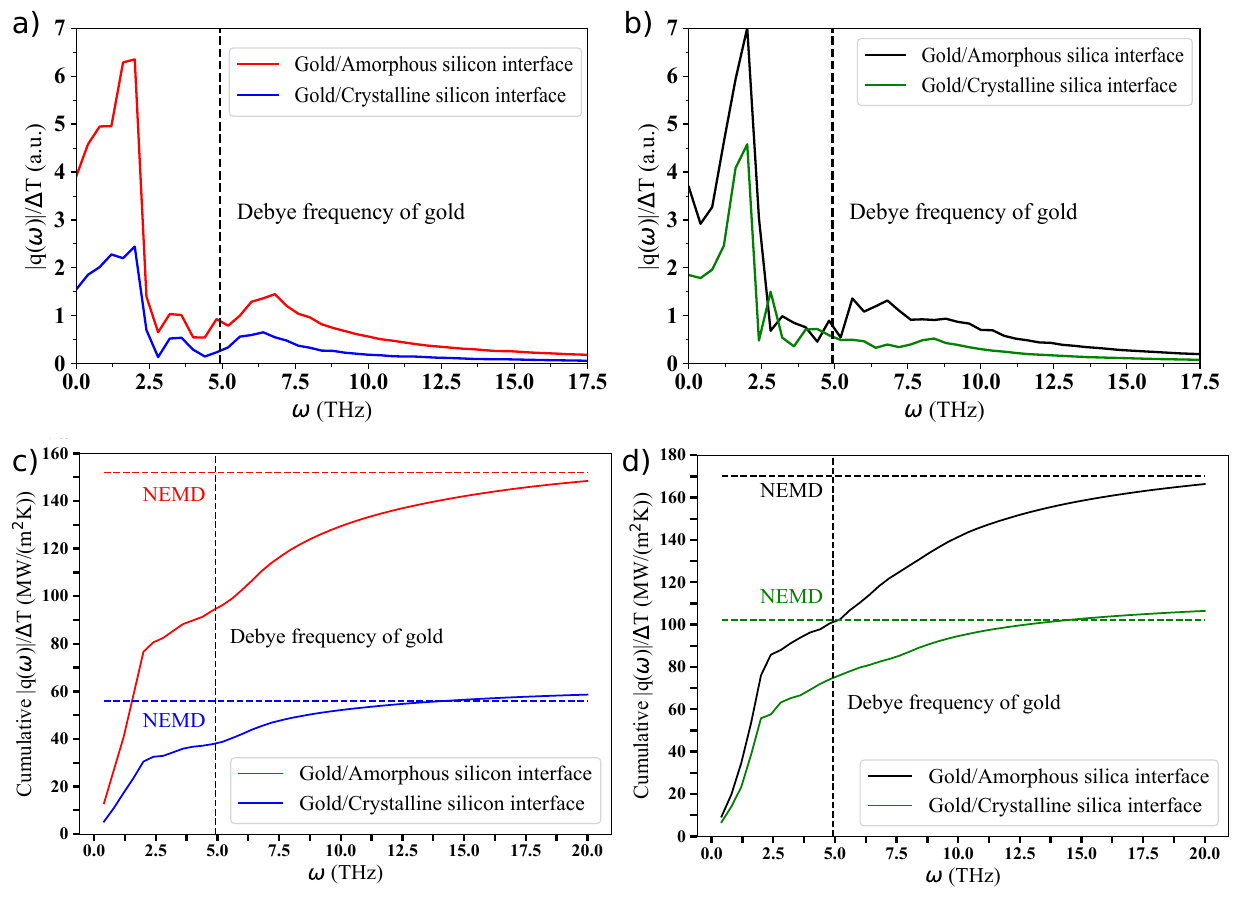}
    \caption{Frequency-dependent thermal flux at $300$ K at the: a) gold-silicon interfaces, \textcolor{black} { b) gold-silica interfaces}. ITC calculated from the cumulative frequency-dependent thermal flux at $300$ K compared with the NEMD values at the: c) gold-silicon interfaces, \textcolor{black}{d) gold-silica interfaces}.}
    \label{fig:TS}
\end{figure*} 
We now quantify the contribution of the vibrational frequency modes to the interfacial thermal transfer at the considered interfaces. To this end, we computed the frequency-dependent thermal flux that encodes the combined information on the elastic and inelastic heat transmission as follows \cite[]{Thermalflux}:
\begin{equation} \label{eq:heatflux}
   q(\omega) = \frac{2}{A} \textbf{Re} [\sum_{i \in Au} \sum_{j \in Si} \int_{0}^{t_{\rm max}} \langle F_{ij}(\tau) \cdot v_i(0) \rangle e^{i\omega\tau} d\tau]
\end{equation}
where $A$ is the surface area, $F_{ij}$ is the force applied on the gold atom \textit{i} as a result of its interaction with silicon atom \textit{j}, $v_i$ is the velocity of atom $i$, $\langle F_{ij}(\tau) \cdot v_i(0) \rangle$ is the correlation function between these two quantities, $\tau$ is the correlation time, and $t_{\rm max}$ is the maximum time.
The frequency-dependent thermal flux measures the heat flux transferred between two adjacent materials as a function of frequency, giving indications on the dominant frequency range for interfacial heat transfer. 
\textcolor{black}{Note that $q(\omega)$ in eq.~\ref{eq:heatflux} includes the contribution of both elastic and inelastic phonon 
scattering.}
\\
Alternately, ITC can be defined as the cumulative integral of the frequency dependent thermal spectrum as: 
\begin{equation} \label{eq:th-int}
    ITC = \frac{1}{\Delta T} \int_{0}^{w_{max}} q(\omega') d\omega'
\end{equation}
with $\Delta T$, the temperature jump at the interface, and $w_{max}$, the maximum calculated frequency.
\noindent Figures \ref{fig:TS}a) and \ref{fig:TS}b) compares the frequency dependent thermal spectrum at the gold silicon interfaces and the gold silica interfaces respectively at $300$ K. The contribution of the vibrational modes at gold amorphous interfaces is significantly greater than their crystalline counterparts, resulting in a higher ITC. Additionally, this result shows that heat transfer across all considered interfaces is primarily driven by low-frequency vibrational modes, specifically those within gold Debye frequency ($\approx$ 5 THz). This insight emphasizes the importance of these vibrational modes in facilitating thermal transport and the overall heat transfer process. Furthermore, Figures \ref{fig:TS}c) and \ref{fig:TS}d) show good agreement between the cumulative integrals of the frequency dependent thermal spectrum calculated using eq.~\ref{eq:th-int} and the NEMD values.

\subsection{Elastic transmission contribution: analysis of interfacial bonding and atomic displacements}
\begin{figure*}[htbp!]
    \centering
    \includegraphics[width=0.96\textwidth]{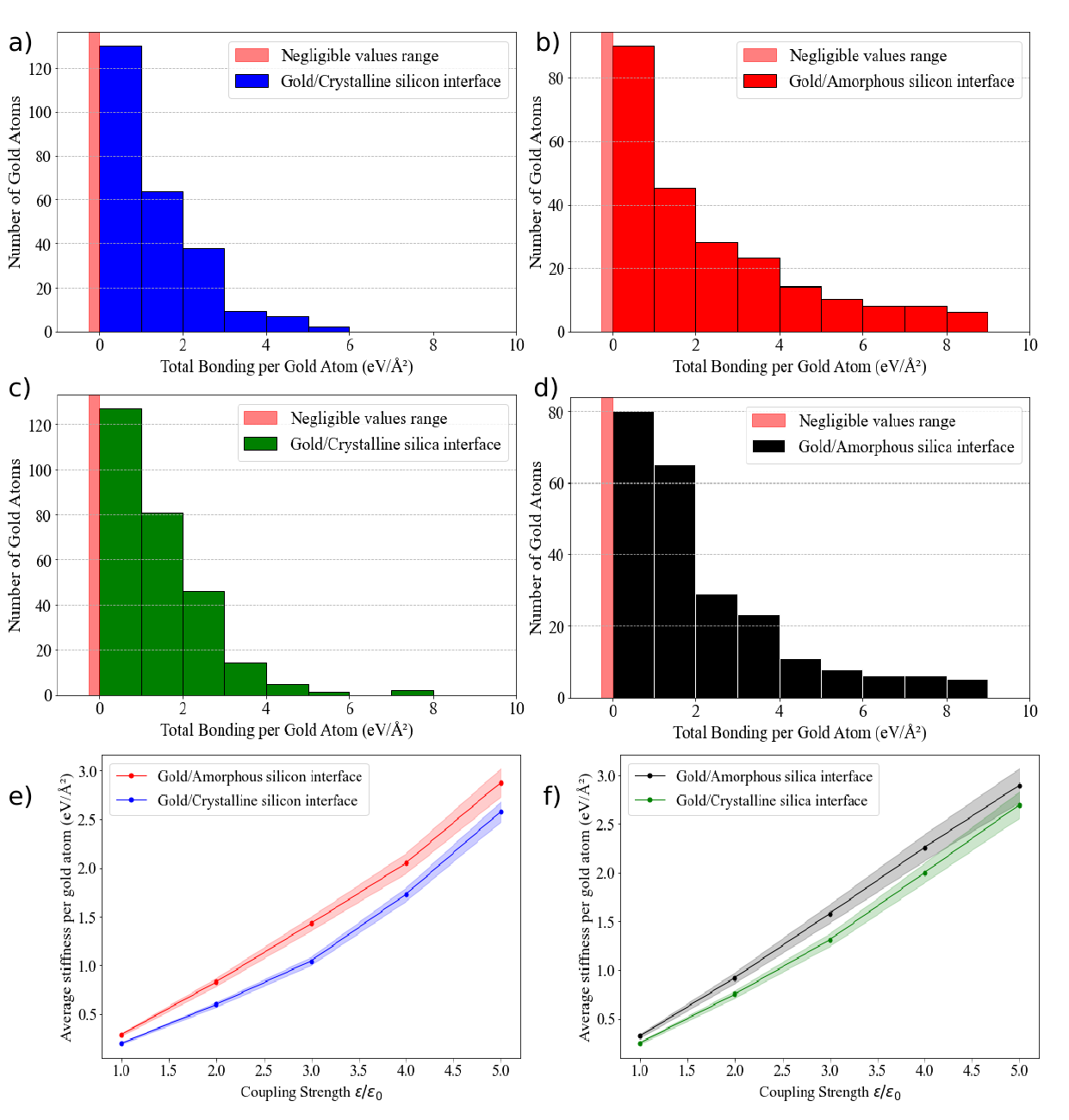}
    \caption{\textcolor{black} {The distribution of gold atoms based on their total interfacial bonding per atom for: a) gold-crystalline silicon interface, b) gold-amorphous silicon interface, c) gold-crystalline silica interface, d) gold-amorphous silica interface. The red shaded area represents a region where gold atoms have a negligible value of the bonding stiffness. The average stiffness per gold atom as function of the normalized coupling strength for: e) gold-crystalline silicon and gold-amorphous silicon interfaces, f) gold-crystalline silica and gold-amorphous silica interfaces, where $\epsilon_0$ corresponds to the reference value of the coupling strength defined between gold and silicon (oxygen) atoms.}}
    \label{fig:4}
\end{figure*} 

\begin{figure*}[htbp!]
    \centering
    \includegraphics[width=0.96\textwidth]{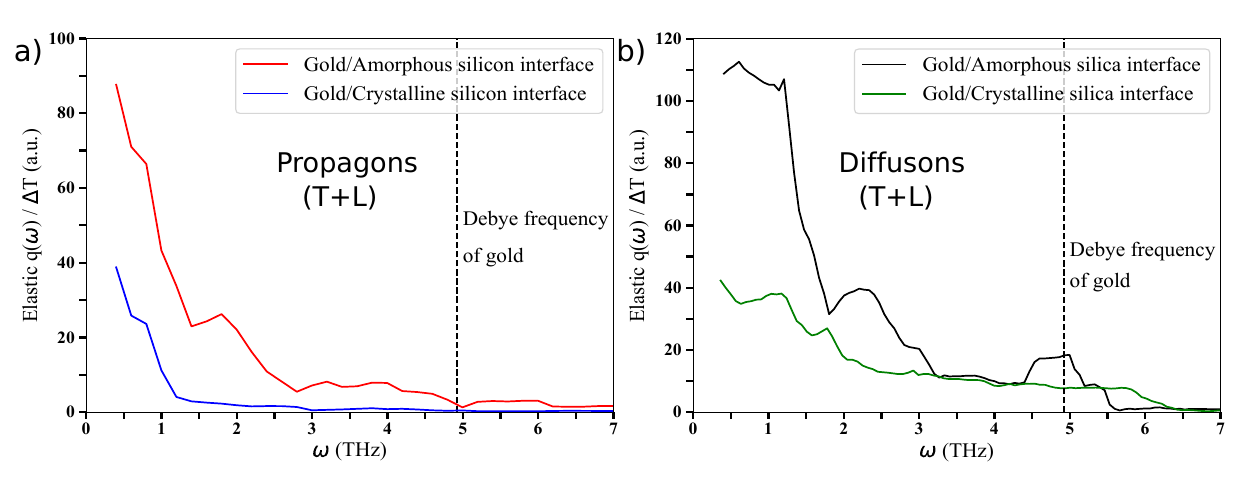}
    \caption{Frequency-dependent spring stiffness and atomic displacements cross-correlations thermal flux at $300$ K, representing the elastic transmission of the total heat flux eq.\ref{Qomega} at the: a) gold/crystalline silicon and gold/amorphous silicon interfaces, \textcolor{black} {b) gold/crystalline silica and gold/amorphous silica interfaces}.}
    \label{fig:5}
\end{figure*} 
\textcolor{black}{To obtain further insight in the physical significance and microscopic quantities underlying the observed threefold (twofold) increase at the gold-silicon (silica) interfaces, we divide now ITC in terms of elastic and inelastic transmission. In order to describe the elastic contributions, we relate the heat flux to the interaction spring stiffness and the atomic displacements at the interface utilizing eq.~\ref{Qomega}, which is comprehensively defined and discussed in section \ref{sec:theory} starting from an harmonic approximation.
We will start by characterizing the statistics of interfacial bonding $K_{ij}$.}
\\
\noindent \textcolor{black}{First, concentrating on harmonic interactions between the gold atoms and the silicon (oxygen) atoms near the interface, the interatomic forces can be approximated by a set of harmonic springs. The effective spring constant, which is proportional to the coupling between neighboring atoms, can be calculated as follows: $k_{ij} =  \frac{\partial^2 U(r_{ij})}{\partial r^2}$, where $U(r_{ij})$ is the inter-atomic potential used to model the interaction between a gold atom \textit{i} and a silicon (oxygen) atom \textit{j} at the interface. As a result, the higher the spring stiffness, the stronger the coupling.}

\textcolor{black}{Figures \ref{fig:4}a) and \ref{fig:4}b) display the histograms of the calculated spring stiffness exerted on a gold atom as a result of its interactions with various silicon atoms within the gold-crystalline silicon and gold-amorphous silicon interfaces respectively, while Figures \ref{fig:4}c) and \ref{fig:4}d) show the corresponding histograms at the gold-crystalline silica and gold-amorphous silica interfaces respectively. 
The calculated average spring stiffness acting on a gold atom at the interface is about 30\% greater at the gold-amorphous silicon interfaces as compared to gold-crystalline silicon interfaces and about $25$\% greater at the gold-amorphous silica interfaces as compared to gold-crystalline silica interfaces. Specifically, the higher spring stiffness at the interfaces between gold-amorphous silicon and gold-amorphous silica than between their crystalline counterparts highlights a fundamental aspect that the disordered state of amorphous silicon and silica strengthens their coupling with gold, leading to better thermal coupling. }

\textcolor{black}{We now assess to which extent an increase in the coupling strength between the gold and silicon (oxygen) atoms will preserve the relative increase in bonding between the gold-amorphous and gold-crystalline interfaces. As such, we study a range of coupling strengths, all of which are given as multiples of $\epsilon_0$, the gold-silicon (oxygen) reference interaction parameter defined in section \ref{sec:meto}. Thereby, Figures \ref{fig:4}e) and \ref{fig:4}f) display the average stiffness per gold atom as a function of the normalized coupling strength $\epsilon/\epsilon_0$ for the gold-silicon and gold-silica interfaces respectively. The increase in the average stiffness per gold atom observed when comparing gold-amorphous interfaces to the gold-crystalline ones is found to be constant throughout the whole coupling strength region under consideration. This consistency suggests that the relative enhancement in stiffness from crystalline to amorphous interfaces is independent of the specific strength of the atomic coupling. Consequently, the corresponding increase in ITC due to this enhanced stiffness remain stable across different coupling strengths as discussed in the Supplemental Material \cite{supplement}.} \\
\noindent \textcolor{black}{Building on these findings, it is crucial to further elucidate the intricate relationship between thermal flux and the microscopic interactions at the considered interfaces. 
Figures~\ref{fig:5}a) and \ref{fig:5}b) compare the elastic thermal spectrum calculated from the spring stiffness and the frequency dependent atomic displacements cross-correlations at the gold-silicon interfaces and at the gold-silica interfaces, respectively. The results show that the elastic heat transfer at the four interfaces is only due to low-frequency vibrational modes ($\omega < 6$ THz). This frequency range aligns well with gold Debye frequency ($\approx$ 5 THz). Based on these calculations, we can determine the total contribution of elastic transmission, which is approximately $70$\% of the total thermal transfer.
Also, we clearly see that, already at the elastic phonon scattering level, the integrated thermal flux is larger 
for gold-amorphous interfaces. More precisely, we report an increase of $250 \%$ for gold-amorphous silicon and $150 \%$ for gold-amorphous silica interfaces. These relative large levels of enhancement suggest that moderate increase in interfacial bonding ($30 \%$) induce strong correlations in the atomic displacements across the interfaces, as evidenced by eq.~\ref{Qomega}, thus contributing to boosting significantly elastic phonon scattering at the interface.}
\noindent \textcolor{black}{To further evidence the strong effect of enhanced bonding on the atomic displacements at the interface}, we calculate the cross-correlation coefficient between atomic vibrational displacements which is defined as \cite[]{stat-crosscoef}: \begin{equation}
r = \frac{\sum\limits_{i=1}^{n} (u_i - \bar{u})(v_i - \bar{v})}{\sqrt{\sum\limits_{i=1}^{n} (u_i - \bar{u})^2} \cdot \sqrt{\sum\limits_{i=1}^{n} (v_i - \bar{v})^2}} \label{r-coef}
\end{equation}
We select the gold-silicon (oxygen) atom pairs \textit{n} that are at their equilibrium distance once the system is fully equilibrated satisfying: $r_{Au-Si} = 2^{\frac{1}{6}} \sigma_{Au-Si}$ and $r_{Au-O} = 2^{\frac{1}{6}} \sigma_{Au-O}$ with a variation of $0.5$ {\AA}, where $u$ and $v$ represent the atomic displacement of the gold and silicon (oxygen) atoms respectively. The cross-correlation coefficient between atomic vibrational displacements in the ideal cases is close to one, when $u$ and $v$ belong to the identical atom type, indicating thermal resonance, whereas it is close to zero when the displacements between them are uncorrelated. As a result, high interfacial thermal conductance can be associated with a larger thermal flux, which is directly influenced by thermal resonance \cite[]{cross-cor}. For gold/crystalline silicon interfaces and gold/amorphous silicon interfaces, the calculated cross-correlation coefficient, $r$, between atomic vibrational displacements is $0.128$ and $0.23$, respectively. This corresponds to an increase by a factor of approximately $2$. Similarly, for gold/crystalline silica interfaces and gold/amorphous silica interfaces, it is $0.18$ and $0.29$, respectively, corresponding to an approximately 40\% increase. A higher spring stiffness implies stronger bonding between atoms, resulting in smaller atomic displacements from their equilibrium positions and higher cross-correlation coefficient, $r$. \\
\begin{figure*}[htbp!]
    \centering
    \includegraphics[width=0.96\textwidth]{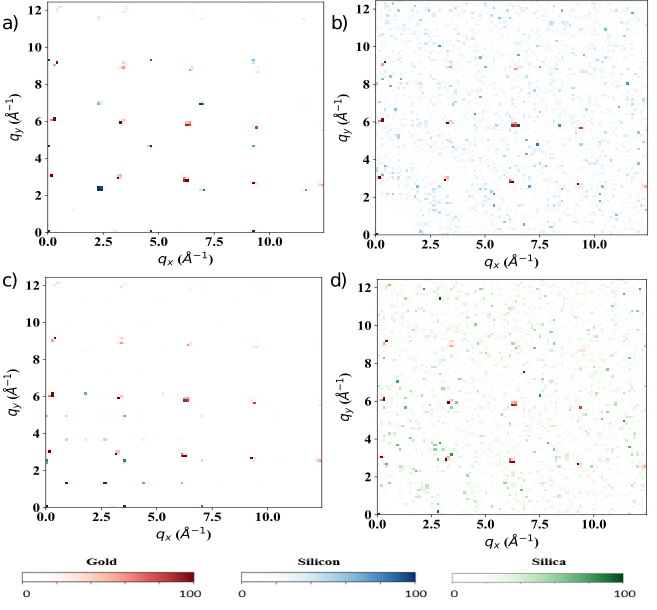}
    \caption{\textcolor{black}{ 2D static structure factors of the first interfacial layers of gold and the first interfacial layers of: a) crystalline silicon, b) amorphous silicon c) crystalline silica d) amorphous silica. The structure factor patterns of gold and amorphous structures exhibit a higher degree of overlap compared to those of gold and crystalline structures.}}
    \label{fig:com}
\end{figure*}

\begin{figure*}[htbp!]
    \centering
    \includegraphics[width=0.96\textwidth]{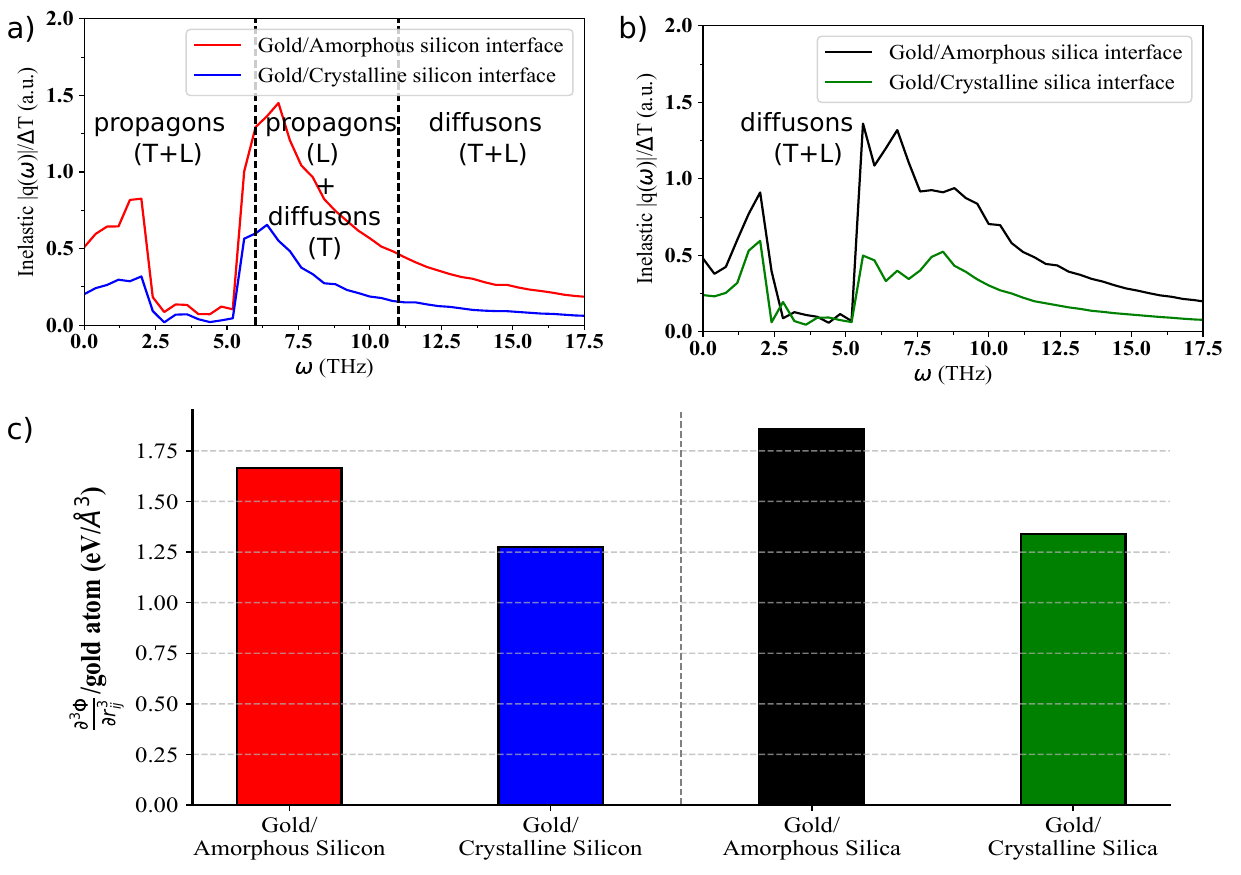}
    \caption{\textcolor{black}{ The frequency-dependent inelastic spectral thermal flux at $300$ K calculated using the difference between the total spectral thermal flux and the spectral thermal flux calculated using a harmonic approximation at the: a) gold-silicon interfaces, b) gold-silica interfaces. c) Histogram illustrating the value of the third derivative of the interatomic potential at the gold-silicon interfaces and gold silica interfaces.}}
    \label{fig:inelastic}
\end{figure*} 

\noindent \textcolor{black}{To further investigate the bonding enhancement and understand the primary reason for the stiffer bonds between gold and amorphous silicon (silica), we calculate now the two-dimensional structure factor S($q$) of the first interfacial layers of gold and silicon (silica) at both amorphous and crystalline interfaces and is defined as \cite[]{PhysRevE.Lafon}:}  
\begin{equation}
    S(q) = \frac{1}{N}\left[ \left(\sum_{i=0}^{N}  \cos({\vec{r}}_i \cdot \vec{q})\right)^2 + \left(\sum_{i=0}^{N} \sin({\vec{r}}_i \cdot \vec{q})\right)^2 \right]
\end{equation}
\textcolor{black}{where $\vec{r}_i = x_i \vec{e}_x + y_i \vec{e}_y$ is the position of atom $i$, $N$ is the total number of atoms considered in the calculation, and $q$ are the wave vectors at which the structure factor is calculated, being multiples of 2$\pi$/$L$ with $L$ the size of the simulation box in the $x$ and $y$ directions.}

\noindent \textcolor{black}{Plotting the static structure factor in terms of wave-vector components for two materials allows one to see the distribution of scattering intensities corresponding to the different atomic arrangements. Figures \ref{fig:com}a) and \ref{fig:com}b) show the structure factors for gold-crystalline silicon and gold-amorphous silicon, respectively and Figures \ref{fig:com}c) and \ref{fig:com}d) show the structure factor for gold-crystalline silica and gold-amorphous silica. We quantify the overlap between the structure factors $S_1$ and $S_2$ by calculating: }
\begin{equation}
\text{Overlap} = \frac{\int_0^{q_{\text{max}}} S_{\text{1}}(q) \cdot S_{\text{2}}(q) \, dq}{\sqrt{\int_0^{q_{\text{max}}} S_{\text{1}}^2(q) \, dq} \cdot \sqrt{\int_0^{q_{\text{max}}} S_{\text{2}}^2(q) \, dq}}
\end{equation}
\textcolor{black}{where 1 and 2 are the left and right of the interface, respectively, and \( q_{\text{max}} \) is the maximum $q$ magnitude considered in the integration, with $q$ magnitude is defined as \( q = \sqrt{q_x^2 + q_y^2} \) as discussed in the Supplemental Material \cite{supplement}. 
The calculated overlap at the gold-crystalline silicon interface is $0.73$, at the gold-amorphous silicon interface is $0.98$, at the gold-crystalline silica interface is $0.77$, and at the gold-amorphous silica interface is $0.99$. The structure factor patterns of gold and amorphous silicon exhibit a higher degree of overlap compared to those of gold and crystalline silicon. We obtain qualitatively the same results for gold with amorphous and crystalline silica systems. This suggests that gold atomic arrangements and amorphous structures are more compatible. The greater the alignment between the periodicity or spatial correlations, the better matching between their atomic structures. Consequently, higher commensurability is expected to result in stronger interatomic interactions at the interface, leading to higher effective spring stiffness between gold and the amorphous structures. Conversely, the lower number of common points between gold and the crystalline structure factors indicates poorer alignment and weaker interatomic forces, resulting in lower spring stiffness at their interface. In conclusion, the stiffer bonds observed can be attributed to a higher commensurability between gold and the amorphous structures investigated. }
\begin{figure*}[htbp!]
    \centering
    \includegraphics[width=0.96\textwidth]{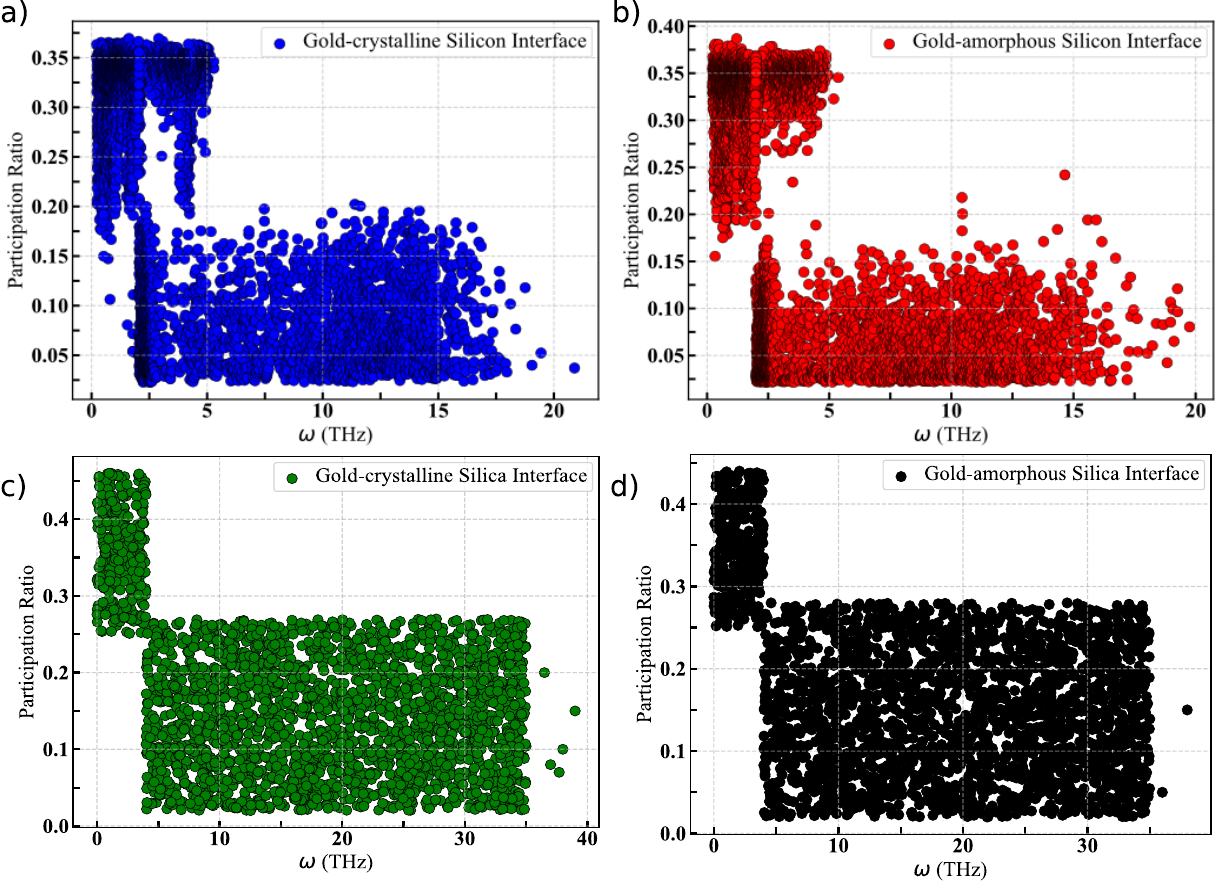}
    \caption{\textcolor{black}{Modal participation ratio $P(\omega)$ as a function of frequency obtained from diagonalizing the respective system dynamical matrix at the: a) gold-crystalline silicon interface , b) gold-amorphous silicon interface, c) gold-crystalline silica interface, d) gold-amorphous silica interface.}}
    \label{fig:PR}
\end{figure*} 
\subsection{Inelastic phonon transmission contribution}
\textcolor{black}{We now focus on the contribution of the inelastic transmission to the total interfacial heat flux.  We quantify this contribution as the difference between the total heat flux calculated using  eq.~\ref{eq:heatflux} and the heat flux calculated using the harmonic approximation in eq.\ref{Qomega}. Figures \ref{fig:inelastic}a) and b) show the resulting contribution for gold silicon interfaces and gold silica interfaces, respectively. The findings show that, in comparison to the gold crystalline interfaces, the inelastic thermal flux is likewise significantly higher at the gold amorphous interfaces, suggesting a higher contribution from inelastic scattering processes. The higher inelastic thermal flux observed at the gold amorphous interfaces can be attributed to the increased anharmonicity of the vibrational modes of the amorphous silicon/silica. This was quantified by calculating the third derivative of the interatomic potential employed in the simulations, which was found to be approximately $30$\% higher at gold amorphous interfaces than crystalline ones as shown in  Figures \ref{fig:inelastic}c). Increased anharmonicity indicates greater deviations from harmonic behavior, which leads to stronger vibrational mode interactions and, as a result, higher inelastic scattering rates. These interactions are stronger in the disordered structure of amorphous silicon/silica, where the lack of periodicity results in a broader range of vibrational modes and enhanced scattering mechanisms, contributing to the higher inelastic thermal flux observed.}

\subsection{Interface phonon localization}
Another possible explanation of the difference in ITC is the localization of phonons in the vicinity of amorphous structures.
This phenomenon has been indeed invoked to explain the relative high ITCs at amorphous-amorphous interfaces~\cite[]{gordizHopkins}.
To assess any localization effect at the interface, we investigate the localization properties of phonons in the vicinity of crystalline and amorphous structures. Unlike in crystalline structures, thermal transport in amorphous materials is related to propagons (propagating modes) and diffusons (nonpropagating modes), both of which are delocalized modes \cite[]{amor-phonons}. Delocalized phonon modes are the main heat carriers in amorphous interfaces whereas in the crystalline counterparts, spatially extended modes contribute primarily to thermal transport. \\
\noindent In order to investigate the properties of phonons in the different systems considered in this study, the modal participation ratio of phonons is calculated as follows \cite[]{pop-ration}:
\begin{equation} \label{eq:MPR}
P(\omega)=\frac{1}{N_{\mathrm{b}} \sum_{1}^{N_{\mathrm{b}}}\left(\sum_{\alpha}^{3} e_{i, \alpha}^{*}(\omega) e_{i, \alpha}(\omega)\right)^{2}}
\end{equation}
where $N_b$ is the total number of atoms in the system, and $e_{i, \alpha}$$(\omega)$ is the component of the eigenmode relative to atom $i$'s coordinate along the direction $\alpha$. \textcolor{black}{This quantity can only be calculated if the corresponding eigenvalues and eigenmodes are known. This entails computing and diagonalizing the dynamical matrix of the whole system. The dynamical matrix, denoted as $\mathbf{D}$, is derived from the second derivatives of the potential energy $U$ with respect to the atomic displacements $\mathbf{u}_i$ and $\mathbf{u}_j$ and is given by $D_{ij} = \frac{1}{\sqrt{m_i m_j}} \frac{\partial^2 U}{\partial \mathbf{u}_i \partial \mathbf{u}_j}$, where $m_i$ and $m_j$ are the masses of atoms $i$ and $j$. Therefore, we calculate the dynamical matrix by investigating the forces generated by interactions between a gold (Au) atom and nearby silicon (Si) and oxygen (O) atoms, as well as taking into account the interactions with neighbour gold atoms.} \textcolor{black}{This calculation is performed using a combination of LAMMPS and Phonopy \cite[]{LAMMPS,phonopy-phono3py-JPCM}. The process begins by methodically generating random, but controlled, displacements of 0.1 \AA \, for all relevant atoms along the $x$, $y$, and $z$ directions. These displacements are applied to the static, relaxed structure of the system obtained during the simulation. This is a critical step as it simulates perturbations, allowing for the observation and quantification of the forces generated by these displacements.} The resulting dynamical matrix represents the system's response to atomic displacements, capturing the complex interplay of forces within the heterogeneous system. To obtain the phonon modes and frequencies, we solve the eigenvalue problem $\mathbf{D} \mathbf{e}_k = \omega_k^2 \mathbf{e}_k$, where $\omega_k$ are the eigenfrequencies and $\mathbf{e}_k$ are the $3 N_b$ dimensional eigenvectors, which correspond to the vibrational modes of the system and their respective eigenmodes, where $k$ is an integer running from $1$ to $3 N_b$ indexing the vibrational modes. The square roots of the eigenvalues give the phonon frequencies $\omega_k = \sqrt{\lambda_k}$. As a result, in the computation of eq.\ref{eq:MPR}, each mode accounts for the contributions of the atoms across the whole system. The smaller $P(\omega)$, the fewer atoms involved in the motion of a specific mode. More precisely, $P(\omega)$ equals unity when all atoms participate in a specific mode. When only one atom contributes to a mode, $P(\omega)$ equals $1/N_b \ll 1$. \\
\noindent \textcolor{black}{Figure \ref{fig:PR}a) and b) show the modal participation ratio as a function of the  frequency at the gold-crystalline silicon and the gold-amorphous silicon interfaces respectively, while \ref{fig:PR}c) and d) shows the same quantity at the gold silica interfaces. As can be seen, all the interfaces show delocalized vibrational modes and there are no significant differences between the modal participation ratio calculated at the gold-crystalline interfaces and gold-amorphous interfaces. We further verified that all the modes at all the studied interfaces have a modal participation ratio $P(\omega) >> 1/N_{int}$, where $N_{int}$ is the total number of atoms at the interface. This observation confirms that all of the vibrational modes are delocalized throughout the entire gold-amorphous systems and none of them is localized close to the interface.}
\begin{figure}[htbp!]
    \centering
    \includegraphics[width=0.48\textwidth]{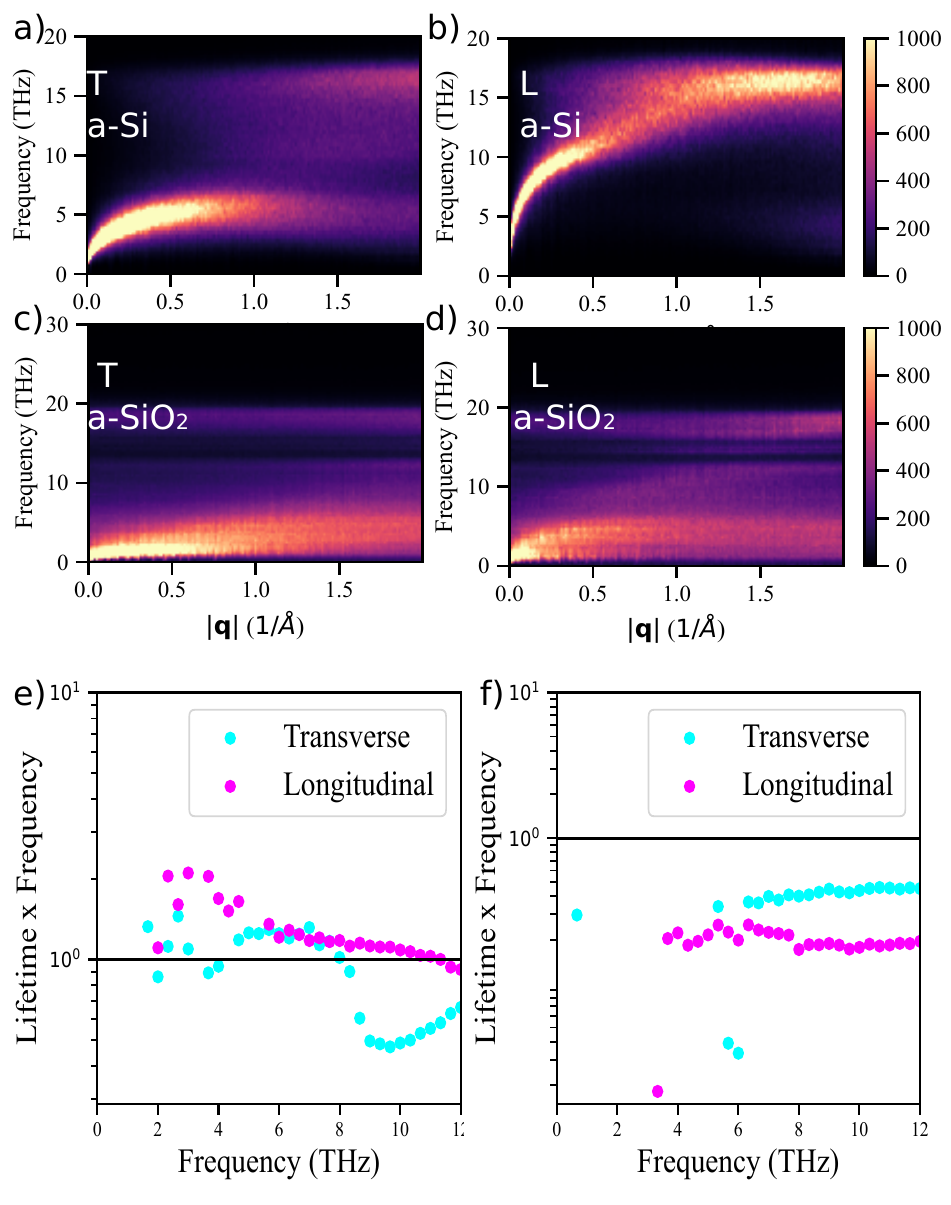}
    \caption{\textcolor{black}{The longitudinal (L) and transverse (T) dynamical structure factors as a function of the wavenumber and the frequency of the: a) b) first wall of silicon atoms at the gold-amorphous silicon interface, c) d) first wall of silica atoms at the gold-amorphous silica interface. The lifetime multiplied by frequency versus frequency for longitudinal and transverse wave for: e) amorphous silicon, f) amorphous silica. The Ioffe-Regel criterion occurs when lifetime multiplied by frequency equals $1$.}}
    \label{fig:SF-IR}
\end{figure} 
\\
\subsection{Contribution of propagons and diffusons}
\textcolor{black}{Now in order to distinguish whether the delocalized modes at the amorphous side of the interface are propagating (propagons) or non-propagating (diffusons), we analyze the dynamical structure factor of the first amorphous wall of atoms at the interfaces. This analysis helps determine the dispersion law and the mean free path for longitudinal and transverse vibrational modes. The Ioffe-Regel criterion plays a crucial role in this analysis. It is based on the comparison between the phonon mean free path and the phonon wavelength. When the mean free path is larger than half the wavelength, phonons can be considered to have well-defined dispersion relations, and they behave as propagons. Conversely, when the mean free path is shorter than half the wavelength, the phonons cannot maintain coherent propagation, they are diffusons \cite[]{Moon-PhysRevB.97.024201,Parshin-PhysRevB.87.134203,Tanguy-PhysRevE.93.023006}. In other words, a vibrational mode is considered to be propagating if its lifetime is larger than its vibrational period; on the other hand, if a mode lifetime is smaller than its vibrational period, it is considered to be non-propagating and classified as a diffuson. We used Dynasor library \cite[]{dynasor} to calculate the dynamical structure factors of the first layer of silicon and silica atoms at the gold-amorphous interfaces. The longitudinal and transverse current correlation functions are defined as:}
\begin{equation}
C_{L}(\boldsymbol{q}, \omega)=\int_{-\infty}^{\infty} \frac{1}{N}\left\langle\boldsymbol{j}_{L}(\boldsymbol{q}, t) \cdot \boldsymbol{j}_{L}(-\boldsymbol{q}, 0)\right\rangle \mathrm{e}^{-\mathrm{i} \omega t} \mathrm{dt}
\end{equation}
\begin{equation}
C_{T}(\boldsymbol{q}, \omega)=\int_{-\infty}^{\infty} \frac{1}{N}\left\langle\boldsymbol{j}_{T}(\boldsymbol{q}, t) \cdot \boldsymbol{j}_{T}(-\boldsymbol{q}, 0)\right\rangle \mathrm{e}^{-\mathrm{i} \omega t} \mathrm{dt}
\end{equation}

\noindent \textcolor{black}{where $\boldsymbol{j}_{T}(\boldsymbol{q}, t)$ and $\boldsymbol{j}_{L}(\boldsymbol{q}, t)$ represents the transverse and the longitudinal current densities respectively and defined as: }
\begin{equation}
\boldsymbol{j}_{L}(\boldsymbol{q}, t)=\sum_{i}^{N}\left(\boldsymbol{v}_{\boldsymbol{i}}(t) \cdot \hat{\boldsymbol{q}}\right) \hat{\boldsymbol{q}} \mathrm{e}^{\mathrm{i} \boldsymbol{q} \cdot \boldsymbol{r}_{i}(t)}
\end{equation}
\begin{equation}
\boldsymbol{j}_{T}(\boldsymbol{q}, t)=\sum_{i}^{N}\left[\boldsymbol{v}_{\boldsymbol{i}}(t)-\left(\boldsymbol{v}_{\boldsymbol{i}}(t) \cdot \hat{\boldsymbol{q}}\right) \hat{\boldsymbol{q}}\right] \mathrm{e}^{\mathrm{i} \boldsymbol{q} \cdot \boldsymbol{r}_{i}(t)}
\end{equation}
\noindent \textcolor{black}{where $N$ is the number of atoms considered, $\boldsymbol{v}_{i}(t)$ and $\boldsymbol{r}_{i}(t)$ are the velocity and the position of atom $i$ at time $t$, respectively. 
The calculated dynamical structure factor of the first amorphous silicon wall is shown in Figure \ref{fig:SF-IR}a) for transverse waves and \ref{fig:SF-IR}b) for longitudinal ones. Despite the atomic disorder of the wall, there is a discernible dispersion for transverse waves up to a frequency of $6$ THz. A distinct dispersion with broadening is also noted up to $11$ THz along the longitudinal direction. By contrast, the dynamical structure factor of the first amorphous silica wall shown in Figure \ref{fig:SF-IR}c) for transverse waves and \ref{fig:SF-IR}d) for longitudinal waves, does not display such dispersion even for frequencies as low as $1$ THz in both transverse and longitudinal directions. To extract the phonon lifetime from the dynamical structure factor, for each wavevector, we fit the resulting spectral peaks both in the longitudinal and transverse components to a Gaussian function to obtain the full-width at half-maximum (FWHM) of the peaks. The phonon lifetime \(\tau\) is then calculated from the FWHM \(\Gamma\) as \(\tau = \frac{1}{\pi \Gamma}\). The results are shown as plots of the product of lifetime and vibrational frequency as a function of the vibrational frequencies in Figure \ref{fig:SF-IR}e) for amorphous silicon and \ref{fig:SF-IR}f) for amorphous silica. The Ioffe-Regel criterion is defined when the period of vibration is equal to the lifetime and is presented by the horizontal line separating the transition from propagons to diffusons. The IR crossover for the amorphous silicon wall occurs at around $7$ THz for transverse waves. For longitudinal waves, the IR crossover occurs at approximately $12$ THz (for comparison with bulk amorphous silicon and bulk silica} \textcolor{black}{see the Supplemental Material \cite{supplement}, see also
references \cite[]{Moon-PhysRevB.97.024201,Tanguy-PhysRevE.93.023006,SNTaraskin_1999} therein)}. Both of these values are consistent with the qualitative estimate of the transition frequency based on the structure factor shown in Figure \ref{fig:SF-IR}a) and b) and with previously reported values for bulk amorphous silicon \cite[]{Moon-PhysRevB.97.024201,Tanguy-PhysRevE.93.023006}. For the amorphous silica wall, the lifetime \(\times\) vibrational frequency values are all less than the IR criterion, indicating that all modes are non-propagating as predicted by our structural factor calculations. This behavior is also observed in bulk amorphous silica, where the crossover IR occurs at very low frequencies, approximately $1$ THz, for both longitudinal and transverse polarizations\cite[]{SNTaraskin_1999}. \\
We now discuss the contribution of propagons and diffusons to the thermal spectra displayed in Figures \ref{fig:5} and \ref{fig:inelastic}. For gold/amorphous silicon interfaces, elastic transmission at low frequencies ($\omega \approx 6$ THz) is due to propagating modes (propagons) for both transverse and longitudinal channels, as shown in Figure \ref{fig:5}a). Inelastic phonon transmission, which covers both low and high frequency ranges, is also due to propagons both transverse and longitudinal at low frequencies. However, in the intermediate frequency range, longitudinal propagons coexist with transverse diffusons. At high frequencies, the transmission is dominated by diffusons both transverse and longitudinal, as illustrated in Figure \ref{fig:inelastic}a). In contrast, the gold/amorphous silica interface exhibits elastic transmission at low frequencies ($\omega \approx 6$ THz) due to non-propagating modes (diffusons) for both transverse and longitudinal channels, as illustrated in Figure \ref{fig:5}b). Inelastic transmission, which extends to high frequencies, is entirely mediated by diffusons both transverse and longitudinal, as illustrated in Figure \ref{fig:inelastic}b).

\section{Conclusion}
We investigated the interfacial thermal conductance at interfaces involving crystalline or amorphous silicon and silica with gold, using non-equilibrium molecular dynamics simulations. We report a substantial threefold (twofold) increase in interfacial thermal conductance specifically for gold-amorphous silicon (silica) interfaces. This noticeable enhancement was observed across both planar and atomistically rough interfaces. \textcolor{black}{
The enhancement of interfacial thermal conductance at gold/amorphous interfaces can be attributed to both elastic and inelastic phonon scattering. Elastic thermal conductance is notably increased due to the combined effects of enhanced bonding and atomic displacement cross-correlations. Specifically, the stiffer bonds between atoms at gold/amorphous interfaces, where gold is more commensurate with the amorphous structures, contribute to this increase compared to gold crystalline interfaces. Inelastic transmission is also increased due to the higher level of anharmonicity in the vibrational modes of the amorphous structures. 
We have characterized the nature of the vibrational modes
carrying energy at the interfaces between gold and the amorphous materials. First, we showed that, for all the interfaces investigated, all the vibrational modes are delocalized. Secondly, we classified the modes at the gold-amorphous structures in terms of propagons and diffusons. 
For gold-amorphous silicon, we conclude that elastic phonon scattering involves low frequencies ($\omega < 6$ THz), lower than the Ioffe-Regel criterion for both transverse and longitudinal excitations. Therefore, only propagons participate to elastic phonon scattering. Inelastic phonon transmission at gold-amorphous silicon interfaces spans a broader range of frequencies, and as a consequence involves a mixture of propagons~(both transverse and longitudinal) and diffusons.
By contrast, at gold-amorphous interfaces, only diffusons participate to interfacial phonon scattering wether elastic or inelastic. This is the consequence of the very low Ioffe-Regel frequencies in silica in comparison with those of amorphous silicon.}

\textcolor{black}{Several perspectives are open by this work. First, it would be interesting to investigate different metal/amorphous silicon interfaces to further understand the role of the metal structure factor on interfacial heat transfer. Secondly, while we saw that electron-phonon processes play a negligible role at gold-silicon interfaces, it remains to characterize these coupled processes at metal-silica interfaces. 
}

\section*{Conflicts of interest}
There are no conflicts of interest to declare.

\section*{Acknowledgement}
We thank interesting discussions with A. Rajabpour, F. Banfi, N. Horny, X. Ruan and K. Termentzidis. This work is supported by the ANR, France project "CASTEX" ANR-21-CE30-0027-01. 
%

\section*{Data availability}

The data that support the findings of this study are available from the corresponding author upon reasonable request.


\begin{thebibliography}{0}%
\makeatletter
\providecommand \@ifxundefined [1]{%
 \@ifx{#1\undefined}
}%
\providecommand \@ifnum [1]{%
 \ifnum #1\expandafter \@firstoftwo
 \else \expandafter \@secondoftwo
 \fi
}%
\providecommand \@ifx [1]{%
 \ifx #1\expandafter \@firstoftwo
 \else \expandafter \@secondoftwo
 \fi
}%
\providecommand \natexlab [1]{#1}%
\providecommand \enquote  [1]{``#1''}%
\providecommand \bibnamefont  [1]{#1}%
\providecommand \bibfnamefont [1]{#1}%
\providecommand \citenamefont [1]{#1}%
\providecommand \href@noop [0]{\@secondoftwo}%
\providecommand \href [0]{\begingroup \@sanitize@url \@href}%
\providecommand \@href[1]{\@@startlink{#1}\@@href}%
\providecommand \@@href[1]{\endgroup#1\@@endlink}%
\providecommand \@sanitize@url [0]{\catcode `\\12\catcode `\$12\catcode `\&12\catcode `\#12\catcode `\^12\catcode `\_12\catcode `\%12\relax}%
\providecommand \@@startlink[1]{}%
\providecommand \@@endlink[0]{}%
\providecommand \url  [0]{\begingroup\@sanitize@url \@url }%
\providecommand \@url [1]{\endgroup\@href {#1}{\urlprefix }}%
\providecommand \urlprefix  [0]{URL }%
\providecommand \Eprint [0]{\href }%
\providecommand \doibase [0]{https://doi.org/}%
\providecommand \selectlanguage [0]{\@gobble}%
\providecommand \bibinfo  [0]{\@secondoftwo}%
\providecommand \bibfield  [0]{\@secondoftwo}%
\providecommand \translation [1]{[#1]}%
\providecommand \BibitemOpen [0]{}%
\providecommand \bibitemStop [0]{}%
\providecommand \bibitemNoStop [0]{.\EOS\space}%
\providecommand \EOS [0]{\spacefactor3000\relax}%
\providecommand \BibitemShut  [1]{\csname bibitem#1\endcsname}%
\let\auto@bib@innerbib\@empty
\end{thebibliography}%


\begin{thebibliography}{66}%
\makeatletter
\providecommand \@ifxundefined [1]{%
 \@ifx{#1\undefined}
}%
\providecommand \@ifnum [1]{%
 \ifnum #1\expandafter \@firstoftwo
 \else \expandafter \@secondoftwo
 \fi
}%
\providecommand \@ifx [1]{%
 \ifx #1\expandafter \@firstoftwo
 \else \expandafter \@secondoftwo
 \fi
}%
\providecommand \natexlab [1]{#1}%
\providecommand \enquote  [1]{``#1''}%
\providecommand \bibnamefont  [1]{#1}%
\providecommand \bibfnamefont [1]{#1}%
\providecommand \citenamefont [1]{#1}%
\providecommand \href@noop [0]{\@secondoftwo}%
\providecommand \href [0]{\begingroup \@sanitize@url \@href}%
\providecommand \@href[1]{\@@startlink{#1}\@@href}%
\providecommand \@@href[1]{\endgroup#1\@@endlink}%
\providecommand \@sanitize@url [0]{\catcode `\\12\catcode `\$12\catcode `\&12\catcode `\#12\catcode `\^12\catcode `\_12\catcode `\%12\relax}%
\providecommand \@@startlink[1]{}%
\providecommand \@@endlink[0]{}%
\providecommand \url  [0]{\begingroup\@sanitize@url \@url }%
\providecommand \@url [1]{\endgroup\@href {#1}{\urlprefix }}%
\providecommand \urlprefix  [0]{URL }%
\providecommand \Eprint [0]{\href }%
\providecommand \doibase [0]{https://doi.org/}%
\providecommand \selectlanguage [0]{\@gobble}%
\providecommand \bibinfo  [0]{\@secondoftwo}%
\providecommand \bibfield  [0]{\@secondoftwo}%
\providecommand \translation [1]{[#1]}%
\providecommand \BibitemOpen [0]{}%
\providecommand \bibitemStop [0]{}%
\providecommand \bibitemNoStop [0]{.\EOS\space}%
\providecommand \EOS [0]{\spacefactor3000\relax}%
\providecommand \BibitemShut  [1]{\csname bibitem#1\endcsname}%
\let\auto@bib@innerbib\@empty
\bibitem [{\citenamefont {Shen}\ \emph {et~al.}(2017)\citenamefont {Shen}, \citenamefont {Gaskins}, \citenamefont {Xie}, \citenamefont {Foley}, \citenamefont {Cheaito}, \citenamefont {Hopkins},\ and\ \citenamefont {Campbell}}]{shrinkdevices1}%
  \BibitemOpen
  \bibfield  {author} {\bibinfo {author} {\bibfnamefont {Y.}~\bibnamefont {Shen}}, \bibinfo {author} {\bibfnamefont {J.}~\bibnamefont {Gaskins}}, \bibinfo {author} {\bibfnamefont {X.}~\bibnamefont {Xie}}, \bibinfo {author} {\bibfnamefont {B.}~\bibnamefont {Foley}}, \bibinfo {author} {\bibfnamefont {R.}~\bibnamefont {Cheaito}}, \bibinfo {author} {\bibfnamefont {P.}~\bibnamefont {Hopkins}},\ and\ \bibinfo {author} {\bibfnamefont {J.}~\bibnamefont {Campbell}},\ }\bibfield  {title} {{\selectlanguage {English (US)}\bibinfo {title} {Thermal analysis of high-power flip-chip-bonded photodiodes}},\ }\href {https://doi.org/10.1109/JLT.2017.2736884} {\bibfield  {journal} {\bibinfo  {journal} {Journal of Lightwave Technology}\ }\textbf {\bibinfo {volume} {35}},\ \bibinfo {pages} {4242} (\bibinfo {year} {2017})}\BibitemShut {NoStop}%
\bibitem [{\citenamefont {Wong}\ \emph {et~al.}(2010)\citenamefont {Wong}, \citenamefont {Raoux}, \citenamefont {Kim}, \citenamefont {Liang}, \citenamefont {Reifenberg}, \citenamefont {Rajendran}, \citenamefont {Asheghi},\ and\ \citenamefont {Goodson}}]{shrinkdevices2}%
  \BibitemOpen
  \bibfield  {author} {\bibinfo {author} {\bibfnamefont {H.-S.~P.}\ \bibnamefont {Wong}}, \bibinfo {author} {\bibfnamefont {S.}~\bibnamefont {Raoux}}, \bibinfo {author} {\bibfnamefont {S.}~\bibnamefont {Kim}}, \bibinfo {author} {\bibfnamefont {J.}~\bibnamefont {Liang}}, \bibinfo {author} {\bibfnamefont {J.~P.}\ \bibnamefont {Reifenberg}}, \bibinfo {author} {\bibfnamefont {B.}~\bibnamefont {Rajendran}}, \bibinfo {author} {\bibfnamefont {M.}~\bibnamefont {Asheghi}},\ and\ \bibinfo {author} {\bibfnamefont {K.~E.}\ \bibnamefont {Goodson}},\ }\bibfield  {title} {\bibinfo {title} {Phase change memory},\ }\href {https://doi.org/10.1109/JPROC.2010.2070050} {\bibfield  {journal} {\bibinfo  {journal} {Proceedings of the IEEE}\ }\textbf {\bibinfo {volume} {98}},\ \bibinfo {pages} {2201} (\bibinfo {year} {2010})}\BibitemShut {NoStop}%
\bibitem [{\citenamefont {Cahill}\ \emph {et~al.}(2014)\citenamefont {Cahill}, \citenamefont {Braun}, \citenamefont {Chen}, \citenamefont {Clarke}, \citenamefont {Fan}, \citenamefont {Goodson}, \citenamefont {Keblinski}, \citenamefont {King}, \citenamefont {Mahan}, \citenamefont {Majumdar}, \citenamefont {Maris}, \citenamefont {Phillpot}, \citenamefont {Pop},\ and\ \citenamefont {Shi}}]{shrinkdevices3}%
  \BibitemOpen
  \bibfield  {author} {\bibinfo {author} {\bibfnamefont {D.~G.}\ \bibnamefont {Cahill}}, \bibinfo {author} {\bibfnamefont {P.~V.}\ \bibnamefont {Braun}}, \bibinfo {author} {\bibfnamefont {G.}~\bibnamefont {Chen}}, \bibinfo {author} {\bibfnamefont {D.~R.}\ \bibnamefont {Clarke}}, \bibinfo {author} {\bibfnamefont {S.}~\bibnamefont {Fan}}, \bibinfo {author} {\bibfnamefont {K.~E.}\ \bibnamefont {Goodson}}, \bibinfo {author} {\bibfnamefont {P.}~\bibnamefont {Keblinski}}, \bibinfo {author} {\bibfnamefont {W.~P.}\ \bibnamefont {King}}, \bibinfo {author} {\bibfnamefont {G.~D.}\ \bibnamefont {Mahan}}, \bibinfo {author} {\bibfnamefont {A.}~\bibnamefont {Majumdar}}, \bibinfo {author} {\bibfnamefont {H.~J.}\ \bibnamefont {Maris}}, \bibinfo {author} {\bibfnamefont {S.~R.}\ \bibnamefont {Phillpot}}, \bibinfo {author} {\bibfnamefont {E.}~\bibnamefont {Pop}},\ and\ \bibinfo {author} {\bibfnamefont {L.}~\bibnamefont {Shi}},\ }\bibfield  {title} {\bibinfo {title} {{Nanoscale thermal transport. II. 2003–2012}},\ }\href
  {https://doi.org/10.1063/1.4832615} {\bibfield  {journal} {\bibinfo  {journal} {Applied Physics Reviews}\ }\textbf {\bibinfo {volume} {1}},\ \bibinfo {pages} {011305} (\bibinfo {year} {2014})}\BibitemShut {NoStop}%
\bibitem [{\citenamefont {Kapitza}(1941)}]{kapitza}%
  \BibitemOpen
  \bibfield  {author} {\bibinfo {author} {\bibfnamefont {P.~L.}\ \bibnamefont {Kapitza}},\ }\bibfield  {title} {\bibinfo {title} {Heat transfer and superfluidity of helium ii},\ }\href {https://doi.org/10.1103/PhysRev.60.354} {\bibfield  {journal} {\bibinfo  {journal} {Phys. Rev.}\ }\textbf {\bibinfo {volume} {60}},\ \bibinfo {pages} {354} (\bibinfo {year} {1941})}\BibitemShut {NoStop}%
\bibitem [{\citenamefont {Cahill}\ \emph {et~al.}(2003)\citenamefont {Cahill}, \citenamefont {Ford}, \citenamefont {Goodson}, \citenamefont {Mahan}, \citenamefont {Majumdar}, \citenamefont {Maris}, \citenamefont {Merlin},\ and\ \citenamefont {Phillpot}}]{ITCbyexp1}%
  \BibitemOpen
  \bibfield  {author} {\bibinfo {author} {\bibfnamefont {D.}~\bibnamefont {Cahill}}, \bibinfo {author} {\bibfnamefont {W.}~\bibnamefont {Ford}}, \bibinfo {author} {\bibfnamefont {K.}~\bibnamefont {Goodson}}, \bibinfo {author} {\bibfnamefont {G.}~\bibnamefont {Mahan}}, \bibinfo {author} {\bibfnamefont {A.}~\bibnamefont {Majumdar}}, \bibinfo {author} {\bibfnamefont {H.}~\bibnamefont {Maris}}, \bibinfo {author} {\bibfnamefont {R.}~\bibnamefont {Merlin}},\ and\ \bibinfo {author} {\bibfnamefont {S.}~\bibnamefont {Phillpot}},\ }\bibfield  {title} {{\selectlanguage {English (US)}\bibinfo {title} {Nanoscale thermal transport}},\ }\href {https://doi.org/10.1063/1.1524305} {\bibfield  {journal} {\bibinfo  {journal} {Journal of Applied Physics}\ }\textbf {\bibinfo {volume} {93}},\ \bibinfo {pages} {793} (\bibinfo {year} {2003})}\BibitemShut {NoStop}%
\bibitem [{\citenamefont {Lyeo}\ and\ \citenamefont {Cahill}(2006)}]{ITCbyexp2}%
  \BibitemOpen
  \bibfield  {author} {\bibinfo {author} {\bibfnamefont {H.-K.}\ \bibnamefont {Lyeo}}\ and\ \bibinfo {author} {\bibfnamefont {D.~G.}\ \bibnamefont {Cahill}},\ }\bibfield  {title} {\bibinfo {title} {Thermal conductance of interfaces between highly dissimilar materials},\ }\href {https://doi.org/10.1103/PhysRevB.73.144301} {\bibfield  {journal} {\bibinfo  {journal} {Phys. Rev. B}\ }\textbf {\bibinfo {volume} {73}},\ \bibinfo {pages} {144301} (\bibinfo {year} {2006})}\BibitemShut {NoStop}%
\bibitem [{\citenamefont {Juvé}\ \emph {et~al.}(2009)\citenamefont {Juvé}, \citenamefont {Scardamaglia}, \citenamefont {Maioli}, \citenamefont {Crut}, \citenamefont {Merabia}, \citenamefont {Joly}, \citenamefont {Fatti},\ and\ \citenamefont {Vallée}}]{expITCamorphous}%
  \BibitemOpen
  \bibfield  {author} {\bibinfo {author} {\bibfnamefont {V.}~\bibnamefont {Juvé}}, \bibinfo {author} {\bibfnamefont {M.}~\bibnamefont {Scardamaglia}}, \bibinfo {author} {\bibfnamefont {P.}~\bibnamefont {Maioli}}, \bibinfo {author} {\bibfnamefont {A.}~\bibnamefont {Crut}}, \bibinfo {author} {\bibfnamefont {S.}~\bibnamefont {Merabia}}, \bibinfo {author} {\bibfnamefont {L.}~\bibnamefont {Joly}}, \bibinfo {author} {\bibfnamefont {N.}~\bibnamefont {Fatti}},\ and\ \bibinfo {author} {\bibfnamefont {F.}~\bibnamefont {Vallée}},\ }\bibfield  {title} {\bibinfo {title} {Cooling dynamics and thermal interface resistance of glass-embedded metal nanoparticles},\ }\href {https://doi.org/10.1103/PhysRevB.80.195406} {\bibfield  {journal} {\bibinfo  {journal} {Physical Review B}\ }\textbf {\bibinfo {volume} {80}} (\bibinfo {year} {2009})}\BibitemShut {NoStop}%
\bibitem [{\citenamefont {POLLACK}(1969)}]{IntroDMM}%
  \BibitemOpen
  \bibfield  {author} {\bibinfo {author} {\bibfnamefont {G.~L.}\ \bibnamefont {POLLACK}},\ }\bibfield  {title} {\bibinfo {title} {Kapitza resistance},\ }\href {https://doi.org/10.1103/RevModPhys.41.48} {\bibfield  {journal} {\bibinfo  {journal} {Rev. Mod. Phys.}\ }\textbf {\bibinfo {volume} {41}},\ \bibinfo {pages} {48} (\bibinfo {year} {1969})}\BibitemShut {NoStop}%
\bibitem [{\citenamefont {Little}(1959)}]{introAMM}%
  \BibitemOpen
  \bibfield  {author} {\bibinfo {author} {\bibfnamefont {W.~A.}\ \bibnamefont {Little}},\ }\bibfield  {title} {\bibinfo {title} {The transport of heat between dissimilar solids at low temperatures},\ }\href {https://doi.org/10.1139/p59-037} {\bibfield  {journal} {\bibinfo  {journal} {Canadian Journal of Physics}\ }\textbf {\bibinfo {volume} {37}},\ \bibinfo {pages} {334} (\bibinfo {year} {1959})}\BibitemShut {NoStop}%
\bibitem [{\citenamefont {S\"a\"askilahti}\ \emph {et~al.}(2014)\citenamefont {S\"a\"askilahti}, \citenamefont {Oksanen}, \citenamefont {Tulkki},\ and\ \citenamefont {Volz}}]{Saaskilahti_prb2014}%
  \BibitemOpen
  \bibfield  {author} {\bibinfo {author} {\bibfnamefont {K.}~\bibnamefont {S\"a\"askilahti}}, \bibinfo {author} {\bibfnamefont {J.}~\bibnamefont {Oksanen}}, \bibinfo {author} {\bibfnamefont {J.}~\bibnamefont {Tulkki}},\ and\ \bibinfo {author} {\bibfnamefont {S.}~\bibnamefont {Volz}},\ }\bibfield  {title} {\bibinfo {title} {Role of anharmonic phonon scattering in the spectrally decomposed thermal conductance at planar interfaces},\ }\href {https://doi.org/https://doi.org/10.1103/PhysRevB.90.134312} {\bibfield  {journal} {\bibinfo  {journal} {Phys. Rev. B}\ }\textbf {\bibinfo {volume} {90}},\ \bibinfo {pages} {134312} (\bibinfo {year} {2014})}\BibitemShut {NoStop}%
\bibitem [{\citenamefont {Cahill}\ \emph {et~al.}(2002)\citenamefont {Cahill}, \citenamefont {Ford}, \citenamefont {Goodson}, \citenamefont {Mahan}, \citenamefont {Majumdar}, \citenamefont {Maris}, \citenamefont {Merlin},\ and\ \citenamefont {Phillpot}}]{introMD1}%
  \BibitemOpen
  \bibfield  {author} {\bibinfo {author} {\bibfnamefont {D.~G.}\ \bibnamefont {Cahill}}, \bibinfo {author} {\bibfnamefont {W.~K.}\ \bibnamefont {Ford}}, \bibinfo {author} {\bibfnamefont {K.~E.}\ \bibnamefont {Goodson}}, \bibinfo {author} {\bibfnamefont {G.~D.}\ \bibnamefont {Mahan}}, \bibinfo {author} {\bibfnamefont {A.}~\bibnamefont {Majumdar}}, \bibinfo {author} {\bibfnamefont {H.~J.}\ \bibnamefont {Maris}}, \bibinfo {author} {\bibfnamefont {R.}~\bibnamefont {Merlin}},\ and\ \bibinfo {author} {\bibfnamefont {S.~R.}\ \bibnamefont {Phillpot}},\ }\bibfield  {title} {\bibinfo {title} {{Nanoscale thermal transport}},\ }\href {https://doi.org/10.1063/1.1524305} {\bibfield  {journal} {\bibinfo  {journal} {Journal of Applied Physics}\ }\textbf {\bibinfo {volume} {93}},\ \bibinfo {pages} {793} (\bibinfo {year} {2002})}\BibitemShut {NoStop}%
\bibitem [{\citenamefont {Landry}\ and\ \citenamefont {McGaughey}(2009)}]{intromd2}%
  \BibitemOpen
  \bibfield  {author} {\bibinfo {author} {\bibfnamefont {E.~S.}\ \bibnamefont {Landry}}\ and\ \bibinfo {author} {\bibfnamefont {A.~J.~H.}\ \bibnamefont {McGaughey}},\ }\bibfield  {title} {\bibinfo {title} {Thermal boundary resistance predictions from molecular dynamics simulations and theoretical calculations},\ }\href {https://doi.org/10.1103/PhysRevB.80.165304} {\bibfield  {journal} {\bibinfo  {journal} {Phys. Rev. B}\ }\textbf {\bibinfo {volume} {80}},\ \bibinfo {pages} {165304} (\bibinfo {year} {2009})}\BibitemShut {NoStop}%
\bibitem [{\citenamefont {Merabia}\ and\ \citenamefont {Termentzidis}(2012)}]{merabia-2012}%
  \BibitemOpen
  \bibfield  {author} {\bibinfo {author} {\bibfnamefont {S.}~\bibnamefont {Merabia}}\ and\ \bibinfo {author} {\bibfnamefont {K.}~\bibnamefont {Termentzidis}},\ }\bibfield  {title} {\bibinfo {title} {Thermal conductance at the interface between crystals using equilibrium and nonequilibrium molecular dynamics},\ }\href {https://doi.org/10.1103/PhysRevB.86.094303} {\bibfield  {journal} {\bibinfo  {journal} {Phys. Rev. B}\ }\textbf {\bibinfo {volume} {86}},\ \bibinfo {pages} {094303} (\bibinfo {year} {2012})}\BibitemShut {NoStop}%
\bibitem [{\citenamefont {Chen}\ \emph {et~al.}(2022)\citenamefont {Chen}, \citenamefont {Xu}, \citenamefont {Zhou},\ and\ \citenamefont {Li}}]{chen-2022}%
  \BibitemOpen
  \bibfield  {author} {\bibinfo {author} {\bibfnamefont {J.}~\bibnamefont {Chen}}, \bibinfo {author} {\bibfnamefont {X.}~\bibnamefont {Xu}}, \bibinfo {author} {\bibfnamefont {J.}~\bibnamefont {Zhou}},\ and\ \bibinfo {author} {\bibfnamefont {B.}~\bibnamefont {Li}},\ }\bibfield  {title} {\bibinfo {title} {Interfacial thermal resistance: Past, present, and future},\ }\href {https://doi.org/10.1103/RevModPhys.94.025002} {\bibfield  {journal} {\bibinfo  {journal} {Rev. Mod. Phys.}\ }\textbf {\bibinfo {volume} {94}},\ \bibinfo {pages} {025002} (\bibinfo {year} {2022})}\BibitemShut {NoStop}%
\bibitem [{\citenamefont {Giri}\ \emph {et~al.}(2015)\citenamefont {Giri}, \citenamefont {Hopkins}, \citenamefont {Wessel},\ and\ \citenamefont {Duda}}]{highITC1}%
  \BibitemOpen
  \bibfield  {author} {\bibinfo {author} {\bibfnamefont {A.}~\bibnamefont {Giri}}, \bibinfo {author} {\bibfnamefont {P.~E.}\ \bibnamefont {Hopkins}}, \bibinfo {author} {\bibfnamefont {J.~G.}\ \bibnamefont {Wessel}},\ and\ \bibinfo {author} {\bibfnamefont {J.~C.}\ \bibnamefont {Duda}},\ }\bibfield  {title} {\bibinfo {title} {{Kapitza resistance and the thermal conductivity of amorphous superlattices}},\ }\href {https://doi.org/10.1063/1.4934511} {\bibfield  {journal} {\bibinfo  {journal} {Journal of Applied Physics}\ }\textbf {\bibinfo {volume} {118}},\ \bibinfo {pages} {165303} (\bibinfo {year} {2015})}\BibitemShut {NoStop}%
\bibitem [{\citenamefont {Allen}\ \emph {et~al.}(1999{\natexlab{a}})\citenamefont {Allen}, \citenamefont {Feldman}, \citenamefont {Fabian},\ and\ \citenamefont {Wooten}}]{Allen_1999}%
  \BibitemOpen
  \bibfield  {author} {\bibinfo {author} {\bibfnamefont {P.~B.}\ \bibnamefont {Allen}}, \bibinfo {author} {\bibfnamefont {J.~L.}\ \bibnamefont {Feldman}}, \bibinfo {author} {\bibfnamefont {J.}~\bibnamefont {Fabian}},\ and\ \bibinfo {author} {\bibfnamefont {F.}~\bibnamefont {Wooten}},\ }\bibfield  {title} {\bibinfo {title} {Diffusons, locons and propagons: Character of atomic vibrations in amorphous si},\ }\href {https://doi.org/10.1080/13642819908223054} {\bibfield  {journal} {\bibinfo  {journal} {Philosophical Magazine B}\ }\textbf {\bibinfo {volume} {79}},\ \bibinfo {pages} {1715–1731} (\bibinfo {year} {1999}{\natexlab{a}})}\BibitemShut {NoStop}%
\bibitem [{\citenamefont {Feldman}\ \emph {et~al.}(1993)\citenamefont {Feldman}, \citenamefont {Kluge}, \citenamefont {Allen},\ and\ \citenamefont {Wooten}}]{Fledman-PhysRevB.48.12589}%
  \BibitemOpen
  \bibfield  {author} {\bibinfo {author} {\bibfnamefont {J.~L.}\ \bibnamefont {Feldman}}, \bibinfo {author} {\bibfnamefont {M.~D.}\ \bibnamefont {Kluge}}, \bibinfo {author} {\bibfnamefont {P.~B.}\ \bibnamefont {Allen}},\ and\ \bibinfo {author} {\bibfnamefont {F.}~\bibnamefont {Wooten}},\ }\bibfield  {title} {\bibinfo {title} {Thermal conductivity and localization in glasses: Numerical study of a model of amorphous silicon},\ }\href {https://doi.org/10.1103/PhysRevB.48.12589} {\bibfield  {journal} {\bibinfo  {journal} {Phys. Rev. B}\ }\textbf {\bibinfo {volume} {48}},\ \bibinfo {pages} {12589} (\bibinfo {year} {1993})}\BibitemShut {NoStop}%
\bibitem [{\citenamefont {Ioffe}\ and\ \citenamefont {Regel}(1960)}]{ioffe1960non}%
  \BibitemOpen
  \bibfield  {author} {\bibinfo {author} {\bibfnamefont {A.}~\bibnamefont {Ioffe}}\ and\ \bibinfo {author} {\bibfnamefont {A.}~\bibnamefont {Regel}},\ }\bibfield  {title} {\bibinfo {title} {Non-crystalline, amorphous, and liquid electronic semiconductors},\ }in\ \href@noop {} {\emph {\bibinfo {booktitle} {Progress in semiconductors}}}\ (\bibinfo {year} {1960})\ pp.\ \bibinfo {pages} {237--291}\BibitemShut {NoStop}%
\bibitem [{\citenamefont {Gordiz}\ and\ \citenamefont {Henry}(2017)}]{highITC2}%
  \BibitemOpen
  \bibfield  {author} {\bibinfo {author} {\bibfnamefont {K.}~\bibnamefont {Gordiz}}\ and\ \bibinfo {author} {\bibfnamefont {A.}~\bibnamefont {Henry}},\ }\bibfield  {title} {\bibinfo {title} {{Phonon transport at interfaces between different phases of silicon and germanium}},\ }\href {https://doi.org/10.1063/1.4973573} {\bibfield  {journal} {\bibinfo  {journal} {Journal of Applied Physics}\ }\textbf {\bibinfo {volume} {121}},\ \bibinfo {pages} {025102} (\bibinfo {year} {2017})}\BibitemShut {NoStop}%
\bibitem [{\citenamefont {Murakami}\ \emph {et~al.}(2014)\citenamefont {Murakami}, \citenamefont {Hori}, \citenamefont {Shiga},\ and\ \citenamefont {Shiomi}}]{introSi/GeMD1}%
  \BibitemOpen
  \bibfield  {author} {\bibinfo {author} {\bibfnamefont {T.}~\bibnamefont {Murakami}}, \bibinfo {author} {\bibfnamefont {T.}~\bibnamefont {Hori}}, \bibinfo {author} {\bibfnamefont {T.}~\bibnamefont {Shiga}},\ and\ \bibinfo {author} {\bibfnamefont {J.}~\bibnamefont {Shiomi}},\ }\bibfield  {title} {\bibinfo {title} {Probing and tuning inelastic phonon conductance across finite-thickness interface},\ }\href@noop {} {\bibfield  {journal} {\bibinfo  {journal} {Applied Physics Express}\ }\textbf {\bibinfo {volume} {7}},\ \bibinfo {pages} {121801} (\bibinfo {year} {2014})}\BibitemShut {NoStop}%
\bibitem [{\citenamefont {Gordiz}\ and\ \citenamefont {Henry}(2016)}]{introGe/Simd2}%
  \BibitemOpen
  \bibfield  {author} {\bibinfo {author} {\bibfnamefont {K.}~\bibnamefont {Gordiz}}\ and\ \bibinfo {author} {\bibfnamefont {A.}~\bibnamefont {Henry}},\ }\bibfield  {title} {\bibinfo {title} {{Phonon transport at interfaces: Determining the correct modes of vibration}},\ }\href {https://doi.org/10.1063/1.4939207} {\bibfield  {journal} {\bibinfo  {journal} {Journal of Applied Physics}\ }\textbf {\bibinfo {volume} {119}},\ \bibinfo {pages} {015101} (\bibinfo {year} {2016})}\BibitemShut {NoStop}%
\bibitem [{\citenamefont {Giri}\ and\ \citenamefont {Hopkins}(2020)}]{gordizHopkins}%
  \BibitemOpen
  \bibfield  {author} {\bibinfo {author} {\bibfnamefont {A.}~\bibnamefont {Giri}}\ and\ \bibinfo {author} {\bibfnamefont {P.~E.}\ \bibnamefont {Hopkins}},\ }\bibfield  {title} {\bibinfo {title} {A review of experimental and computational advances in thermal boundary conductance and nanoscale thermal transport across solid interfaces},\ }\href {https://doi.org/https://doi.org/10.1002/adfm.201903857} {\bibfield  {journal} {\bibinfo  {journal} {Advanced Functional Materials}\ }\textbf {\bibinfo {volume} {30}},\ \bibinfo {pages} {1903857} (\bibinfo {year} {2020})}\BibitemShut {NoStop}%
\bibitem [{\citenamefont {France-Lanord}\ \emph {et~al.}(2014)\citenamefont {France-Lanord}, \citenamefont {Merabia}, \citenamefont {Albaret}, \citenamefont {Lacroix},\ and\ \citenamefont {Termentzidis}}]{France-Lanord_2014}%
  \BibitemOpen
  \bibfield  {author} {\bibinfo {author} {\bibfnamefont {A.}~\bibnamefont {France-Lanord}}, \bibinfo {author} {\bibfnamefont {S.}~\bibnamefont {Merabia}}, \bibinfo {author} {\bibfnamefont {T.}~\bibnamefont {Albaret}}, \bibinfo {author} {\bibfnamefont {D.}~\bibnamefont {Lacroix}},\ and\ \bibinfo {author} {\bibfnamefont {K.}~\bibnamefont {Termentzidis}},\ }\bibfield  {title} {\bibinfo {title} {Thermal properties of amorphous/crystalline silicon superlattices},\ }\href {https://doi.org/10.1088/0953-8984/26/35/355801} {\bibfield  {journal} {\bibinfo  {journal} {Journal of Physics: Condensed Matter}\ }\textbf {\bibinfo {volume} {26}},\ \bibinfo {pages} {355801} (\bibinfo {year} {2014})}\BibitemShut {NoStop}%
\bibitem [{\citenamefont {Alkurdi}\ \emph {et~al.}(2020)\citenamefont {Alkurdi}, \citenamefont {Lombard}, \citenamefont {Detcheverry},\ and\ \citenamefont {Merabia}}]{alkurdi-2020}%
  \BibitemOpen
  \bibfield  {author} {\bibinfo {author} {\bibfnamefont {A.}~\bibnamefont {Alkurdi}}, \bibinfo {author} {\bibfnamefont {J.}~\bibnamefont {Lombard}}, \bibinfo {author} {\bibfnamefont {F.}~\bibnamefont {Detcheverry}},\ and\ \bibinfo {author} {\bibfnamefont {S.}~\bibnamefont {Merabia}},\ }\bibfield  {title} {\bibinfo {title} {Enhanced heat transfer with metal-dielectric core-shell nanoparticles},\ }\href {https://doi.org/10.1103/PhysRevApplied.13.034036} {\bibfield  {journal} {\bibinfo  {journal} {Phys. Rev. Appl.}\ }\textbf {\bibinfo {volume} {13}},\ \bibinfo {pages} {034036} (\bibinfo {year} {2020})}\BibitemShut {NoStop}%
\bibitem [{\citenamefont {Hatamlee}\ \emph {et~al.}(2022)\citenamefont {Hatamlee}, \citenamefont {Jabbari},\ and\ \citenamefont {Rajabpour}}]{Raj-goldsilica}%
  \BibitemOpen
  \bibfield  {author} {\bibinfo {author} {\bibfnamefont {S.~M.}\ \bibnamefont {Hatamlee}}, \bibinfo {author} {\bibfnamefont {F.}~\bibnamefont {Jabbari}},\ and\ \bibinfo {author} {\bibfnamefont {A.}~\bibnamefont {Rajabpour}},\ }\bibfield  {title} {\bibinfo {title} {Interfacial thermal conductance between gold and sio 2 : A molecular dynamics study},\ }\href {https://doi.org/10.1080/15567265.2022.2066585} {\bibfield  {journal} {\bibinfo  {journal} {Nanoscale and Microscale Thermophysical Engineering}\ }\textbf {\bibinfo {volume} {26}} (\bibinfo {year} {2022})}\BibitemShut {NoStop}%
\bibitem [{\citenamefont {Zhang}\ \emph {et~al.}(2022)\citenamefont {Zhang}, \citenamefont {Yang},\ and\ \citenamefont {Cao}}]{zhang-2022}%
  \BibitemOpen
  \bibfield  {author} {\bibinfo {author} {\bibfnamefont {X.-D.}\ \bibnamefont {Zhang}}, \bibinfo {author} {\bibfnamefont {G.}~\bibnamefont {Yang}},\ and\ \bibinfo {author} {\bibfnamefont {B.-Y.}\ \bibnamefont {Cao}},\ }\bibfield  {title} {\bibinfo {title} {Bonding-enhanced interfacial thermal transport: Mechanisms, materials, and applications},\ }\href {https://doi.org/10.1002/admi.202200078} {\bibfield  {journal} {\bibinfo  {journal} {Adv. Mater. Interfaces}\ }\textbf {\bibinfo {volume} {9}},\ \bibinfo {pages} {2200078} (\bibinfo {year} {2022})}\BibitemShut {NoStop}%
\bibitem [{\citenamefont {Zong}\ \emph {et~al.}(2023)\citenamefont {Zong}, \citenamefont {Deng}, \citenamefont {Qin}, \citenamefont {Wan}, \citenamefont {Zhan}, \citenamefont {Ma},\ and\ \citenamefont {Yang}}]{zong-2023}%
  \BibitemOpen
  \bibfield  {author} {\bibinfo {author} {\bibfnamefont {Z.}~\bibnamefont {Zong}}, \bibinfo {author} {\bibfnamefont {S.}~\bibnamefont {Deng}}, \bibinfo {author} {\bibfnamefont {Y.}~\bibnamefont {Qin}}, \bibinfo {author} {\bibfnamefont {X.}~\bibnamefont {Wan}}, \bibinfo {author} {\bibfnamefont {J.}~\bibnamefont {Zhan}}, \bibinfo {author} {\bibfnamefont {D.}~\bibnamefont {Ma}},\ and\ \bibinfo {author} {\bibfnamefont {N.}~\bibnamefont {Yang}},\ }\bibfield  {title} {\bibinfo {title} {Enhancing the interfacial thermal conductance of si/pvdf by strengthening atomic couplings},\ }\href {https://doi.org/10.1039/d3nr03706a} {\bibfield  {journal} {\bibinfo  {journal} {Nanoscale}\ }\textbf {\bibinfo {volume} {15}},\ \bibinfo {pages} {16472} (\bibinfo {year} {2023})}\BibitemShut {NoStop}%
\bibitem [{\citenamefont {Duda}\ \emph {et~al.}(2013)\citenamefont {Duda}, \citenamefont {Yang}, \citenamefont {Foley}, \citenamefont {Cheaito}, \citenamefont {Medlin}, \citenamefont {Jones},\ and\ \citenamefont {Hopkins}}]{duda}%
  \BibitemOpen
  \bibfield  {author} {\bibinfo {author} {\bibfnamefont {J.~C.}\ \bibnamefont {Duda}}, \bibinfo {author} {\bibfnamefont {C.-Y.~P.}\ \bibnamefont {Yang}}, \bibinfo {author} {\bibfnamefont {B.~M.}\ \bibnamefont {Foley}}, \bibinfo {author} {\bibfnamefont {R.}~\bibnamefont {Cheaito}}, \bibinfo {author} {\bibfnamefont {D.~L.}\ \bibnamefont {Medlin}}, \bibinfo {author} {\bibfnamefont {R.~E.}\ \bibnamefont {Jones}},\ and\ \bibinfo {author} {\bibfnamefont {P.~E.}\ \bibnamefont {Hopkins}},\ }\bibfield  {title} {\bibinfo {title} {Influence of interfacial properties on thermal transport at gold:silicon contacts},\ }\href@noop {} {\bibfield  {journal} {\bibinfo  {journal} {Appl. Phys. Lett.}\ }\textbf {\bibinfo {volume} {102}},\ \bibinfo {pages} {081902} (\bibinfo {year} {2013})}\BibitemShut {NoStop}%
\bibitem [{\citenamefont {Cheaito}\ \emph {et~al.}(2015)\citenamefont {Cheaito}, \citenamefont {Gaskins}, \citenamefont {Caplan}, \citenamefont {Donovan}, \citenamefont {Foley}, \citenamefont {Giri}, \citenamefont {Duda}, \citenamefont {Szwejkowski}, \citenamefont {Constantin}, \citenamefont {Brown-Shaklee}, \citenamefont {Ihlefeld},\ and\ \citenamefont {Hopkins}}]{Au-cSi_expITC}%
  \BibitemOpen
  \bibfield  {author} {\bibinfo {author} {\bibfnamefont {R.}~\bibnamefont {Cheaito}}, \bibinfo {author} {\bibfnamefont {J.~T.}\ \bibnamefont {Gaskins}}, \bibinfo {author} {\bibfnamefont {M.~E.}\ \bibnamefont {Caplan}}, \bibinfo {author} {\bibfnamefont {B.~F.}\ \bibnamefont {Donovan}}, \bibinfo {author} {\bibfnamefont {B.~M.}\ \bibnamefont {Foley}}, \bibinfo {author} {\bibfnamefont {A.}~\bibnamefont {Giri}}, \bibinfo {author} {\bibfnamefont {J.~C.}\ \bibnamefont {Duda}}, \bibinfo {author} {\bibfnamefont {C.~J.}\ \bibnamefont {Szwejkowski}}, \bibinfo {author} {\bibfnamefont {C.}~\bibnamefont {Constantin}}, \bibinfo {author} {\bibfnamefont {H.~J.}\ \bibnamefont {Brown-Shaklee}}, \bibinfo {author} {\bibfnamefont {J.~F.}\ \bibnamefont {Ihlefeld}},\ and\ \bibinfo {author} {\bibfnamefont {P.~E.}\ \bibnamefont {Hopkins}},\ }\bibfield  {title} {\bibinfo {title} {Thermal boundary conductance accumulation and interfacial phonon transmission: Measurements and theory},\ }\href {https://doi.org/10.1103/PhysRevB.91.035432}
  {\bibfield  {journal} {\bibinfo  {journal} {Phys. Rev. B}\ }\textbf {\bibinfo {volume} {91}},\ \bibinfo {pages} {035432} (\bibinfo {year} {2015})}\BibitemShut {NoStop}%
\bibitem [{\citenamefont {Oh}\ \emph {et~al.}(2011)\citenamefont {Oh}, \citenamefont {Kim}, \citenamefont {Rogers}, \citenamefont {Cahill},\ and\ \citenamefont {Sinha}}]{oh-dong-exp3}%
  \BibitemOpen
  \bibfield  {author} {\bibinfo {author} {\bibfnamefont {D.-W.}\ \bibnamefont {Oh}}, \bibinfo {author} {\bibfnamefont {S.}~\bibnamefont {Kim}}, \bibinfo {author} {\bibfnamefont {J.~A.}\ \bibnamefont {Rogers}}, \bibinfo {author} {\bibfnamefont {D.~G.}\ \bibnamefont {Cahill}},\ and\ \bibinfo {author} {\bibfnamefont {S.}~\bibnamefont {Sinha}},\ }\bibfield  {title} {\bibinfo {title} {Interfacial thermal conductance of transfer-printed metal films},\ }\href {https://doi.org/https://doi.org/10.1002/adma.201102994} {\bibfield  {journal} {\bibinfo  {journal} {Advanced Materials}\ }\textbf {\bibinfo {volume} {23}},\ \bibinfo {pages} {5028} (\bibinfo {year} {2011})}\BibitemShut {NoStop}%
\bibitem [{\citenamefont {Thompson}\ \emph {et~al.}(2022)\citenamefont {Thompson}, \citenamefont {Aktulga}, \citenamefont {Berger}, \citenamefont {Bolintineanu}, \citenamefont {Brown}, \citenamefont {Crozier}, \citenamefont {in~'t Veld}, \citenamefont {Kohlmeyer}, \citenamefont {Moore}, \citenamefont {Nguyen}, \citenamefont {Shan}, \citenamefont {Stevens}, \citenamefont {Tranchida}, \citenamefont {Trott},\ and\ \citenamefont {Plimpton}}]{LAMMPS}%
  \BibitemOpen
  \bibfield  {author} {\bibinfo {author} {\bibfnamefont {A.~P.}\ \bibnamefont {Thompson}}, \bibinfo {author} {\bibfnamefont {H.~M.}\ \bibnamefont {Aktulga}}, \bibinfo {author} {\bibfnamefont {R.}~\bibnamefont {Berger}}, \bibinfo {author} {\bibfnamefont {D.~S.}\ \bibnamefont {Bolintineanu}}, \bibinfo {author} {\bibfnamefont {W.~M.}\ \bibnamefont {Brown}}, \bibinfo {author} {\bibfnamefont {P.~S.}\ \bibnamefont {Crozier}}, \bibinfo {author} {\bibfnamefont {P.~J.}\ \bibnamefont {in~'t Veld}}, \bibinfo {author} {\bibfnamefont {A.}~\bibnamefont {Kohlmeyer}}, \bibinfo {author} {\bibfnamefont {S.~G.}\ \bibnamefont {Moore}}, \bibinfo {author} {\bibfnamefont {T.~D.}\ \bibnamefont {Nguyen}}, \bibinfo {author} {\bibfnamefont {R.}~\bibnamefont {Shan}}, \bibinfo {author} {\bibfnamefont {M.~J.}\ \bibnamefont {Stevens}}, \bibinfo {author} {\bibfnamefont {J.}~\bibnamefont {Tranchida}}, \bibinfo {author} {\bibfnamefont {C.}~\bibnamefont {Trott}},\ and\ \bibinfo {author} {\bibfnamefont {S.~J.}\ \bibnamefont {Plimpton}},\
  }\bibfield  {title} {\bibinfo {title} {{LAMMPS} - a flexible simulation tool for particle-based materials modeling at the atomic, meso, and continuum scales},\ }\href@noop {} {\bibfield  {journal} {\bibinfo  {journal} {Comp. Phys. Comm.}\ }\textbf {\bibinfo {volume} {271}},\ \bibinfo {pages} {108171} (\bibinfo {year} {2022})}\BibitemShut {NoStop}%
\bibitem [{\citenamefont {Stukowski}(2010)}]{ovito}%
  \BibitemOpen
  \bibfield  {author} {\bibinfo {author} {\bibfnamefont {A.}~\bibnamefont {Stukowski}},\ }\bibfield  {title} {\bibinfo {title} {{Visualization and analysis of atomistic simulation data with OVITO-the Open Visualization Tool}},\ }\bibfield  {journal} {\bibinfo  {journal} {{MODELLING AND SIMULATION IN MATERIALS SCIENCE AND ENGINEERING}}\ }\textbf {\bibinfo {volume} {{18}}},\ \href {https://doi.org/{10.1088/0965-0393/18/1/015012}} {{10.1088/0965-0393/18/1/015012}} (\bibinfo {year} {{2010}})\BibitemShut {NoStop}%
\bibitem [{\citenamefont {Stillinger}\ and\ \citenamefont {Weber}(1986)}]{SW}%
  \BibitemOpen
  \bibfield  {author} {\bibinfo {author} {\bibfnamefont {F.~H.}\ \bibnamefont {Stillinger}}\ and\ \bibinfo {author} {\bibfnamefont {T.~A.}\ \bibnamefont {Weber}},\ }\bibfield  {title} {\bibinfo {title} {Erratum: Computer simulation of local order in condensed phases of silicon [phys. rev. b 31, 5262 (1985)]},\ }\href@noop {} {\bibfield  {journal} {\bibinfo  {journal} {Phys. Rev. B}\ }\textbf {\bibinfo {volume} {33}},\ \bibinfo {pages} {1451} (\bibinfo {year} {1986})}\BibitemShut {NoStop}%
\bibitem [{\citenamefont {Munetoh}\ \emph {et~al.}(2007)\citenamefont {Munetoh}, \citenamefont {Motooka}, \citenamefont {Moriguchi},\ and\ \citenamefont {Shintani}}]{Tersoff-param}%
  \BibitemOpen
  \bibfield  {author} {\bibinfo {author} {\bibfnamefont {S.}~\bibnamefont {Munetoh}}, \bibinfo {author} {\bibfnamefont {T.}~\bibnamefont {Motooka}}, \bibinfo {author} {\bibfnamefont {K.}~\bibnamefont {Moriguchi}},\ and\ \bibinfo {author} {\bibfnamefont {A.}~\bibnamefont {Shintani}},\ }\bibfield  {title} {\bibinfo {title} {Interatomic potential for si–o systems using tersoff parameterization},\ }\href {https://doi.org/https://doi.org/10.1016/j.commatsci.2006.06.010} {\bibfield  {journal} {\bibinfo  {journal} {Computational Materials Science}\ }\textbf {\bibinfo {volume} {39}},\ \bibinfo {pages} {334} (\bibinfo {year} {2007})}\BibitemShut {NoStop}%
\bibitem [{\citenamefont {Heinz}\ \emph {et~al.}(2008)\citenamefont {Heinz}, \citenamefont {Vaia}, \citenamefont {Farmer},\ and\ \citenamefont {Naik}}]{gold-heinz}%
  \BibitemOpen
  \bibfield  {author} {\bibinfo {author} {\bibfnamefont {H.}~\bibnamefont {Heinz}}, \bibinfo {author} {\bibfnamefont {R.~A.}\ \bibnamefont {Vaia}}, \bibinfo {author} {\bibfnamefont {B.~L.}\ \bibnamefont {Farmer}},\ and\ \bibinfo {author} {\bibfnamefont {R.~R.}\ \bibnamefont {Naik}},\ }\bibfield  {title} {\bibinfo {title} {Accurate simulation of surfaces and interfaces of face-centered cubic metals using 12−6 and 9−6 lennard-jones potentials},\ }\href@noop {} {\bibfield  {journal} {\bibinfo  {journal} {The Journal of Physical Chemistry C}\ }\textbf {\bibinfo {volume} {112}},\ \bibinfo {pages} {17281} (\bibinfo {year} {2008})}\BibitemShut {NoStop}%
\bibitem [{\citenamefont {Rappe}\ \emph {et~al.}(1992)\citenamefont {Rappe}, \citenamefont {Casewit}, \citenamefont {Colwell}, \citenamefont {Goddard},\ and\ \citenamefont {Skiff}}]{LJ-interfaces}%
  \BibitemOpen
  \bibfield  {author} {\bibinfo {author} {\bibfnamefont {A.~K.}\ \bibnamefont {Rappe}}, \bibinfo {author} {\bibfnamefont {C.~J.}\ \bibnamefont {Casewit}}, \bibinfo {author} {\bibfnamefont {K.~S.}\ \bibnamefont {Colwell}}, \bibinfo {author} {\bibfnamefont {W.~A.~I.}\ \bibnamefont {Goddard}},\ and\ \bibinfo {author} {\bibfnamefont {W.~M.}\ \bibnamefont {Skiff}},\ }\bibfield  {title} {\bibinfo {title} {Uff, a full periodic table force field for molecular mechanics and molecular dynamics simulations},\ }\href@noop {} {\bibfield  {journal} {\bibinfo  {journal} {Journal of the American Chemical Society}\ }\textbf {\bibinfo {volume} {114}},\ \bibinfo {pages} {10024} (\bibinfo {year} {1992})}\BibitemShut {NoStop}%
\bibitem [{\citenamefont {Grochola}\ \emph {et~al.}(2005)\citenamefont {Grochola}, \citenamefont {Russo},\ and\ \citenamefont {Snook}}]{EAM-gold}%
  \BibitemOpen
  \bibfield  {author} {\bibinfo {author} {\bibfnamefont {G.}~\bibnamefont {Grochola}}, \bibinfo {author} {\bibfnamefont {S.~P.}\ \bibnamefont {Russo}},\ and\ \bibinfo {author} {\bibfnamefont {I.~K.}\ \bibnamefont {Snook}},\ }\bibfield  {title} {\bibinfo {title} {{On fitting a gold embedded atom method potential using the force matching method}},\ }\href {https://doi.org/10.1063/1.2124667} {\bibfield  {journal} {\bibinfo  {journal} {The Journal of Chemical Physics}\ }\textbf {\bibinfo {volume} {123}},\ \bibinfo {pages} {204719} (\bibinfo {year} {2005})}\BibitemShut {NoStop}%
\bibitem [{\citenamefont {Larkin}\ and\ \citenamefont {McGaughey}(2014)}]{asi-sio2conductivity}%
  \BibitemOpen
  \bibfield  {author} {\bibinfo {author} {\bibfnamefont {J.~M.}\ \bibnamefont {Larkin}}\ and\ \bibinfo {author} {\bibfnamefont {A.~J.~H.}\ \bibnamefont {McGaughey}},\ }\bibfield  {title} {\bibinfo {title} {Thermal conductivity accumulation in amorphous silica and amorphous silicon},\ }\href {https://doi.org/10.1103/PhysRevB.89.144303} {\bibfield  {journal} {\bibinfo  {journal} {Phys. Rev. B}\ }\textbf {\bibinfo {volume} {89}},\ \bibinfo {pages} {144303} (\bibinfo {year} {2014})}\BibitemShut {NoStop}%
\bibitem [{sup()}]{supplement}%
  \BibitemOpen
  \href@noop {} {}\bibinfo {note} {See Supplemental Material at [URL will be inserted by publisher] for results from density functional non-equilibrium Green’s function calculations, radial distribution functions, heat flux steady state conditions, and the influence of system length and temperature differences. Additional discussions cover the failure of the diffuse mismatch model, the effect of coupling strength on thermal conductance, and the projection of static and dynamical structure factors.}\BibitemShut {Stop}%
\bibitem [{\citenamefont {Laaziri}\ \emph {et~al.}(1999)\citenamefont {Laaziri}, \citenamefont {Kycia}, \citenamefont {Roorda}, \citenamefont {Chicoine}, \citenamefont {Robertson}, \citenamefont {Wang},\ and\ \citenamefont {Moss}}]{silicon-exp-rdf}%
  \BibitemOpen
  \bibfield  {author} {\bibinfo {author} {\bibfnamefont {K.}~\bibnamefont {Laaziri}}, \bibinfo {author} {\bibfnamefont {S.}~\bibnamefont {Kycia}}, \bibinfo {author} {\bibfnamefont {S.}~\bibnamefont {Roorda}}, \bibinfo {author} {\bibfnamefont {M.}~\bibnamefont {Chicoine}}, \bibinfo {author} {\bibfnamefont {J.~L.}\ \bibnamefont {Robertson}}, \bibinfo {author} {\bibfnamefont {J.}~\bibnamefont {Wang}},\ and\ \bibinfo {author} {\bibfnamefont {S.~C.}\ \bibnamefont {Moss}},\ }\bibfield  {title} {\bibinfo {title} {High-energy x-ray diffraction study of pure amorphous silicon},\ }\href {https://doi.org/10.1103/PhysRevB.60.13520} {\bibfield  {journal} {\bibinfo  {journal} {Phys. Rev. B}\ }\textbf {\bibinfo {volume} {60}},\ \bibinfo {pages} {13520} (\bibinfo {year} {1999})}\BibitemShut {NoStop}%
\bibitem [{\citenamefont {Lorch}(1969)}]{ELorch_1969}%
  \BibitemOpen
  \bibfield  {author} {\bibinfo {author} {\bibfnamefont {E.}~\bibnamefont {Lorch}},\ }\bibfield  {title} {\bibinfo {title} {Neutron diffraction by germania, silica and radiation-damaged silica glasses},\ }\href@noop {} {\bibfield  {journal} {\bibinfo  {journal} {Journal of Physics C: Solid State Physics}\ }\textbf {\bibinfo {volume} {2}},\ \bibinfo {pages} {229} (\bibinfo {year} {1969})}\BibitemShut {NoStop}%
\bibitem [{\citenamefont {Zhu}\ \emph {et~al.}(2018)\citenamefont {Zhu}, \citenamefont {Zheng}, \citenamefont {Cao},\ and\ \citenamefont {He}}]{cond2silica}%
  \BibitemOpen
  \bibfield  {author} {\bibinfo {author} {\bibfnamefont {W.}~\bibnamefont {Zhu}}, \bibinfo {author} {\bibfnamefont {G.}~\bibnamefont {Zheng}}, \bibinfo {author} {\bibfnamefont {S.}~\bibnamefont {Cao}},\ and\ \bibinfo {author} {\bibfnamefont {H.}~\bibnamefont {He}},\ }\bibfield  {title} {\bibinfo {title} {Thermal conductivity of amorphous sio2 thin film: A molecular dynamics study},\ }\href {https://doi.org/10.1038/s41598-018-28925-6} {\bibfield  {journal} {\bibinfo  {journal} {Scientific Reports}\ }\textbf {\bibinfo {volume} {8}} (\bibinfo {year} {2018})}\BibitemShut {NoStop}%
\bibitem [{\citenamefont {Huang}\ \emph {et~al.}(2009)\citenamefont {Huang}, \citenamefont {Tang}, \citenamefont {Yu},\ and\ \citenamefont {Bai}}]{sio2cond_1}%
  \BibitemOpen
  \bibfield  {author} {\bibinfo {author} {\bibfnamefont {Z.}~\bibnamefont {Huang}}, \bibinfo {author} {\bibfnamefont {Z.}~\bibnamefont {Tang}}, \bibinfo {author} {\bibfnamefont {J.}~\bibnamefont {Yu}},\ and\ \bibinfo {author} {\bibfnamefont {S.}~\bibnamefont {Bai}},\ }\bibfield  {title} {\bibinfo {title} {Thermal conductivity of amorphous and crystalline thin films by molecular dynamics simulation},\ }\href {https://doi.org/https://doi.org/10.1016/j.physb.2009.02.022} {\bibfield  {journal} {\bibinfo  {journal} {Physica B: Condensed Matter}\ }\textbf {\bibinfo {volume} {404}},\ \bibinfo {pages} {1790} (\bibinfo {year} {2009})}\BibitemShut {NoStop}%
\bibitem [{\citenamefont {{Nos{\'e}}}(1984)}]{nosee}%
  \BibitemOpen
  \bibfield  {author} {\bibinfo {author} {\bibfnamefont {S.}~\bibnamefont {{Nos{\'e}}}},\ }\bibfield  {title} {\bibinfo {title} {{A unified formulation of the constant temperature molecular dynamics methods}},\ }\href {https://doi.org/10.1063/1.447334} {\bibfield  {journal} {\bibinfo  {journal} {\jcp}\ }\textbf {\bibinfo {volume} {81}},\ \bibinfo {pages} {511} (\bibinfo {year} {1984})}\BibitemShut {NoStop}%
\bibitem [{\citenamefont {Hoover}(1985)}]{HOOver}%
  \BibitemOpen
  \bibfield  {author} {\bibinfo {author} {\bibfnamefont {W.~G.}\ \bibnamefont {Hoover}},\ }\bibfield  {title} {\bibinfo {title} {Canonical dynamics: Equilibrium phase-space distributions},\ }\href {https://doi.org/10.1103/PhysRevA.31.1695} {\bibfield  {journal} {\bibinfo  {journal} {Phys. Rev. A}\ }\textbf {\bibinfo {volume} {31}},\ \bibinfo {pages} {1695} (\bibinfo {year} {1985})}\BibitemShut {NoStop}%
\bibitem [{\citenamefont {Swartz}\ and\ \citenamefont {Pohl}(1989)}]{DMM}%
  \BibitemOpen
  \bibfield  {author} {\bibinfo {author} {\bibfnamefont {E.~T.}\ \bibnamefont {Swartz}}\ and\ \bibinfo {author} {\bibfnamefont {R.~O.}\ \bibnamefont {Pohl}},\ }\bibfield  {title} {\bibinfo {title} {Thermal boundary resistance},\ }\href {https://doi.org/10.1103/RevModPhys.61.605} {\bibfield  {journal} {\bibinfo  {journal} {Rev. Mod. Phys.}\ }\textbf {\bibinfo {volume} {61}},\ \bibinfo {pages} {605} (\bibinfo {year} {1989})}\BibitemShut {NoStop}%
\bibitem [{\citenamefont {Lombard}\ \emph {et~al.}(2014)\citenamefont {Lombard}, \citenamefont {Detcheverry},\ and\ \citenamefont {Merabia}}]{landau}%
  \BibitemOpen
  \bibfield  {author} {\bibinfo {author} {\bibfnamefont {J.}~\bibnamefont {Lombard}}, \bibinfo {author} {\bibfnamefont {F.}~\bibnamefont {Detcheverry}},\ and\ \bibinfo {author} {\bibfnamefont {S.}~\bibnamefont {Merabia}},\ }\bibfield  {title} {\bibinfo {title} {Influence of the electron-phonon interfacial conductance on the thermal transport at metal/dielectric interfaces},\ }\href {https://doi.org/10.1088/0953-8984/27/1/015007} {\bibfield  {journal} {\bibinfo  {journal} {Journal of physics. Condensed matter : an Institute of Physics journal}\ }\textbf {\bibinfo {volume} {27}},\ \bibinfo {pages} {015007} (\bibinfo {year} {2014})}\BibitemShut {NoStop}%
\bibitem [{\citenamefont {Fransson}\ \emph {et~al.}(2021)\citenamefont {Fransson}, \citenamefont {Slabanja}, \citenamefont {Erhart},\ and\ \citenamefont {Wahnström}}]{dynasor}%
  \BibitemOpen
  \bibfield  {author} {\bibinfo {author} {\bibfnamefont {E.}~\bibnamefont {Fransson}}, \bibinfo {author} {\bibfnamefont {M.}~\bibnamefont {Slabanja}}, \bibinfo {author} {\bibfnamefont {P.}~\bibnamefont {Erhart}},\ and\ \bibinfo {author} {\bibfnamefont {G.}~\bibnamefont {Wahnström}},\ }\bibfield  {title} {\bibinfo {title} {A tool for extracting dynamical structure factors and current correlation functions from molecular dynamics simulations},\ }\bibfield  {journal} {\bibinfo  {journal} {Advanced Theory and Simulations}\ }\textbf {\bibinfo {volume} {4}},\ \href {https://doi.org/10.1002/adts.202000240} {10.1002/adts.202000240} (\bibinfo {year} {2021})\BibitemShut {NoStop}%
\bibitem [{\citenamefont {Soler}\ \emph {et~al.}(2002)\citenamefont {Soler}, \citenamefont {Artacho}, \citenamefont {Gale}, \citenamefont {Garc\'{\i}a}, \citenamefont {Junquera}, \citenamefont {Ordej\'on},\ and\ \citenamefont {S\'anchez-Portal}}]{siesta}%
  \BibitemOpen
  \bibfield  {author} {\bibinfo {author} {\bibfnamefont {J.~M.}\ \bibnamefont {Soler}}, \bibinfo {author} {\bibfnamefont {E.}~\bibnamefont {Artacho}}, \bibinfo {author} {\bibfnamefont {J.~D.}\ \bibnamefont {Gale}}, \bibinfo {author} {\bibfnamefont {A.}~\bibnamefont {Garc\'{\i}a}}, \bibinfo {author} {\bibfnamefont {J.}~\bibnamefont {Junquera}}, \bibinfo {author} {\bibfnamefont {P.}~\bibnamefont {Ordej\'on}},\ and\ \bibinfo {author} {\bibfnamefont {D.}~\bibnamefont {S\'anchez-Portal}},\ }\bibfield  {title} {\bibinfo {title} {The siesta method for ab initio order-{N} materials simulation},\ }\href@noop {} {\bibfield  {journal} {\bibinfo  {journal} {J. Phys.: Condens. Matter}\ }\textbf {\bibinfo {volume} {14}},\ \bibinfo {pages} {2745} (\bibinfo {year} {2002})}\BibitemShut {NoStop}%
\bibitem [{\citenamefont {Perdew}\ \emph {et~al.}(1996)\citenamefont {Perdew}, \citenamefont {Burke},\ and\ \citenamefont {Ernzerhof}}]{perdew-prl-96}%
  \BibitemOpen
  \bibfield  {author} {\bibinfo {author} {\bibfnamefont {J.}~\bibnamefont {Perdew}}, \bibinfo {author} {\bibfnamefont {K.}~\bibnamefont {Burke}},\ and\ \bibinfo {author} {\bibfnamefont {M.}~\bibnamefont {Ernzerhof}},\ }\bibfield  {title} {\bibinfo {title} {Generalized gradient approximation made simple},\ }\href@noop {} {\bibfield  {journal} {\bibinfo  {journal} {Physical Review Letters}\ }\textbf {\bibinfo {volume} {77}},\ \bibinfo {pages} {3865} (\bibinfo {year} {1996})}\BibitemShut {NoStop}%
\bibitem [{\citenamefont {Troullier}\ and\ \citenamefont {Martins}(1991)}]{troullier-prb-91}%
  \BibitemOpen
  \bibfield  {author} {\bibinfo {author} {\bibfnamefont {N.}~\bibnamefont {Troullier}}\ and\ \bibinfo {author} {\bibfnamefont {J.~L.}\ \bibnamefont {Martins}},\ }\bibfield  {title} {\bibinfo {title} {Efficient pseudopotentials for plane-wave calculations},\ }\href@noop {} {\bibfield  {journal} {\bibinfo  {journal} {Physical Review B}\ }\textbf {\bibinfo {volume} {43}},\ \bibinfo {pages} {1993} (\bibinfo {year} {1991})}\BibitemShut {NoStop}%
\bibitem [{\citenamefont {Togo}\ and\ \citenamefont {Tanaka}(2015)}]{phonopy}%
  \BibitemOpen
  \bibfield  {author} {\bibinfo {author} {\bibfnamefont {A.}~\bibnamefont {Togo}}\ and\ \bibinfo {author} {\bibfnamefont {I.}~\bibnamefont {Tanaka}},\ }\bibfield  {title} {\bibinfo {title} {First principles phonon calculations in materials science},\ }\href@noop {} {\bibfield  {journal} {\bibinfo  {journal} {Scr. Mater.}\ }\textbf {\bibinfo {volume} {108}},\ \bibinfo {pages} {1} (\bibinfo {year} {2015})}\BibitemShut {NoStop}%
\bibitem [{\citenamefont {Adessi}\ \emph {et~al.}(2020)\citenamefont {Adessi}, \citenamefont {Pecorario}, \citenamefont {Th\'ebaud},\ and\ \citenamefont {Bouzerar}}]{adessi-pccp-2020}%
  \BibitemOpen
  \bibfield  {author} {\bibinfo {author} {\bibfnamefont {C.}~\bibnamefont {Adessi}}, \bibinfo {author} {\bibfnamefont {S.}~\bibnamefont {Pecorario}}, \bibinfo {author} {\bibfnamefont {S.}~\bibnamefont {Th\'ebaud}},\ and\ \bibinfo {author} {\bibfnamefont {G.}~\bibnamefont {Bouzerar}},\ }\bibfield  {title} {\bibinfo {title} {First principle investigation of the influence of sulfur vacancies on thermoelectric properties of single layered mos$_2$},\ }\href@noop {} {\bibfield  {journal} {\bibinfo  {journal} {Phys. Chem. Chem. Phys.}\ }\textbf {\bibinfo {volume} {22}},\ \bibinfo {pages} {15048} (\bibinfo {year} {2020})}\BibitemShut {NoStop}%
\bibitem [{\citenamefont {Merabia}\ and\ \citenamefont {Termentzidis}(2014)}]{merabia-2014}%
  \BibitemOpen
  \bibfield  {author} {\bibinfo {author} {\bibfnamefont {S.}~\bibnamefont {Merabia}}\ and\ \bibinfo {author} {\bibfnamefont {K.}~\bibnamefont {Termentzidis}},\ }\bibfield  {title} {\bibinfo {title} {Thermal boundary conductance across rough interfaces probed by molecular dynamics},\ }\href {https://doi.org/10.1103/PhysRevB.89.054309} {\bibfield  {journal} {\bibinfo  {journal} {Phys. Rev. B}\ }\textbf {\bibinfo {volume} {89}},\ \bibinfo {pages} {054309} (\bibinfo {year} {2014})}\BibitemShut {NoStop}%
\bibitem [{\citenamefont {Iteney}\ \emph {et~al.}(2024)\citenamefont {Iteney}, \citenamefont {{Gonzalez Joa}}, \citenamefont {{Le Bourlot}}, \citenamefont {Cornelius}, \citenamefont {Thomas},\ and\ \citenamefont {Amodeo}}]{pyrough}%
  \BibitemOpen
  \bibfield  {author} {\bibinfo {author} {\bibfnamefont {H.}~\bibnamefont {Iteney}}, \bibinfo {author} {\bibfnamefont {J.~A.}\ \bibnamefont {{Gonzalez Joa}}}, \bibinfo {author} {\bibfnamefont {C.}~\bibnamefont {{Le Bourlot}}}, \bibinfo {author} {\bibfnamefont {T.~W.}\ \bibnamefont {Cornelius}}, \bibinfo {author} {\bibfnamefont {O.}~\bibnamefont {Thomas}},\ and\ \bibinfo {author} {\bibfnamefont {J.}~\bibnamefont {Amodeo}},\ }\bibfield  {title} {\bibinfo {title} {Pyrough: A tool to build 3d samples with rough surfaces for atomistic and finite-element simulations},\ }\href {https://doi.org/https://doi.org/10.1016/j.cpc.2023.108958} {\bibfield  {journal} {\bibinfo  {journal} {Computer Physics Communications}\ }\textbf {\bibinfo {volume} {295}},\ \bibinfo {pages} {108958} (\bibinfo {year} {2024})}\BibitemShut {NoStop}%
\bibitem [{\citenamefont {S\"a\"askilahti}\ \emph {et~al.}(2016)\citenamefont {S\"a\"askilahti}, \citenamefont {Oksanen}, \citenamefont {Tulkki},\ and\ \citenamefont {Volz}}]{Thermalflux}%
  \BibitemOpen
  \bibfield  {author} {\bibinfo {author} {\bibfnamefont {K.}~\bibnamefont {S\"a\"askilahti}}, \bibinfo {author} {\bibfnamefont {J.}~\bibnamefont {Oksanen}}, \bibinfo {author} {\bibfnamefont {J.}~\bibnamefont {Tulkki}},\ and\ \bibinfo {author} {\bibfnamefont {S.}~\bibnamefont {Volz}},\ }\bibfield  {title} {\bibinfo {title} {Spectral mapping of heat transfer mechanisms at liquid-solid interfaces},\ }\href {https://doi.org/10.1103/PhysRevE.93.052141} {\bibfield  {journal} {\bibinfo  {journal} {Phys. Rev. E}\ }\textbf {\bibinfo {volume} {93}},\ \bibinfo {pages} {052141} (\bibinfo {year} {2016})}\BibitemShut {NoStop}%
\bibitem [{\citenamefont {Rodgers}\ and\ \citenamefont {Nicewander}(1988)}]{stat-crosscoef}%
  \BibitemOpen
  \bibfield  {author} {\bibinfo {author} {\bibfnamefont {J.~L.}\ \bibnamefont {Rodgers}}\ and\ \bibinfo {author} {\bibfnamefont {W.~A.}\ \bibnamefont {Nicewander}},\ }\bibfield  {title} {\bibinfo {title} {Thirteen ways to look at the correlation coefficient},\ }\href {https://doi.org/10.1080/00031305.1988.10475524} {\bibfield  {journal} {\bibinfo  {journal} {The American Statistician}\ }\textbf {\bibinfo {volume} {42}},\ \bibinfo {pages} {59} (\bibinfo {year} {1988})}\BibitemShut {NoStop}%
\bibitem [{\citenamefont {Li}\ \emph {et~al.}(2023)\citenamefont {Li}, \citenamefont {Chen},\ and\ \citenamefont {Nagayama}}]{cross-cor}%
  \BibitemOpen
  \bibfield  {author} {\bibinfo {author} {\bibfnamefont {X.}~\bibnamefont {Li}}, \bibinfo {author} {\bibfnamefont {W.}~\bibnamefont {Chen}},\ and\ \bibinfo {author} {\bibfnamefont {G.}~\bibnamefont {Nagayama}},\ }\bibfield  {title} {\bibinfo {title} {Interfacial thermal resonance in an sic–sic nanogap with various atomic surface terminations},\ }\href {https://doi.org/10.1039/D3NR00533J} {\bibfield  {journal} {\bibinfo  {journal} {Nanoscale}\ }\textbf {\bibinfo {volume} {15}},\ \bibinfo {pages} {8603} (\bibinfo {year} {2023})}\BibitemShut {NoStop}%
\bibitem [{\citenamefont {Lafon}\ \emph {et~al.}(2023)\citenamefont {Lafon}, \citenamefont {Chennevi\`ere}, \citenamefont {Restagno}, \citenamefont {Merabia},\ and\ \citenamefont {Joly}}]{PhysRevE.Lafon}%
  \BibitemOpen
  \bibfield  {author} {\bibinfo {author} {\bibfnamefont {S.}~\bibnamefont {Lafon}}, \bibinfo {author} {\bibfnamefont {A.}~\bibnamefont {Chennevi\`ere}}, \bibinfo {author} {\bibfnamefont {F.}~\bibnamefont {Restagno}}, \bibinfo {author} {\bibfnamefont {S.}~\bibnamefont {Merabia}},\ and\ \bibinfo {author} {\bibfnamefont {L.}~\bibnamefont {Joly}},\ }\bibfield  {title} {\bibinfo {title} {Giant slip length at a supercooled liquid-solid interface},\ }\href {https://doi.org/10.1103/PhysRevE.107.025101} {\bibfield  {journal} {\bibinfo  {journal} {Phys. Rev. E}\ }\textbf {\bibinfo {volume} {107}},\ \bibinfo {pages} {025101} (\bibinfo {year} {2023})}\BibitemShut {NoStop}%
\bibitem [{\citenamefont {Allen}\ \emph {et~al.}(1999{\natexlab{b}})\citenamefont {Allen}, \citenamefont {Feldman}, \citenamefont {Fabian},\ and\ \citenamefont {Wooten}}]{amor-phonons}%
  \BibitemOpen
  \bibfield  {author} {\bibinfo {author} {\bibfnamefont {P.~B.}\ \bibnamefont {Allen}}, \bibinfo {author} {\bibfnamefont {J.~L.}\ \bibnamefont {Feldman}}, \bibinfo {author} {\bibfnamefont {J.}~\bibnamefont {Fabian}},\ and\ \bibinfo {author} {\bibfnamefont {F.}~\bibnamefont {Wooten}},\ }\bibfield  {title} {\bibinfo {title} {Diffusons, locons and propagons: Character of atomie yibrations in amorphous si},\ }\href {https://doi.org/10.1080/13642819908223054} {\bibfield  {journal} {\bibinfo  {journal} {Philosophical Magazine B}\ }\textbf {\bibinfo {volume} {79}},\ \bibinfo {pages} {1715} (\bibinfo {year} {1999}{\natexlab{b}})}\BibitemShut {NoStop}%
\bibitem [{\citenamefont {Donadio}\ and\ \citenamefont {Galli}(2009)}]{pop-ration}%
  \BibitemOpen
  \bibfield  {author} {\bibinfo {author} {\bibfnamefont {D.}~\bibnamefont {Donadio}}\ and\ \bibinfo {author} {\bibfnamefont {G.}~\bibnamefont {Galli}},\ }\bibfield  {title} {\bibinfo {title} {Atomistic simulations of heat transport in silicon nanowires},\ }\href {https://doi.org/10.1103/PhysRevLett.102.195901} {\bibfield  {journal} {\bibinfo  {journal} {Phys. Rev. Lett.}\ }\textbf {\bibinfo {volume} {102}},\ \bibinfo {pages} {195901} (\bibinfo {year} {2009})}\BibitemShut {NoStop}%
\bibitem [{\citenamefont {Togo}\ \emph {et~al.}(2023)\citenamefont {Togo}, \citenamefont {Chaput}, \citenamefont {Tadano},\ and\ \citenamefont {Tanaka}}]{phonopy-phono3py-JPCM}%
  \BibitemOpen
  \bibfield  {author} {\bibinfo {author} {\bibfnamefont {A.}~\bibnamefont {Togo}}, \bibinfo {author} {\bibfnamefont {L.}~\bibnamefont {Chaput}}, \bibinfo {author} {\bibfnamefont {T.}~\bibnamefont {Tadano}},\ and\ \bibinfo {author} {\bibfnamefont {I.}~\bibnamefont {Tanaka}},\ }\bibfield  {title} {\bibinfo {title} {Implementation strategies in phonopy and phono3py},\ }\href {https://doi.org/10.1088/1361-648X/acd831} {\bibfield  {journal} {\bibinfo  {journal} {J. Phys. Condens. Matter}\ }\textbf {\bibinfo {volume} {35}},\ \bibinfo {pages} {353001} (\bibinfo {year} {2023})}\BibitemShut {NoStop}%
\bibitem [{\citenamefont {Moon}\ \emph {et~al.}(2018)\citenamefont {Moon}, \citenamefont {Latour},\ and\ \citenamefont {Minnich}}]{Moon-PhysRevB.97.024201}%
  \BibitemOpen
  \bibfield  {author} {\bibinfo {author} {\bibfnamefont {J.}~\bibnamefont {Moon}}, \bibinfo {author} {\bibfnamefont {B.}~\bibnamefont {Latour}},\ and\ \bibinfo {author} {\bibfnamefont {A.~J.}\ \bibnamefont {Minnich}},\ }\bibfield  {title} {\bibinfo {title} {Propagating elastic vibrations dominate thermal conduction in amorphous silicon},\ }\href {https://doi.org/10.1103/PhysRevB.97.024201} {\bibfield  {journal} {\bibinfo  {journal} {Phys. Rev. B}\ }\textbf {\bibinfo {volume} {97}},\ \bibinfo {pages} {024201} (\bibinfo {year} {2018})}\BibitemShut {NoStop}%
\bibitem [{\citenamefont {Beltukov}\ \emph {et~al.}(2013)\citenamefont {Beltukov}, \citenamefont {Kozub},\ and\ \citenamefont {Parshin}}]{Parshin-PhysRevB.87.134203}%
  \BibitemOpen
  \bibfield  {author} {\bibinfo {author} {\bibfnamefont {Y.~M.}\ \bibnamefont {Beltukov}}, \bibinfo {author} {\bibfnamefont {V.~I.}\ \bibnamefont {Kozub}},\ and\ \bibinfo {author} {\bibfnamefont {D.~A.}\ \bibnamefont {Parshin}},\ }\bibfield  {title} {\bibinfo {title} {Ioffe-regel criterion and diffusion of vibrations in random lattices},\ }\href {https://doi.org/10.1103/PhysRevB.87.134203} {\bibfield  {journal} {\bibinfo  {journal} {Phys. Rev. B}\ }\textbf {\bibinfo {volume} {87}},\ \bibinfo {pages} {134203} (\bibinfo {year} {2013})}\BibitemShut {NoStop}%
\bibitem [{\citenamefont {Beltukov}\ \emph {et~al.}(2016)\citenamefont {Beltukov}, \citenamefont {Fusco}, \citenamefont {Parshin},\ and\ \citenamefont {Tanguy}}]{Tanguy-PhysRevE.93.023006}%
  \BibitemOpen
  \bibfield  {author} {\bibinfo {author} {\bibfnamefont {Y.~M.}\ \bibnamefont {Beltukov}}, \bibinfo {author} {\bibfnamefont {C.}~\bibnamefont {Fusco}}, \bibinfo {author} {\bibfnamefont {D.~A.}\ \bibnamefont {Parshin}},\ and\ \bibinfo {author} {\bibfnamefont {A.}~\bibnamefont {Tanguy}},\ }\bibfield  {title} {\bibinfo {title} {Boson peak and ioffe-regel criterion in amorphous siliconlike materials: The effect of bond directionality},\ }\href {https://doi.org/10.1103/PhysRevE.93.023006} {\bibfield  {journal} {\bibinfo  {journal} {Phys. Rev. E}\ }\textbf {\bibinfo {volume} {93}},\ \bibinfo {pages} {023006} (\bibinfo {year} {2016})}\BibitemShut {NoStop}%
\bibitem [{\citenamefont {Taraskin}\ and\ \citenamefont {Elliott}(1999)}]{SNTaraskin_1999}%
  \BibitemOpen
  \bibfield  {author} {\bibinfo {author} {\bibfnamefont {S.~N.}\ \bibnamefont {Taraskin}}\ and\ \bibinfo {author} {\bibfnamefont {S.~R.}\ \bibnamefont {Elliott}},\ }\bibfield  {title} {\bibinfo {title} {Low-frequency vibrational excitations in vitreous silica: the ioffe-regel limit},\ }\href {https://doi.org/10.1088/0953-8984/11/10A/018} {\bibfield  {journal} {\bibinfo  {journal} {Journal of Physics: Condensed Matter}\ }\textbf {\bibinfo {volume} {11}},\ \bibinfo {pages} {A219} (\bibinfo {year} {1999})}\BibitemShut {NoStop}%
\end{thebibliography}
\end{document}


\begin{center}
    \Large{Supplemental Material:}
\end{center}
\begin{center}
    \LARGE{\textbf{Enhanced thermal conductance at interfaces between gold and amorphous
    silicon and amorphous silica }}
\end{center}
\begin{center}
    \large{Julien El Hajj,\textit{$^{a}$} Christophe Adessi,\textit{$^{a}$} Michaël de San Feliciano,\textit{$^{a}$} Gilles Le Doux,\textit{$^{a}$} Samy Merabia\textit{$^{a}$}}
\end{center}
\begin{center}
    \small{\textit{$^{a}$Univ Lyon, Univ Claude Bernard Lyon 1, CNRS, Institut Lumi\`ere Mati\`ere, F-69622, VILLEURBANNE, France}}
\end{center}
\tableofcontents
\clearpage
\section{Interfacial thermal conductance using density functional non-equilibrium Green’s function}
All density functional theory (DFT) calculations have been performed using the \textit{ab initio} package SIESTA~\cite{siesta} with the generalized gradient approximation parameterized by Perdew, Burke, and Ernzerhof~\cite{perdew-prl-96}. We have employed the norm-conserving relativistic pseudo-potential proposed by Troullier and Martin~\cite{troullier-prb-91} and an optimized single-zeta-polarized basis for both gold and silicon. The energy cutoff for the real-space grid used for the electronic density (and for the Hartree and exchange-correlation potentials) is $600$~Ry, combined with a Monkhorst-Pack grid of $3 \times 3 \times 1$ k-points. The atomic structures have been relaxed with a threshold of $10^{-3}$~eV/\AA\ for the forces.\

The system is oriented along the (111) direction ($z$ direction) and consists of 6 gold atomic layers and 6 silicon layers with an ABCABC stacking for both. To reduce the lattice mismatch between gold and silicon, we have used a $4 \times 4$ supercell for gold and a $3 \times 3$ supercell for silicon in the $x$ and $y$ directions. The system comprises a total of $204$ atoms ($96$ Au atoms and $108$ Si atoms), and the cell vectors (hexagonal Bravais lattice) have magnitudes of $a = b = 11.64$\AA\ and $c = 33.29$\AA.\

\begin{figure}[h]
\centering
\includegraphics[width=1.\linewidth]{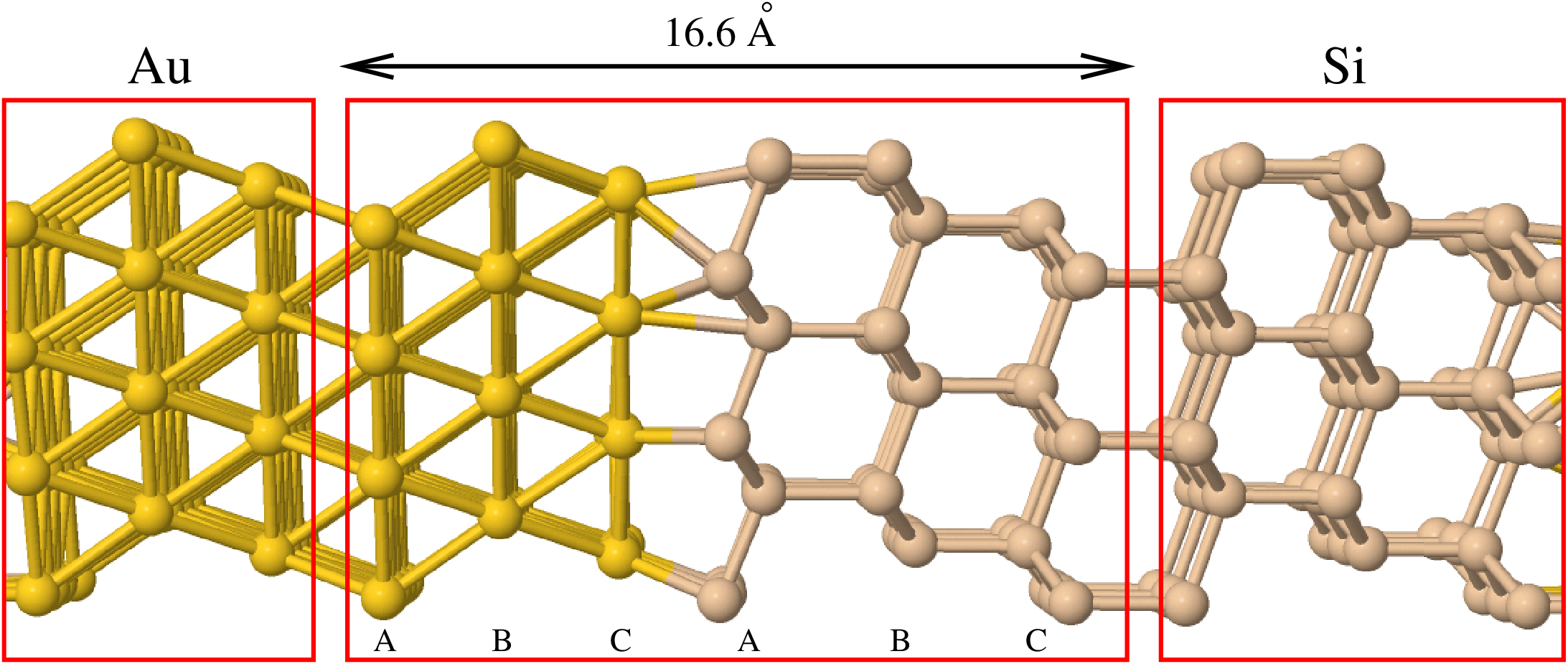}
\caption{Molecular representation of the gold-silicon interface used for the \textit{ab initio} calculation of the interfacial thermal conductance. The central part corresponds to the device used for the atomic Green's function calculation.}
\label{device}
\end{figure}

The force constants and the dynamic matrix used for the transport calculations are computed using the phonopy package~\cite{phonopy}. The interfacial thermal conductance is computed within the Landauer formalism using the atomic Green's function technique~\cite{adessi-pccp-2020}. In this formalism, the central region (the device) corresponds to the gold-silicon interface, the left contact to bulk gold, and the right contact to bulk silicon, as depicted in Fig.~\ref{device}.

The interfacial thermal conductance is derived at quasi-equilibrium and is computed using the formula:

\begin{equation}
G=\frac{k_B^2T}{h}\bigintss\limits_{\Scale[0.75]{0}}^{\Scale[0.75]{+\infty}}
\left(\frac{\hbar \omega}{k_BT}\right)^2 \,
\pazocal{T}_{ph}(\omega)
\left(\frac{\partial f_{\Scale[0.6]{BE}}}{\partial \omega}\right)\,,
\end{equation}

where $h$ is the Planck constant, $k_B$ the Boltzmann constant, $T$
the temperature and $f_{\Scale[0.6]{BE}}(\omega)$ the Bose-Einstein
distribution.  $\pazocal{T}_{ph}(\omega)$ is the phonon
transmission. It is computed within the AGF formalism using the 
equation:

\begin{equation}
\pazocal{T}_{ph}(\omega)=tr\left[\Gamma_L G^r \Gamma_R G^a\right]\,,
\end{equation}

where $G^{r(a)}$ is the retarded (advanced) Green's function and
$\Gamma_{L(R)}$ the coupling between the left (right) lead and the
device.
\section{Radial distribution function for amorphous silicon/silica}
\begin{figure}[H]
        \centering
        \includegraphics[width=.6\linewidth]{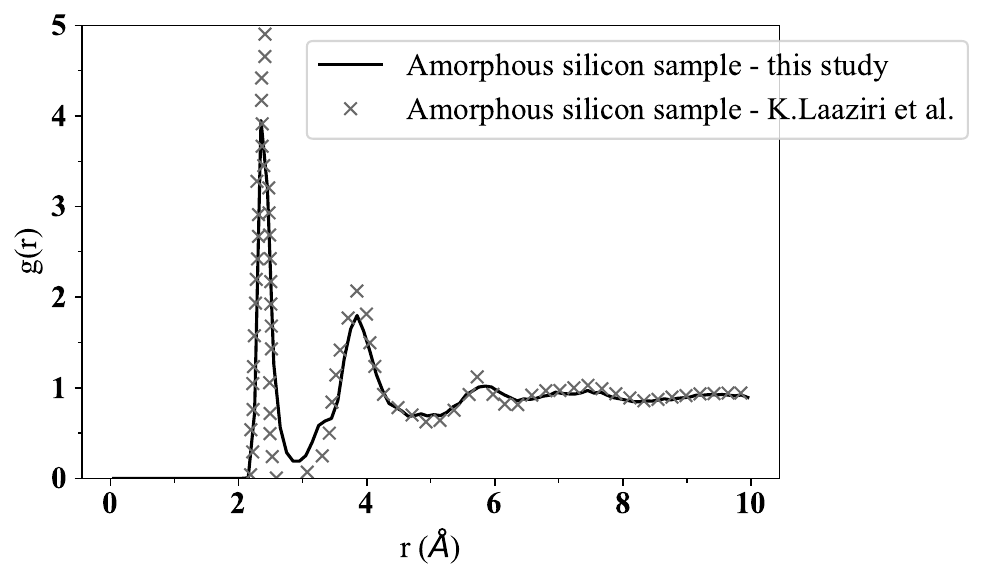}
        \includegraphics[width=.6\linewidth]{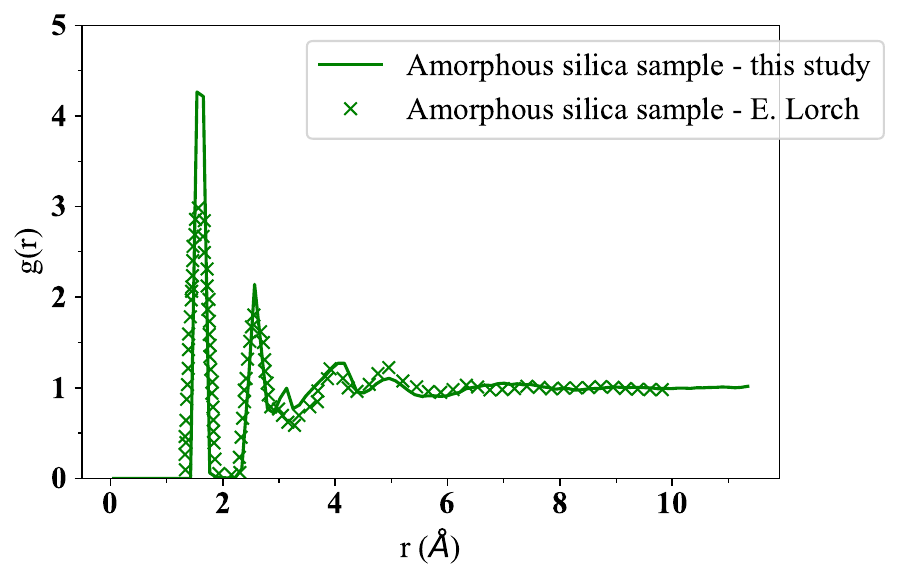}
        \caption{Radial distribution function of a bulk amorphous silicon/silica sample obtained using the melt-quench technique compared to the one obtained from the experimental study \parencite[]{silicon-exp-rdf,ELorch_1969}, showing that both our samples and previously reported experimental ones are in reasonable agreement.}
        \label{fig:rdf}
\end{figure}

Amorphous structures are generated using the melt-quench technique as explained in the main text. We validate our method by comparing the computed Radial Distribution Function to previous measurements as shown in Figure \ref{fig:rdf}.

\section{Heat flux steady state situation}
Each system represents two cubic structures of gold and crystalline/amorphous silicon or silica previously prepared, 50 {\AA} side length each, positioned next to one another for a total system length of 100 {\AA} along the z direction.
\noindent In order to generate and calculate ITC, a temperature gradient within the system should be created. First, the entire system is equilibrated within 100 ps at constant temperature and constant pressure. Then, the system is divided into a few sections in the direction of the temperature gradient. Hot and cold baths are placed in the middle of the left and right sections with a width equal to the lattice parameter of the respective material unit cell at ambient conditions along z, and are connected to the thermostat to create a temperature gradient and heat flow. We simulate all systems for temperature differences of 20 to 50 K between hot and cold baths, which lead to a variation in heat flux in the longitudinal direction.
\noindent Once the thermal baths are fixed, a known amount of energy enters the hot bath and a corresponding amount exits the cold bath. The average temperature remains constant at or near 300 K, and the net energy balance between input and output remains close to zero, indicating that the system has reached a steady state.

\section{Influence of the system length}

\begin{table}[h]
    \centering
    \begin{tabular}{ccc}
    \toprule
         &\textbf{System length (nm)}&\textbf{ITC (MW/m$^2$K)}\\
        \midrule
         \textbf{Gold/Crystalline Silicon} &\textbf{10}&  56 $\pm$ 8 \\
    
         &\textbf{20} & 60  $\pm$ 7 \\  
        \midrule
         \textbf{Gold/Amorphous Silicon} &\textbf{10}&  152 $\pm$ 13 \\
           
         &\textbf{20} & 160 $\pm$ 10 \\   
         \midrule 
         \textbf{Gold/Crystalline Silica} &\textbf{10}&  102 $\pm$ 12 \\
    
         &\textbf{20} & 105  $\pm$ 9 \\  
        \midrule
         \textbf{Gold/Amorphous Silica} &\textbf{10}&  170 $\pm$ 14 \\
    
         &\textbf{20} & 177 $\pm$ 12 \\  
    \bottomrule
    \end{tabular}
    \captionsetup{width=0.9\linewidth}
    \caption{Values of the interfacial thermal conductance for a 10 nm and 20 nm gold silicon/silica systems.}
    \label{tab:lentsable}
\end{table} 
\noindent Table \ref{tab:lentsable} shows results for two different system lengths in order to investigate the variation in ITC as a function of system length. The presented results show that doubling the system length only increases ITC by 5-6\% for the four interfaces. This implies that the effect of system length is minor, and thus the results are almost size independent.

\section{Temperature difference between cold and hot baths}

\begin{figure}[H]
    \centering
    \includegraphics[width=0.6\textwidth]{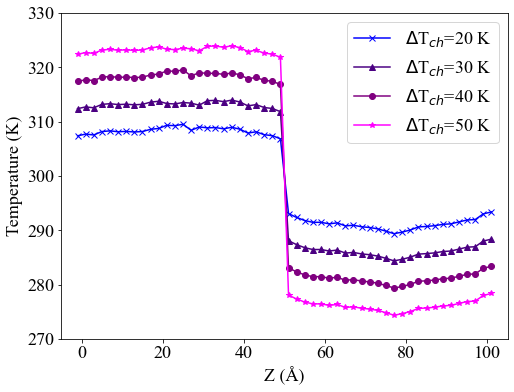}
    \caption{Temperature profiles for different  temperature difference between the cold and hot baths at the crystalline silicon/gold interface.}
    \label{fig:compdeltatemp}
\end{figure}

\begin{figure}[H]
    \centering
    \includegraphics[width=0.9\textwidth]{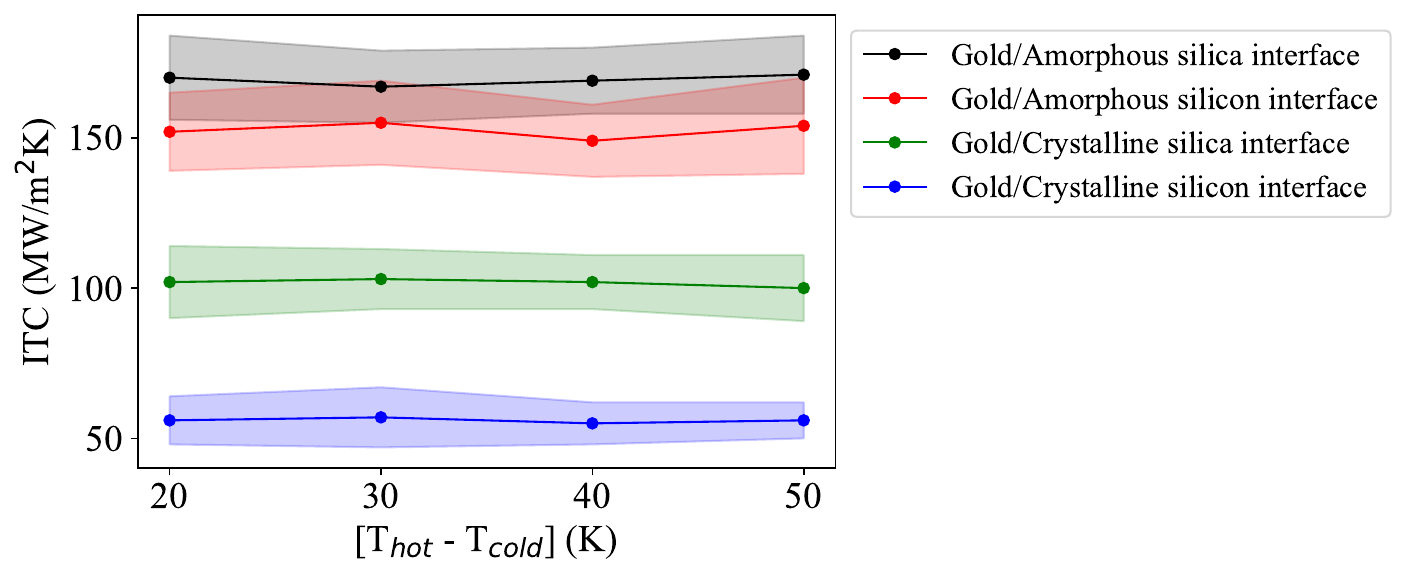}
    \caption{ITC variation as a function of the temperature difference between the cold and hot baths.}
    \label{fig:deltatemp}
\end{figure}

All systems are simulated at different values of the temperature difference between cold and hot baths ranging from 20 to 50 K, with Figure\ref{fig:compdeltatemp} showing the temperature profiles of the crystalline silicon/gold systems for various applied heat fluxes, where $\Delta$T$_{ch}$ represents the difference of temperature between the hot and cold thermostats. The ITC values are displayed in Figure \ref{fig:deltatemp}, in order to investigate the effects of the temperature difference between the cold and hot bath on ITC. It is clear that the changes in ITC are minor as temperature difference values between cold and hot baths varies, making the ITC's value independent of the temperature difference between the cold and hot baths.


\section{Failure of the diffuse mismatch model} \label{section:DMM}
\begin{figure}[H]
    \centering
    \includegraphics[width=0.96\textwidth]{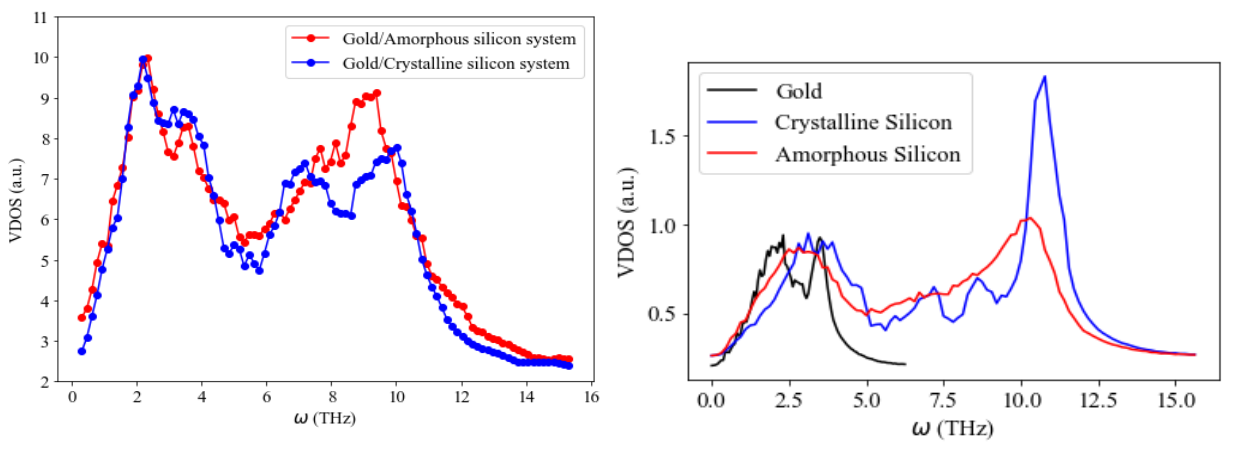}

    \caption{The vibrational density of states in the vicinity of the gold/crystalline silicon and gold/amorphous silicon interfaces, and for gold, crystalline silicon and amorphous silicon bulks as a function of the angular frequency.}
    \label{fig:4}
\end{figure}

We discuss now whether the ITC enhancement observed at the gold/silicon interfaces can be captured by the diffuse mismatch model (DMM) \cite[]{DMM} and compare the obtained results with those calculated from NEMD simulations. 
In Landauer framework of interfacial transport, the phonon–phonon interfacial thermal conductance is given by~\cite[]{landau}:
$$
\begin{aligned}
& G_{p}=\frac{1}{2} \sum_{p} \int_{0}^{\omega_{A u, p}^{c}} v_{A u, p} \hbar \omega D_{A u, p}(\omega) \frac{\partial f_{e q}(\omega, T)}{\partial T} \\
& \times \int_{0}^{\pi / 2} \alpha_{p}(\omega, \cos \theta) \cos \theta d(\cos \theta) d \omega
\end{aligned}
$$

\noindent where the sum runs over the polarizations \textit{p}, $\omega_{A u, p}^{c}$ denotes the cutoff frequency for gold, $v_{\rm Au, p}$ is the group velocity of gold in the mode $p$, $D_{\rm Au, p}(\omega)$ represents the bulk density of states of gold, $f_{\rm eq}$ signifies the equilibrium Bose-Einstein distribution, $\alpha$ refers to phonon transmission coefficient, and in the frame of diffuse interfacial scattering, the transmission coefficient no longer depends on the scattering angle and is assumed to be polarization independent, it can be calculated as \cite[]{DMM}:

$$
\alpha=\frac{\sum_{p} \frac{1}{v_{S i}^{2}}}{\sum_{p} \frac{1}{v_{S i}^{2}}+\sum_{p} \frac{1}{v_{A u}^{2}}}
$$
The longitudinal and transverse values of gold mode group velocity are $3390$ m/s and $1290$ m/s, respectively. The mode group velocities of bulk silicon structures, both crystalline and amorphous, are extracted from the calculation of the dynamical structure factor of each bulk using the Dynasor library \cite[]{dynasor}. Using the DMM formalism described above, the ITC of gold/crystalline silicon is found to be equal to $82.7$ (MW/m$^2$K) which is in fair agreement with previous calculated value using DMM ($\approx$ $86$ MW/m$^2$K) \cite[]{landau}, while the ITC of gold/amorphous silicon is $110$ (MW/m$^2$K). This small difference between the ITC values is expected since we found there are no significant differences in the computed vibrational density of states at the two interfaces, see Figure \ref{fig:4} \\
\noindent The findings show a significant difference in the outcomes obtained using the NEMD and DMM approaches. DMM not only predicts a higher ITC value for the gold/crystalline silicon interface, but it also does not predict a threefold difference in comparison to the gold/amorphous silicon interface.  This significant disparity can be attributed to the simplicity of DMM assumptions, particularly the use of a transmission coefficient which ignores the interface structure, as opposed to NEMD's comprehensive treatment of the interface and inelastic phonon transport. 
The vibrational density of state (VDOS) of the materials involved in these interfaces is calculated to understand the ITC increase. The VDOS is calculated from the velocity auto-correlation function computed from the MD trajectories as: \begin{equation}
     Z(\omega) \propto \int_{0}^{\tau} C_{v}(t) e^{i\omega t} \hspace{0.1 cm} dt
\end{equation}
Where $\tau$ is the time range of the simulation and $C_{v}(t)$ is the velocity auto-correlation function defined as:
\begin{equation}
    C_{v}(t) = < \Vec{v}_{i}(0) . \Vec{v}_{i}(t) >
\end{equation}
Where v(t) is the velocity at time t and the origin of the autocorrelation is 0. \\
\noindent  Figure \ref{fig:4} shows that the VDOS at both interfaces are quite similar across the entire frequency range, despite the significant differences in the VDOS of the bulk materials. This similarity at the interfaces explains why the ITC values calculated by the diffuse mismatch model are fairly comparable.

\section{Coupling strength effect on the interfacial thermal conductance}
\begin{figure}[H]
    \centering
    \includegraphics[width=0.45\textwidth]{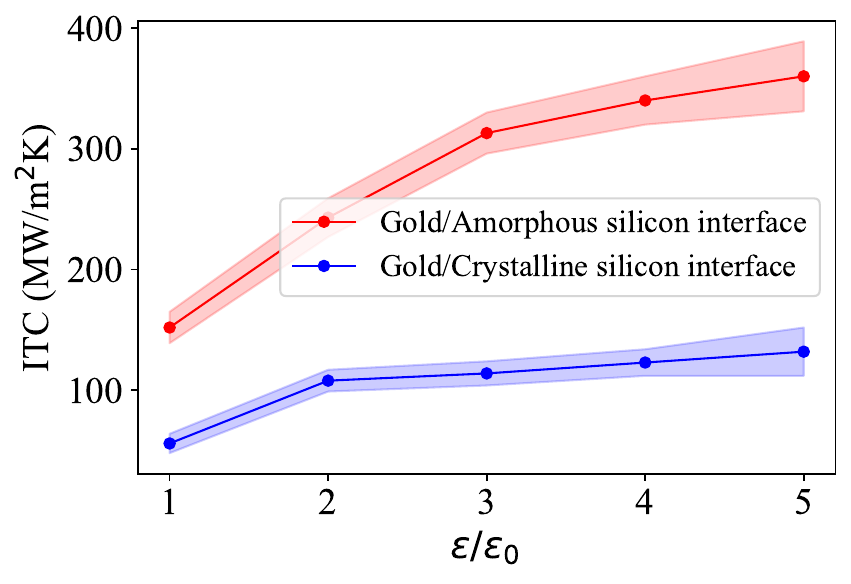}
    \includegraphics[width=0.45\textwidth]{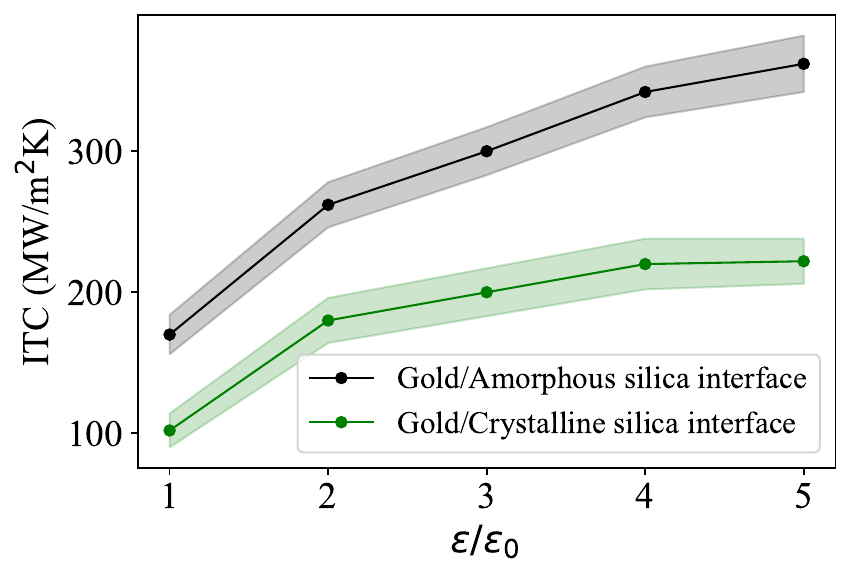}
    \caption{ITC as a function of the normalized dispersion energy $\epsilon/{\epsilon_0}$ for the gold/silicon and gold/silica systems.}
    \label{fig:ep-change}
\end{figure}

To investigate the effect of coupling strength between gold and both silicon and oxygen atoms, Figure \ref{fig:ep-change} presents the interfacial thermal conductance as a function of the normalized dispersion energy $\epsilon/{\epsilon_0}$ for the considered systems. The results clearly show that ITC is directly influenced by the coupling strength. Notably, the threefold increase in ITC for the gold-silicon interface and the twofold increase for the gold-silica interface are continuous across the entire range of coupling strengths considered.

\section{Projection of Static Structure Factor from 2D to 1D}
\begin{figure}[H]
    \centering
    \includegraphics[width=0.45\textwidth]{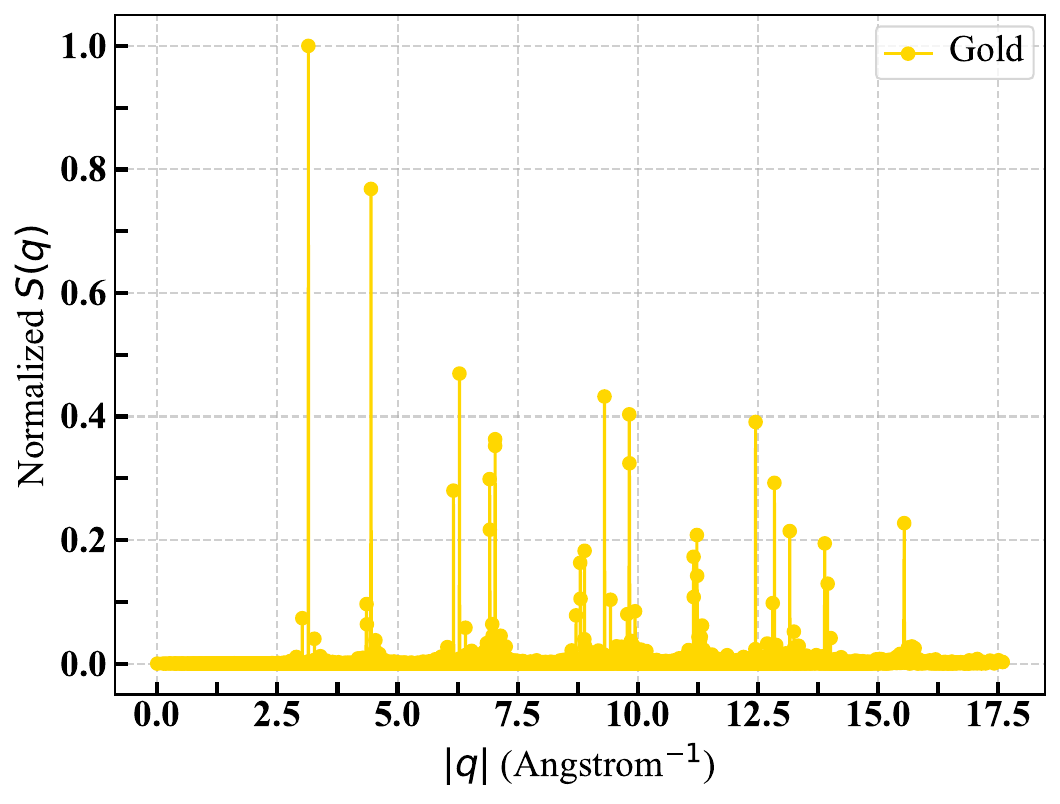}
    \includegraphics[width=0.45\textwidth]{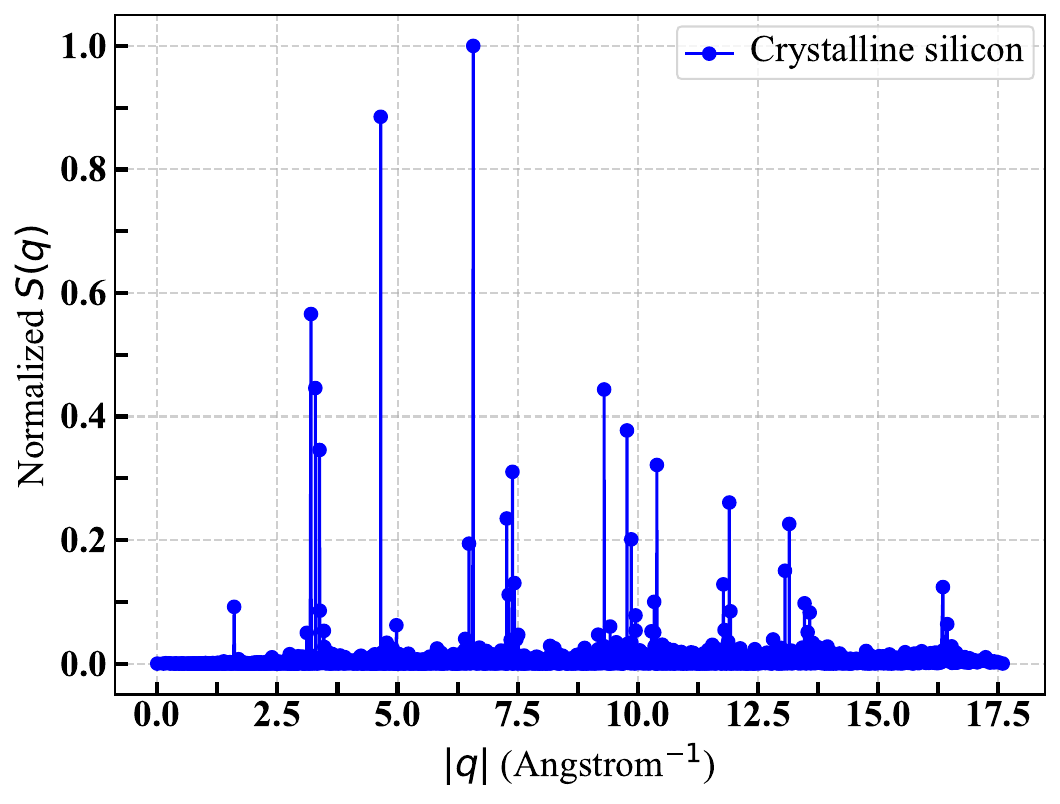}
    \includegraphics[width=0.45\textwidth]{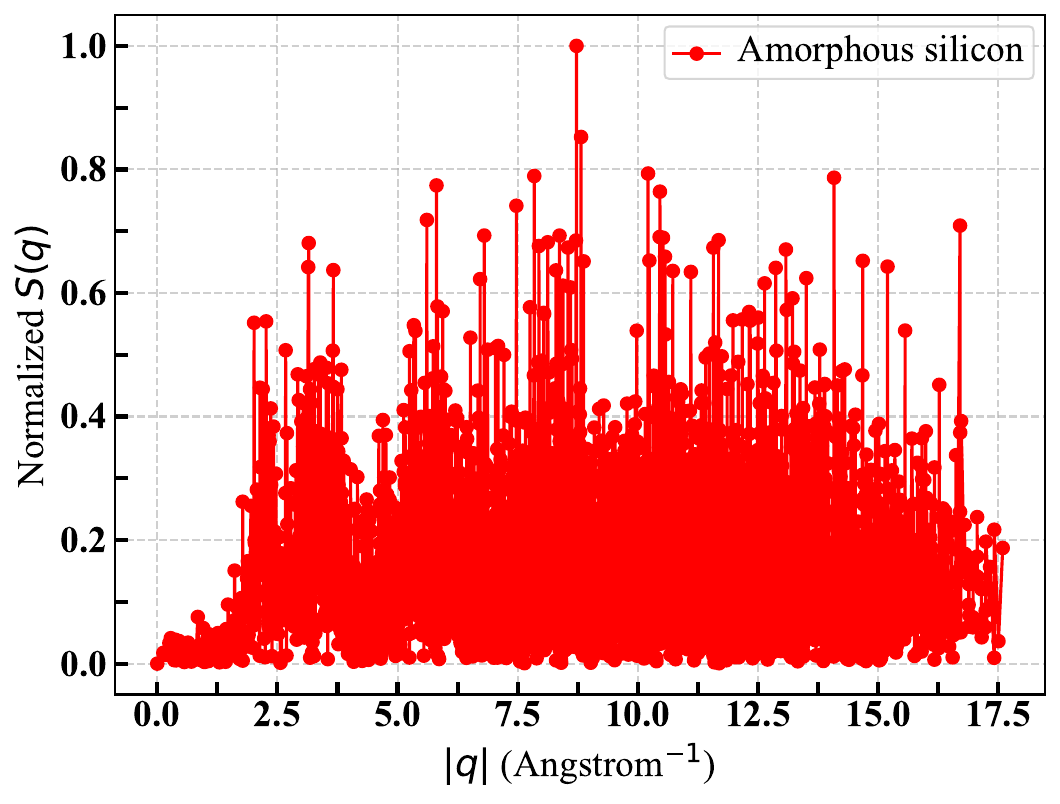}
    \includegraphics[width=0.45\textwidth]{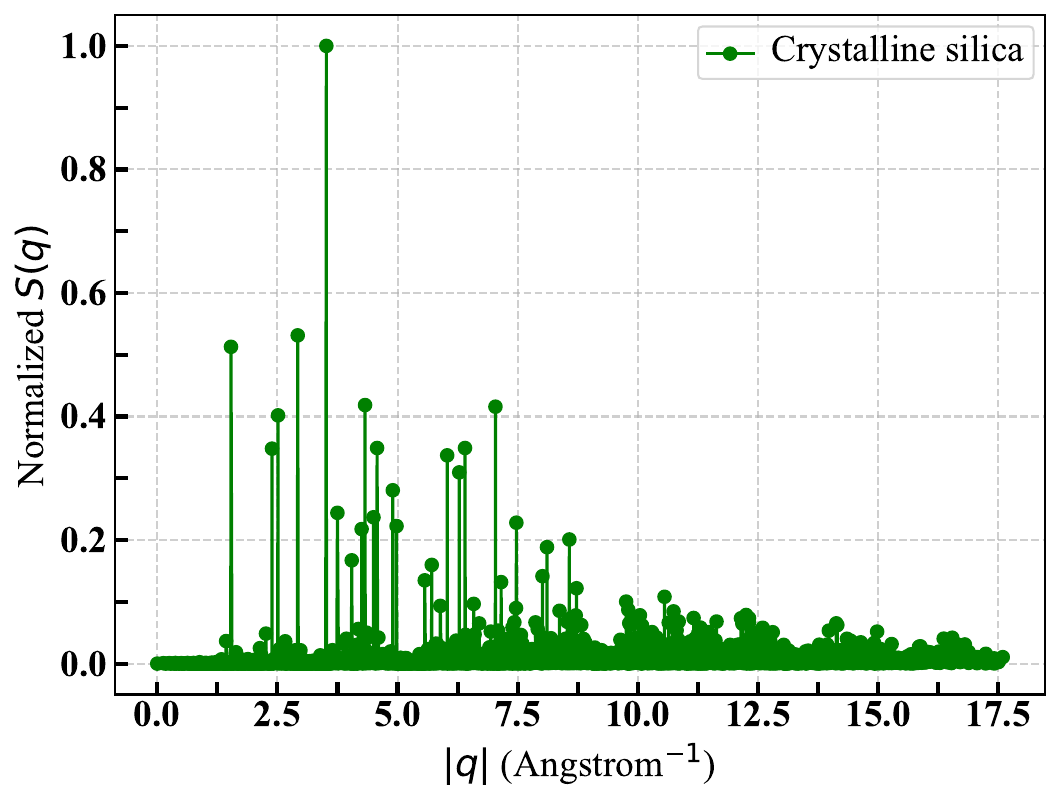}
    \includegraphics[width=0.45\textwidth]{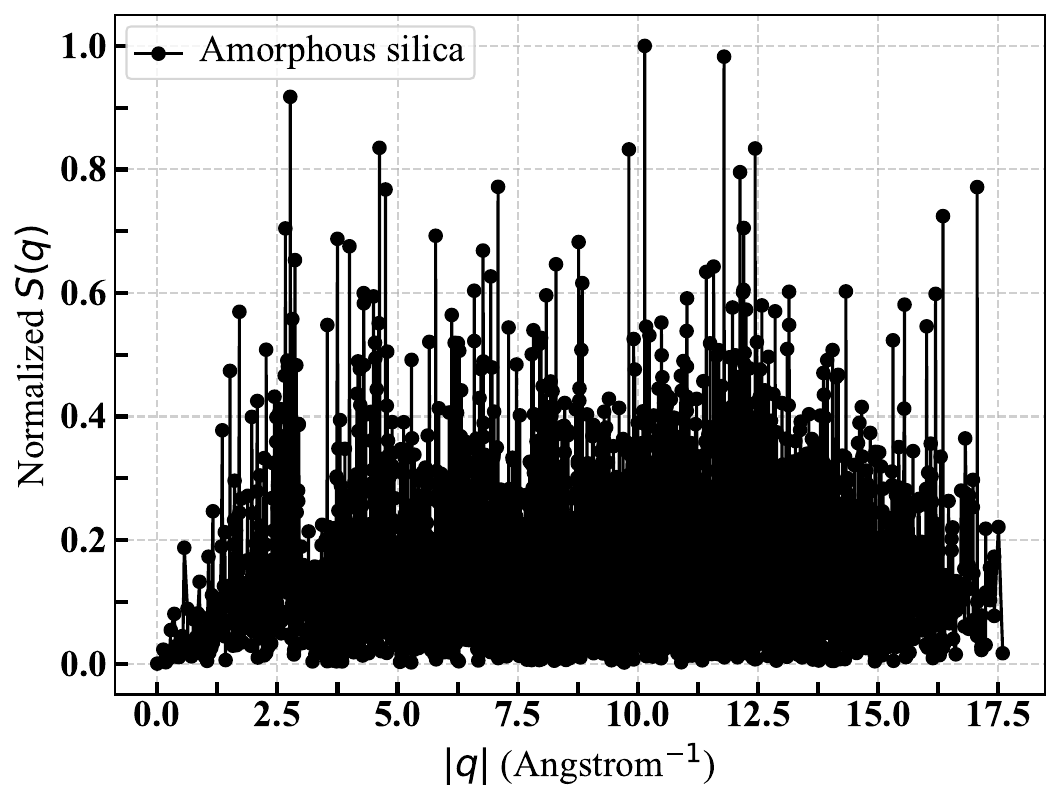}
    \caption{Normalized Static Structure Factor derived from the two-dimensional structure factor S(q) of the first wall of atoms at the interfacial layers of gold, crystalline silicon, amorphous silicon, crystalline silica, and amorphous silica}
    \label{fig:!D-static}
\end{figure}

The one-dimensional static structure factor, $S(q)$, is derived from the two-dimensional data using the following computational approach. We begin by defining the system size, and calculating $q_x$ and $q_y$ vectors using $q_x = n \cdot \frac{2\pi}{L}$ and $q_y = m \cdot \frac{2\pi}{L}$, where $n$ and $m$ are positive integers and $L$ the size of the simulation box in the $x$ and $y$ directions. For each $q$ vector, the static structure factor $S(q)$ is computed using the function $S_q(x, y, qx, qy)$, which sums the cosine and sine components of the dot product between position vectors $(x, y)$ and wave vectors $(qx, qy)$. The resulting structure factor matrix $S_{\text{matrix}}$ is filled by iterating over all combinations of $q_x$ and $q_y$. To obtain the one-dimensional projection, we calculate the magnitude $|q| = \sqrt{q_x^2 + q_y^2}$ for each $q$ vector and store the maximum $S(q)$ value for each unique $|q|$. The final one-dimensional $S(q)$ values are plotted as a function of their corresponding $|q|$ magnitudes in Figure \ref{fig:!D-static} for the first wall of atoms at the interfacial layers of gold, crystalline silicon, amorphous silicon, crystalline silica, and amorphous silica, respectively. 

\section{Dynamical structure factor and the Ioffe-Regel criterion for bulk amorphous silicon and silica.}
\begin{figure}[H]
    \centering
    \includegraphics[width=0.45\textwidth]{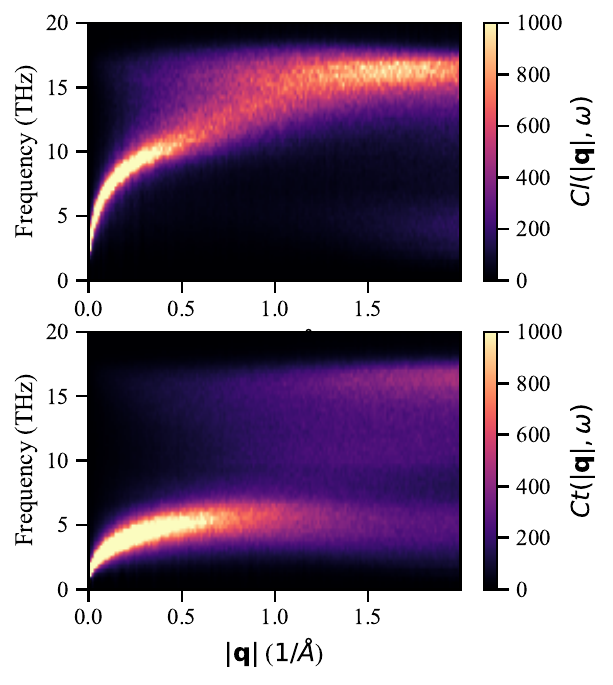}
    \includegraphics[width=0.45\textwidth]{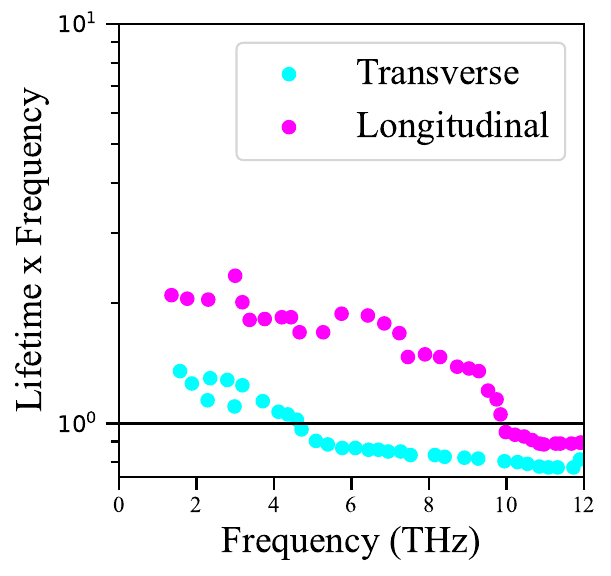}
    \caption{Dynamical structure factor and the lifetime multiplied by frequency versus frequency for transverse and longitudinal waves for the 6346-atoms amorphous silicon bulk.}
    \label{fig:B-IR-silicon}
\end{figure}

\begin{figure}[H]
    \centering
    \includegraphics[width=0.45\textwidth]{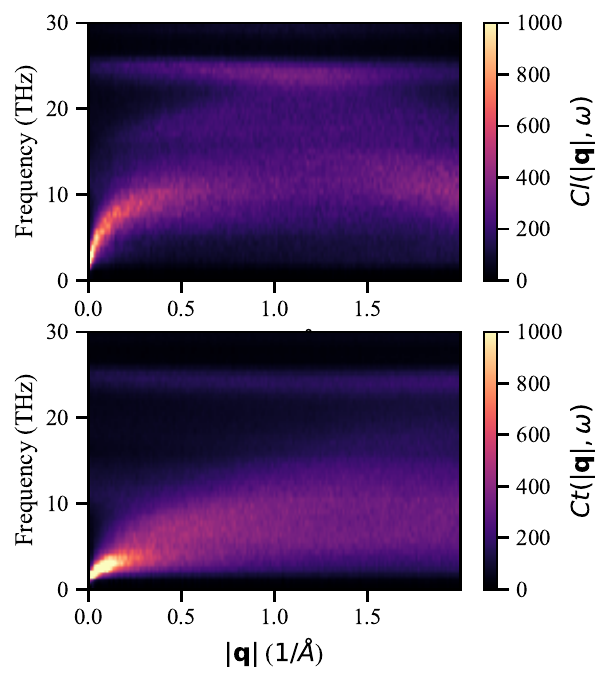}
    \includegraphics[width=0.45\textwidth]{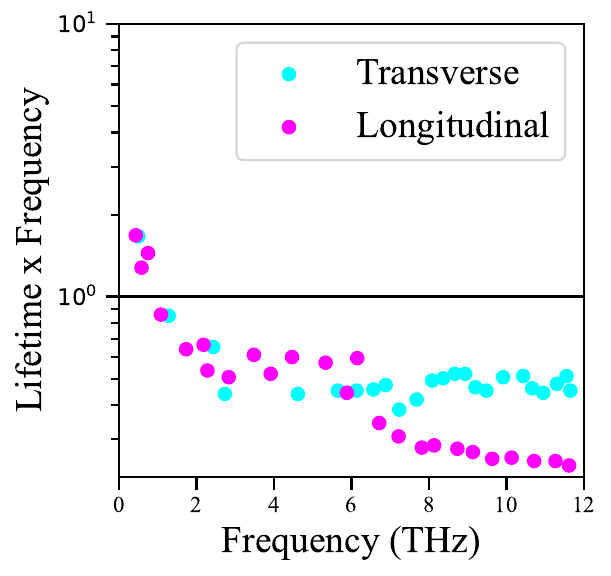}
    \caption{Dynamical structure factor and the lifetime multiplied by frequency versus frequency for transverse and longitudinal waves for the 8400-atom amorphous silica bulk.}
    \label{fig:B-IR-silica}
\end{figure}

Here, we compare the dynamical structure factor and the Ioffe-Regel criterion between amorphous walls and bulk materials. Specifically, we consider a bulk amorphous silicon consisting of 6346 atoms and an amorphous silica bulk consisting of 8400 atoms. Our analysis reveals distinct differences between the bulk and interfacial properties. For the bulk amorphous silicon, as depicted in Figure \ref{fig:B-IR-silicon}, the discernible dispersion occurs at lower frequencies compared to the interfacial regions, with transverse waves showing dispersion up to 5 THz and longitudinal waves up to 10 THz. The Ioffe-Regel (IR) crossover for transverse waves in bulk amorphous silicon is approximately 5 THz, while for longitudinal waves, it is around 10 THz. These findings align well with previously reported values \cite[]{Moon-PhysRevB.97.024201,Tanguy-PhysRevE.93.023006}. Similarly, for the bulk amorphous silica, shown in Figure \ref{fig:B-IR-silica}, both transverse and longitudinal dispersions are narrower compared to their interfacial counterparts. The IR crossover for both transverse and longitudinal waves in bulk amorphous silica occurs at about 1 THz, which also agrees closely with previously reported data\cite[]{SNTaraskin_1999}. These observations highlight the differences in dynamical properties between bulk and interfacial regions in the amorphous materials considered. 

\printbibliography